\documentclass[12pt]{article}

\usepackage{classeJBArxive}
\usepackage{amssymb,amsmath,latexsym,array,verbatim}
\usepackage{nomencl}
\makenomenclature
\usepackage{graphicx}
\usepackage{subfigure}
\usepackage{verbatim} 
\usepackage{mathtools}
\usepackage{wrapfig,lipsum}
\usepackage{booktabs}
\usepackage[colorlinks]{hyperref}
\usepackage{hyperref}
\usepackage{placeins}

\usepackage[font={small,it}]{caption}

\renewcommand{\leq}{\leqslant}
\renewcommand{\geq}{\geqslant}
 
\let\norm\relax
\DeclarePairedDelimiterX{\norm}[1]{\lVert}{\rVert}{#1}
\definecolor{darkviolet}{rgb}{0.31, 0, 0.51}
\newcommand*\ainagul[1]{\leavevmode\textcolor{black}{#1}}
\newcommand*\ajumabek[1]{\leavevmode\textcolor{black}{#1}}

\newcommand{\DF}{\textsc{Dufort}--\textsc{Frankel}}
\definecolor{orcidlogocol}{HTML}{A6CE39}

\usepackage{etoolbox}
\renewcommand\nomgroup[1]{%
  \item[\bfseries
  \ifstrequal{#1}{A}{Latin letters}{%
  \ifstrequal{#1}{B}{Greek letters}{%
  \ifstrequal{#1}{C}{Dimensionless parameters}{}}}%
]}

\newenvironment{answer}{%
    \setlength{\parindent}{0pt}
    \color{black}
}{}

\newenvironment{BS_Answer}{%
    \setlength{\parindent}{0pt}
    \color{black}
}{}

\title{\vspace{-1.5cm}
\ajumabek{An efficient sensitivity analysis for energy performance of a building envelope}: a continuous derivative based approach\\	
\vspace{4pt}}

\author{Ainagul Jumabekova\textsuperscript{a,b}$^{\ast}$, Julien Berger\textsuperscript{a}, 
{Aurélie Foucquier\textsuperscript{b}} }

\begin{document}
\maketitle

\begin{center}
\small
\textsuperscript{a} Univ. Grenoble Alpes, Univ. Savoie Mont Blanc, UMR 5271 CNRS, LOCIE, 73000 Chambéry, France,  \\
\textsuperscript{b} LaSIE, La Rochelle University, CNRS, UMR 7356, 17000 La Rochelle, France,\\ 
\textsuperscript{b} Univ. Grenoble Alpes, CEA, LITEN, DTS, INES, F-38000, Grenoble, France,\\
$^{\ast}$corresponding author, e-mail address : ainagul.jumabekova@univ-smb.fr,\\ ORCiD : 0000-0001-5554-4249\\
\end{center}

\begin{abstract}

Within the framework of building energy assessment, this article proposes \ajumabek{to use a derivative based sensitivity analysis of heat transfer models in a building envelope. Two, global and local, estimators are \ajumabek{obtained at low computational cost, } to evaluate the influence of the parameters on the model outputs.  Ranking of these estimators values allows to reduce the number of model unknown parameters by excluding non-significant parameters.} A comparison with variance and regression-based methods is carried out and the results highlight the satisfactory accuracy of the continuous-based approach. Moreover, for the carried investigations the approach \ajumabek{is $100$ times faster compared to the variance-based methods.} A case study applies the method to a real-world building wall. The sensitivity of the thermal loads to local or global variations of the \ajumabek{wall thermal is investigated. Additionally, a case study of wall with window is analyzed.}

parameter estimation problem ; \DF ~numerical scheme; sensitivity coefficients

\end{abstract}

\section{Introduction}

Within the context of environmental sustainability, retrofit solutions help to decrease the heating and cooling demand of building stocks. In order to evaluate the efficiency of retrofit actions, engineers usually conduct energy audits of buildings. Numerical simulations are often used to analyze the energy performance of buildings according to the retrofit scenarios. However, despite the recent advances in building energy simulation programs, the gap between the actual and the predicted energy consumption still remains a challenging issue.
In existing buildings, this discrepancy may arise from uncertainties in the thermophysical properties of building materials~\citep{Hughes_2015}. The properties change due to \textit{(i)} time degradation, \textit{(ii)} exposure to weather conditions, and \textit{(iii)} traditional processes of construction. Therefore, the estimation of the \ainagul{actual} material properties will improve the accuracy and reliability of building simulation software.

\ajumabek{It is essential to have models that evaluate the impact of input parameters variability on the physical phenomena predictions. It enables to assess the sensitivity of the important fields such as the temperature, the heat flux or the thermal conduction loads according to uncertain material properties. This is particularly relevant in the context of inverse problems and the objective of retrieving the unknown parameters. The number of parameters to be estimated using \emph{in-situ} measurement can be reduced by focusing on the most influencing ones. Besides, it reduces significantly the computational requirements of the inverse problems~\citep{beck_1977}.}

\ajumabek{The use of sensitivity analysis is an attractive approach to highlight the main parameters and their impact on the interesting model outputs.} In the literature, there are several studies that applied sensitivity analysis methods for building energy simulation. An extensive overview is given in \citep{TIAN_2013}. Among others, uncertainty analysis is carried out in \citep{Goffart_2017} to assess the influence of brick material properties on the cooling energy demand of a building. In \citep{Mechri_2010} the design variables that have the most impact on the building energy performance of a typical office building are determined. Sensitivity analysis for an office building in Denmark is conducted in \citep{Heiselberg_2009} to identify the most important design parameters for heating, cooling, and total building energy demand. Multiple building parameters are selected to examine their impact on annual building energy consumption, peak design loads, and building load profiles of office buildings in Hong Kong in \citep{LAM_1996}. An airflow network model~\citep{Monari_2017} is characterized through the combination of the \textsc{Morris} method and the \textsc{Sobol} method: this approach reduced the dimensionality of the model in a more informative way than the standard methods. Uncertainty propagation from over $1000$ parameters is studied by simulation--based sensitivity analysis in~\citep{Eisenhower_2012} to identify the most influential parameters using support vector regression. 

One may note that variance--based sensitivity analysis methods are the most popular choice quantifying the influence of a parameter, while the \textsc{Sobol} sensitivity indices are often used as reference values~\citep{Borgonovo2017}. However, the variance--based methods require a high number of model evaluations. To calculate sensitivity indicators of $M$ parameters through variance decomposition, one may need to make $\mathcal{O}\,(\,100\,\cdot\,M\,)$ model evaluations~\citep{Iooss2015}. \ajumabek{It can be remarked that a promising method, called the derivative based approach, has not been investigated yet for building energy simulation~\citep{Sobol2009,Kucherenko_2015,Kucherenko_2016,Molkenthin_2017}. It is based on the computation of the so-called sensitivity coefficients obtained from direct differentiation of the governing equations according to the input parameters.}
\ajumabek{A global estimator has been proposed and linked to the total \textsc{Sobol} indices by \textsc{Kucherenko and Sobol} in \citep{Sobol2009}. Several studies \citep{DUNKER1981,DUNKER1984,yang1997} across different disciplines suggest the efficiency of such approach. Thus, this article aims at investigating the derivative based approach to carry out sensitivity analysis of the heat transfer models in  buildings components based on the variability of several input parameters. Its efficiency in terms of accuracy and computational cost is assessed through a comparison to standard methods.}

The article is organized as follows. Section~\ref{sec:methodology} presents the governing equations and the methodology for computing the sensitivity coefficients using the continuous approach. The discrete differential approach from the literature is also referred to. In Section~\ref{sec:validation}, a case is considered to validate the approach using a reference solution. Last, in Section~\ref{sec:case_study}, a detailed case study of a wall is analyzed. The issue is to determine the influence of the thermal conductivity or volumetric capacity of each layer of the wall on the thermal loads. \ajumabek{Additionally, in Section~\ref{sec:window_eq}, an envelope composed of a single glazed window and a wall, is considered, and influence of wall and window properties on the overall energy performance is studied. }


\section{Methodology}
\label{sec:methodology}

\subsection{Physical model}
\begin{figure}[!ht] 
\centering
\includegraphics[width=0.5\linewidth]{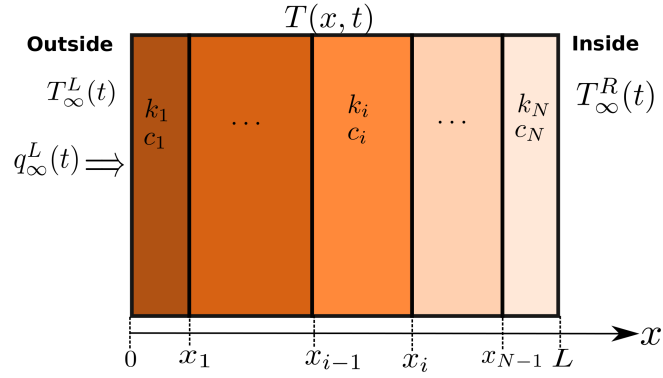}
\caption{Illustration of the wall construction.}
\label{fig:wall_1}
\end{figure}

The physical problem considers one--dimensional heat conduction transfer through a building wall. The wall is composed of $N$ layers, and each layer differs from the other by its thermal properties and thickness, as shown in Figure~\ref{fig:wall_1}. The temperature in the wall is defined on the domains  $\Omega_{\,x}\,:\, x \in [\,0\,,L\,]\,$ and $ \Omega_{\,t}\,:\,t \in [\,0\,,\tau_{\,\mathrm{max}}\,]\,$, where $L\,\mathsf{[\,m\,]}$ is the length of the wall and $\tau_{\,\mathrm{max}} \,\mathsf{[\,s\,]}$ is the duration of the simulation:
\begin{align*}
T\,:\, [\,0\,,L\,]\, \times \,[\,0\,,\tau_{\,\mathrm{max}}\,] \, \longrightarrow \,\mathbb{R}\,.
\end{align*}
The mathematical formulation of the heat transfer process is given as: 
\begin{align}
\label{eq:heat_eq_ph}
c\,\cdot\,\pd{T}{t} \egal \pd{}{x}\,\Biggl(\,k\,\cdot\,\pd{T}{x}\,\Biggr)\,,
\end{align}
where $c\,\mathsf{[\,J/(m^3 \cdot K)\,]}$ is the volumetric heat capacity, \ainagul{ or $c \egal \rho \,\cdot \, c_{\,p}\,$, corresponding to the product between the material density $\rho\,\mathsf{[\,kg/m^3\,]}$  and the specific heat $c_{\,p}\,\mathsf{[\,J/(kg \cdot K)\,]}$,} which varies depending on space:
\begin{align*}
c\,:\, x \, \longmapsto \,\sum_{i=1}^N\,c_{\,i}\,\cdot\,\varphi_{\,i}\,(\,x\,)\,,
\end{align*} 
where $\bigl\{\,\varphi_{\,i}\,(\,x\,)\,\bigr\}_{\,i \egal 1\,,\ldots\,,N\, }$ are piecewise functions and can be written as:
\begin{align*}
\varphi_{\,i}\,(\,x\,) \egal \begin{cases}
1 \,, & x_{\,i-1}\leqslant x \leqslant x_{\,i} \,, \quad i \egal {1\,,\ldots\,,N}\,, \\
0 \,, & \mathrm{otherwise}\,,
\end{cases} \,
\end{align*}
where $x_{\,i-1}$ and $x_{\,i}$ are the left and right interfaces of layer $i$, respectively. A similar definition is assumed for the thermal conductivity $k\,\mathsf{[\,W/(m \cdot K)\,]}$: 
\begin{align*}
k\,:\, x \, \longmapsto \,\sum_{i=1}^N\,k_{\,i}\,\cdot\,\varphi_{\,i}\,(\,x\,)\,.
\end{align*} 
Therefore, equation~\eqref{eq:heat_eq_ph} becomes:
\begin{align}
\sum_{i=1}^N\,c_{\,i}\,\cdot\,\varphi_{\,i}\,(\,x\,)\,\cdot\,\pd{T}{t} \egal \pd{}{x}\,\Biggl(\,\sum_{i=1}^N\,k_{\,i}\,\cdot\,\varphi_{\,i}\,(\,x\,)\,\cdot\,\pd{T}{x}\,\Biggr)\,.
\end{align}
The heat balance on the \ainagul{exterior} wall  includes the convective exchange between the air outside and the wall surface, as well as the absorbed radiation. It can be written as:
\begin{align}
k\,\cdot\,\pd{T}{x} \egal & h_{\,L}\,\cdot\,\Bigl(\,T \moins T_{\,\infty}^{\,L}\,(\,t\,)\,\Bigr)  \moins \alpha\,\cdot\,q_{\,\infty}^{\,L}\,(\,t\,)\,,\qquad x\egal 0\,, 
\end{align}
where $T_{\,\infty}^{\,L}\,\mathsf{[\,K\,]}$ is the temperature of the outside air that varies over time, $h_{\,L}\,\mathsf{[\,W/(m^2 \cdot K)\,]}$ is the exterior convective heat transfer coefficient, and $q_{\,\infty}\,\mathsf{[\,W/m^2\,]}$ is the total incident radiation\ainagul{, which includes the direct, diffuse, and reflexive radiations, and $\alpha$ is the surface absorptivity. The infrared radiation and solar radiation passing through the windows are not considered.}
The interior heat balance consists of the convective exchange between the air inside $T_{\,\infty}^{\,R}\,\mathsf{[\,K\,]}$ and the wall surface, and is given by the following expression:
\begin{align}
k\,\cdot\,\pd{T}{x}  \egal & \moins h_{\,R}\,\cdot\,\Bigl(\,T \moins T_{\,\infty}^{\,R}\,(\,t\,) \,\Bigr)\,,\qquad x\egal L\,,
\end{align}
where $h_{\,R}\,\mathsf{[\,W/(m^2 \cdot K)\,]}$ is the convective heat transfer coefficient on the inside boundary. The initial temperature in the wall is given according to:
\begin{align}
T \egal T_{\,0}\,(\,x\,)\,, \qquad t \egal 0 \,,
\end{align}
\ainagul{where the initial temperature distribution will be set in case studies.}
The outputs of the given physical model are twofold. First, the heat flux  $\boldsymbol{\mathrm{j}}\;\mathsf{[\,W/m^2 \,]}$ is computed according to:
\begin{align}
\mathrm{j} \egal - \, \Biggl(\, k\,\cdot\,\pd{T}{x}\,\Biggr)\,\Bigg|_{\,x \egal L} \,.
\end{align}

In addition, the thermal loads $\boldsymbol{\mathrm{E}}\;\mathsf{[\,W\cdot s/m^2 \,]}$  \ainagul{are computed by integrating the heat flux at the inner surface over the chosen time interval~\citep{ALSANEA_2012,BERGER_2017_E}}:
\begin{align}
\label{eq:E_wall}
\mathrm{E} \egal  \int_{\,t_{}}^{\,t\plus \delta\,t} \, \mathrm{j}(\,\tau\,) \, \mathrm{d\tau} \,,
\end{align}
where $\delta\,t$ is a time interval such as day, week, or month.

\ainagul{The dimensionless presentation of these equations is illustrated in Appendix~\ref{sec:dim_eq}. All methodology, described in further sections, is presented using dimensionless variables. They are indicated by a superscript $^{\star}$. }

\subsection{Sensitivity analysis}

The aim of this section is to describe the methods that identify how the variation in the model parameters affects the variation in the model output. In case of the model describing the heat conduction transfer in the building wall, several outputs can be assessed. \ainagul{The temperature field $T$ is obtained directly from the model. Other variables of interest can be computed using $T$, such as  the heat flux $\mathrm{j}$ and the thermal loads $\mathrm{E}$. Further, let us operate with the dimensionless temperature $u$.} In the framework of sensitivity analysis, the variable $u$ is declared as a function of spatial and time coordinates $\bigl(\,x^{\,\star}\,,t^{\,\star}\,\bigr)$ as well as the parameter $p^{\,\star}\,$. In our case, the parameter belongs to the set of sensitive parameters $\bigl\{\,k^{\,\star}_{\,i}\,,c^{\,\star}_{\,i}\,\bigr\}_{\,i \egal 1\,,\ldots\,,N\, }\,$. Thus, one may formulate the temperature as follows:
\begin{align*}
    u\,:\,  \Omega_{\,x} \, \times \, \Omega_{\,t} \, \times \, \Omega_{\,p} & \longrightarrow \mathbb{R}\\[4pt]
    \bigl(\,x^{\,\star}\,,t^{\,\star}\,,p^{\,\star}\,\bigr) & \longmapsto\, u\,\bigl(\,x^{\,\star}\,,t^{\,\star}\,,p^{\,\star}\,\bigr)\,.
\end{align*}
The sensitivity of the temperature to an individual parameter $p$ is defined as the partial derivative of the model output and called the sensitivity coefficient $X_{\,p^{\,\star}}$ \citep{Saltelli_2004}:
\begin{align*}
    X_{\,p^{\,\star}}\,:\,\bigl(\,x^{\,\star}\,,t^{\,\star}\,,p^{\,\star}\,\bigr) \, \longmapsto\, \pd{u}{p^{\,\star}}\,.
\end{align*}
High values of $X_{\,p^{\,\star}}$ indicate that $u$ is more sensitive to $p^{\,\star}\,$, \emph{i.e.}, small changes in $p^{\,\star}$ cause large changes in the model output.

The key to estimate $X_{\,p^{\,\star}}$ is the calculation of partial derivatives. The partial derivative $ \displaystyle \pd{u}{p^{\,\star}}$ can be found either through discrete approximation or through the direct differentiation of model--governing equations. These procedures are described in the following parts.  

In addition, to explore the parameter domain, its discretization is proposed. Let us discretize uniformly the interval on which parameter $p^{\,\star}$ is defined. Thus, one may obtain the following definition:
\begin{align*}
  \Omega_{\,p} \egal \Bigl\{\,p^{\,\star}_{\,r}\,:\,p^{\,\star}_{\,r}\moins p^{\,\star}_{\,r-1}\egal \delta p^{\,\star}\,\Bigr\}_{r\egal1\,,\ldots\,,N_{\,p}}\,,  
\end{align*}
where $\delta p^{\,\star}$ is a discretization step and $N_{\,p}$ is the number of parameters in the set.



\subsubsection{Discrete approach}
\label{dicr_approx}

The partial derivative $\displaystyle \pd{u}{p^{\,\star}}$ can be expressed through the \emph{forward difference approximation}:
\begin{align}
    X_{\,p^{\,\star}} \egal \dfrac{u\,(\,p^{\,\star} \plus \Delta p^{\,\star}\,)\moins u\,(\,p^{\,\star}\,) }{\Delta p^{\,\star}} \plus \mathcal{O}\,\Bigl(\,\Delta p^{\,\star}\,\Bigr)\,,
\end{align}
or through the \emph{backward difference approximation}:
\begin{align}
    X_{\,p^{\,\star}}  \egal & \dfrac{u\,(\,p^{\,\star}\,)\moins u\,(\,p^{\,\star} \moins \Delta p^{\,\star}) }{\Delta p^{\,\star}} \plus \mathcal{O}\,\Bigl(\,\Delta p^{\,\star}\,\Bigr)\,,
\end{align}
where $\Delta p^{\,\star}$ is a small change in parameter $p^{\,\star}$.  These approximations require only two model evaluations for the different values of the parameter. However, the use of a discrete approach involves \emph{the step--size dilemma}, as use of the small step size $\Delta p^{\,\star}$ may increase the round-off error while decreasing the truncation error \citep{olver2013}. Therefore, the value of $\Delta p^{\,\star}$ should be carefully chosen. The formulation of an importance parameter factor based on the discrete partial derivative can be found in \citep{HELTON_1993}. 

The aforementioned partial derivative estimators have a first--order accuracy. One may also compute the sensitivity coefficients with higher--order accuracy, which may demand more model output evaluations. \ainagul{The second- and higher--order discrete approximations are presented in Appendix~\ref{sec:discr_diff}. } 

The discrete approximation of the partial derivative $\displaystyle \pd{u}{p^{\,\star}}$ has several advantages such as an easy implementation and a low computational cost. To compute all first--order sensitivity coefficients for $N$ parameters, only $N+1$ model output evaluations are required using the classic \emph{forward} and \emph{backward} approaches. However, an inappropriate choice of the step size value $\Delta p^{\,\star}$ may yield inaccurate results and lead to wrong conclusions about parameter importance. To avoid this problem, a continuous approach is suggested.

\subsubsection{Continuous approach using the sensitivity equations}

The continuous approach to estimate the partial derivative $\displaystyle \pd{u}{p^{\,\star}}$ assumes a  direct differentiation of model--governing equations \citep{DICKINSON_1976}. Thus, sensitivity coefficients $X_{\,p^{\,\star}}$  are obtained as a solution of a differential equation also called a sensitivity equation, which is a result of partial differentiation of a model equation. To illustrate this technique on the heat transfer model, let us find the sensitivity equations for the parameters $\bigl\{\,k^{\,\star}_{\,1}\,,c^{\,\star}_{\,1}\,\bigr\}$. The following new variables are presented:
\begin{align*}
X_{\,k^{\,\star}_{\,1}}\,:\, &\bigl(\,x^{\,\star}\,,t^{\,\star}\,,k^{\,\star}_{\,1}\,\bigr) \, \longmapsto\, \dfrac{\partial{u}}{\partial{k^{\,\star}_{\,1}}}\,, \\ 
X_{\,c^{\,\star}_{\,1}} \,: \, &\bigl(\,x^{\,\star}\,,t^{\,\star}\,,c^{\,\star}_{\,1}\,\bigr) \, \longmapsto\,   \dfrac{\partial{u}}{\partial{c^{\,\star}_{\,1}}}\,.
\end{align*}
The differentiation of Eq.~\eqref{eq:dim_heat_eq} with respect to parameter $k^{\,\star}_{\,1}$ provides the following differential equation for $X_{\,k^{\,\star}_{\,1}}$:
\begin{align}
\pd{X_{\,k^{\,\star}_{\,1}}}{t^{\,\star}} = \frac{\mathrm{Fo}}{c^{\,\star}}\,\cdot\,\pd{}{x^{\,\star}}\,\biggl(\,\pd{k^{\,\star}}{k^{\,\star}_{\,1}}\,\cdot\,\pd{u}{x^{\,\star}} + \,k^{\,\star}\,\cdot\,\pd{X_{\,k^{\,\star}_{\,1}}}{x^{\,\star}}\,\biggr)\,,
\end{align}
where 
\begin{align*}
    \pd{k^{\,\star}}{k^{\,\star}_{\,1}} \egal \begin{cases}
    1\,,\, 0\ \leq \ x \ < \ x_{\,1}\,,\\
    0\,, \,\text{otherwise}\,.
    \end{cases}
\end{align*}
Similarly, the following sensitivity equation is obtained for $X_{\,c^{\,\star}_{\,1}}$:
\begin{align}
   \pd{X_{\,c^{\,\star}_{\,1}}}{t^{\,\star}} = & -\frac{\mathrm{Fo}}{c^{\star\,2}}\,\cdot\,\pd{c^{\,\star}}{c^{\,\star}_{\,1}}\,\cdot\,\pd{}{x^{\,\star}}\,\Biggl(k^{\,\star}\,\cdot\,\pd{u}{x^{\,\star}}\Biggr) \plus  \frac{\mathrm{Fo}}{c^{\,\star}}\,\cdot\,\pd{}{x^{\,\star}}\,\Biggl(k^{\,\star}\,\cdot\,\pd{X_{\,c^{\,\star}_{\,1}}}{x}\Biggr)  \,,
\end{align}
where 
\begin{align*}
    \pd{c^{\,\star}}{c^{\,\star}_{\,1}} \egal \begin{cases}
    1\,,\, 0\ \leq \ x \ < \ x_{\,1}\,,\\
    0\,, \,\text{otherwise}\,.
    \end{cases}
\end{align*}
\ainagul{For the sake of compactness, the equations of the second--order sensitivity coefficients are demonstrated in Appendix~\ref{sec:cont_diff}.}
These sensitivity coefficients can also be used to measure how the variation of the model parameters will impact other model outputs such as the heat flux and the thermal loads. Let us provide the sensitivity equations of the heat flux sensitivity coefficients:
\begin{align}
\pd{\mathrm{j^{\,\star}}}{k^{\,\star}_{\,1}} \egal & 
\moins \Biggl(\,\pd{k^{\,\star}}{k^{\,\star}_{\,1}}\,\cdot\,\pd{u}{x^{\,\star}} \plus k^{\,\star}\,\cdot\,\pd{X_{\,k^{\,\star}_{\,1}}}{x^{\,\star}}\,\Biggr)\,\Bigg|_{x^{\,\star} \egal x^{\,\star}_{\,s}}\,,\\
\pd{\mathrm{j^{\,\star}}}{c^{\,\star}_{\,1}} \egal &
 \moins \Biggl(\,k^{\,\star}\,\cdot\,\pd{X_{\,c^{\,\star}_{\,1}}}{x^{\,\star}}\,\Biggr)\,\Bigg|_{x^{\,\star} \egal x^{\,\star}_{\,s}}\,,\\
\frac{\partial^{2}{\mathrm{j^{\,\star}}}}{\partial{k^{\,\star}_{\,1}}^2}\egal &
 \moins \Biggl(\,\frac{\partial^{\,2}\,k^{\,\star} }{\partial\,k^{\,\star\,2}_{\,1}}\,\cdot\,\pd{u}{x^{\,\star}} \plus 2\,\cdot\,\pd{k^{\,\star}}{k^{\,\star}_{\,1}}\,\cdot\,\pd{X_{\,k^{\,\star}_{\,1}}}{x^{\,\star}} \plus k^{\,\star}\,\cdot\,\pd{X_{\,k^{\,\star}_{\,1}\,k^{\,\star}_{\,1}}}{x^{\,\star}}\,\Biggr)\,\Bigg|_{x^{\,\star} \egal x^{\,\star}_{\,s}}\,,\\
\frac{\partial^{2}{\mathrm{j^{\,\star}}}}{\partial{c^{\,\star\,2}_{\,1}}} \egal & 
\moins \Biggl(\,k^{\,\star}\,\cdot\,\pd{X_{\,c^{\,\star}_{\,1}\,c^{\,\star}_{\,1}}}{x^{\,\star}}\,\Biggr)\,\Bigg|_{x^{\,\star} \egal x^{\,\star}_{\,s}}\,,\\
\frac{\partial^{2}{\mathrm{j^{\,\star}}}}{\partial{k^{\,\star}_{\,1}\,\partial{c^{\,\star}_{\,1}}}} \egal & 
\moins \Biggl(\,\pd{k^{\,\star}}{k^{\,\star}_{\,1}}\,\cdot\,\pd{X_{\,c^{\,\star}_{\,1}}}{x^{\,\star}} \plus k^{\,\star}\,\cdot\,\pd{X_{\,k^{\,\star}_{\,1}\,c^{\,\star}_{\,1}}}{x^{\,\star}}\,\Biggr)\,\Bigg|_{x^{\,\star} \egal x^{\,\star}_{\,s}}\,, 
\quad x^{\,\star}_{\,s} \ \in \ \{\,0\,,1\,\} \,.
\end{align}
Finally, the integration of the heat flux sensitivity coefficients during a time period will provide expressions to calculate the sensitivity coefficients for the thermal loads $\mathrm{E^{\,\star}}$.

The magnitudes of the sensitivity coefficients reflect how a parameter influences the chosen model output. Low--magnitude values mean that a parameter does not have a strong impact on the output. However, it is difficult to compare a parameter effect by considering only the sensitivity coefficients values. It is important to define a variable importance measure, which quantifies a parameter effect relative to others. The next section introduces various metrics that are used in the article.
     
\subsubsection{Sensitivity coefficient metrics}
\label{sec:metrics}

To evaluate how a change in a parameter value contributes to the change in a model output, several estimators of the individual parameter importance may be used. They include local and global approaches. The local one, \ainagul{ from a mathematical point of view,} is defined as:
\begin{equation} \label{eq:metrics}
\eta_{\,p_{\,i}} \ \eqdef \ \displaystyle \dfrac{\displaystyle\int_{\Omega_{\,x}} \int_{\Omega_{\,t}}\,\Bigl(\,X_{\,p_{\,i}}\,(\,\chi\,,\tau\,)\,\mathrm{d}\tau\,\Bigr)^{\,2}\,\mathrm{d}\chi}{\displaystyle\sum_{\,j\egal1}^{\,M} \int_{\Omega_{\,x}} \int_{\Omega_{\,t}}\,\Bigl(\,X_{\,p_{\,j}}\,(\,\chi\,,\tau\,)\,\mathrm{d}\tau\,\Bigr)^{\,2}\,\mathrm{d}\chi}\,,
\end{equation}
where $X_{\,p_{\,i}}$ is the sensitivity coefficient of the parameter $p_{\,i}$ obtained through the solution of the corresponding sensitivity equation, and $M$ is the number of input parameters, which influence the output of the model. A large value of this metric means a significant influence on the output, while a small value indicates the nonsensitive parameters. However, this quantity does not include a variation over the parameter domain, therefore the metric $\eta_{\,p_{\,i}}$ can be used as the local estimator only. It is defined as local since the parameter of interest $p_{\,i}$ remains constant. 

Another important estimator, used for the derivative based approach, is the following:
\begin{align*}
\nu_{\,p_{\,i}} \ \eqdef \  \int_{\,\Omega_p}\,\int_{\Omega_{\,x}} \int_{\Omega_{\,t}}\,\Bigl(\,X_{\,p_{\,i}}\,(\,\chi\,,\tau\,)\,\mathrm{d}\tau\,\Bigr)^{\,2}\,\mathrm{d}\chi\,\mathrm{d}p \,.
\end{align*}
According to the results of \textsc{Sobol} in \citep{Sobol2009}, this quantity verifies:
\begin{align}
\label{eq:metrics_pi}
\mathrm{D}^{\,\mathrm{tot}}_{\,p_{\,i}} \, \leq \,\frac{1}{\pi^{\,2}}\,\nu_{\,p_{\,i}} 
\end{align}
where $\mathrm{D}^{\,\mathrm{tot}}_{\,p_{\,i}}$ is the total partial variance. Additionally, the following approximation is also retrieved from \citep{Sobol2009}:
\begin{align}
\label{eq:metrics_12}
\mathrm{D}^{\,\mathrm{tot}}_{\,p_{\,i}} \, \approx \,\frac{1}{12}\,\nu_{\,p_{\,i}} \,.
\end{align}

Last, we define the global metric:
\begin{align}
\gamma_{\,p_{\,i}} \ \eqdef \  \frac{\nu_{\,p_{\,i}}}{\displaystyle \sum_{\,j\egal1}^{\,M}\, \nu_{\,p_{\,j}}}
\end{align}
This metric is global since it examines the solution variations over an interval of variation of parameter $p_{\,i}\,$.

Two properties of the global metric can be noted \citep{Touzani_2014}. First, the following inequality is observed:
\begin{align*}
0 \ < \ \gamma_{\,p_{\,i}} \ < \ 1 \,.
\end{align*}
Then, the sum of the global metrics of each parameter is equal to unity:
\begin{align*}
\sum_{\,j\egal1}^{\,M}\, \gamma_{\,p_{\,j}} \egal 1 \,.
\end{align*}

\subsubsection{Primary Identifiability for parameter estimation problem}
~\begin{BS_Answer}
The aforementioned metrics and sensitivity coefficients values are very efficient in the procedure of the parameter estimation problem. Since it is a difficult optimization problem, whose computational cost depends on the number of unknown parameters in a direct model. For example, in a wall composed of four layers, {to identify the thermal conductivity and the volumetric heat capacity of each layer, theoretically, at least eight parameters should be estimated}. Thus, it is of crucial importance to decrease the number of uncertain model parameters. However, the parameters might not be theoretically identifiable, {lacking the uniqueness of the parameter set that may have produced the measurements}. Therefore, it is futile to try to estimate such parameters. 

The use of sensitivity analysis is an attractive approach to address this issue. It highlights the main parameters impacting on the output of interest of the model. Thus, the number of parameters to be estimated can be reduced to a few numbers, and the computational cost of solving the parameter estimation problem can be reduced. The sensitivity analysis also is a useful tool to arrange an experimental set--up since measurements focus only on significant material properties. This notion of \emph{primary identifiability} is illustrated in Figure~\ref{fig:scheme_primary_identifiability}. 
The \emph{primary identifiability} is a proposed empirical concept that indicates which parameters are more valuable and have more influence on the model output based on their sensitivity coefficients, and it comes before the theoretical and practical identifiability analysis~\citep{Walter_1982,Walter_1990}. It can be noted that the computation of the aforementioned metrics is directly included in the practical identifiability analysis in the framework of the parameter estimation problem.   
\begin{figure}[!ht] 
\centering
\includegraphics[width=0.5\linewidth]{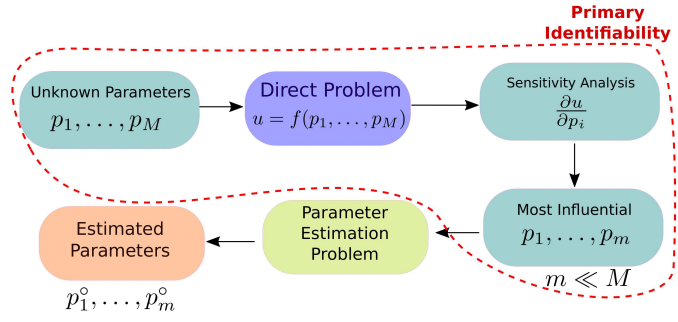}
\caption{{Scheme of the methodology of primary identifiability}.}
\label{fig:scheme_primary_identifiability}
\end{figure}
\end{BS_Answer}

\subsection{\textsc{Taylor} series expansion}

Knowledge of the partial derivatives facilitates the approximation of the model output by utilizing the \textsc{Taylor} series. Indeed, since we are in the context of the parameter estimation problem, the model output can be represented in the neighborhood of the \emph{a priori} values of parameters with high accuracy. The \textsc{Taylor} approximation investigates how the respective model output changes according to the variation in model parameters and analyzes the influence of the parameters. \ainagul{The \textsc{Taylor} series expansion is used to predict how the temperature, the heat flux, and the thermal loads vary according to the changes in the parameters (namely, the thermal conductivity or the heat capacity).} \ajumabek{For the sake of compactness, expressions of the \textsc{Taylor} series expansion for the temperature, the interior heat flux and the thermal loads are presented in Appendix~\ref{sec:taylor_pt2}.}

\subsection{Numerical solution}

After defining the governing and sensitivity equations, this section details the construction of the numerical model. The latter aims at computing the solution and its variation through the sensitivity coefficients. Let us discretize uniformly space and time intervals, with the parameters $\Delta x^{\star}$ and $\Delta t^{\star}$, respectively. The discrete values of the function $u\,(\,x^{\star}\,,t^{\star}\,)$ are defined as $\,u_{\,j}^{\,n} \ \eqdef \ u\,(\,x^{\star}_{\,j}\,,t^{\star}_{\,n}\,)\,$, where $\,j \, \in \, \{\,1,\,\ldots\,,N\, \}\, $ and $\,n\, \in \, \{\,1,\,\ldots\,,N_t\, \} $. The solution  $u\,(\,x^{\star}\,,t^{\star}\,)$ is obtained using the  \textsc{Dufort}--\textsc{Frankel} numerical scheme \citep{Du_Fort_1953,Gasparin_2017a,Gasparin_2017b}. {It is an explicit numerical scheme with a relaxed stability condition.} 
\ajumabek{First, the numerical solution of the temperature is obtained. Next, in order to calculate sensitivity coefficients efficiently, the direct differentiation of the discrete numerical solution is used. Equations are differentiated according to the required parameter. In the end, the numerical model is composed of six algebraic equations to be solved.}

\ajumabek{For the sake of compactness, the equations of the temperature and the sensitivity coefficients are demonstrated in Appendix~\ref{sec:DF_pt2}.}

\subsection{Standards for evaluating the efficiency of the method}
\label{sec:sec_ref}
The efficiency of the method to compute the solution and its sensitivity to input parameters can be evaluated using different metrics. It is important to distinguish the various approximations of the solution that can be computed. The numerical solution, obtained with the \DF ~scheme, is denoted using the superscript "$\mathrm{num}$." Then, the approximation of the solution using the \textsc{Taylor} expansion is written with the superscript "$\mathrm{tay}$."

In order to validate the numerical model, the error between the solution $u^{\,\mathrm{num}}(x,t)$ and a reference $u^{\mathrm{ref}}(x,t)$ is evaluated as a function of $x$ according to the formula:
\begin{align}
\varepsilon_2(x^{\star}) \, \eqdef \, \sqrt{\,\nicefrac{1}{N_t} \ \cdot \ \sum_{j=1}^{N_t} \Bigg(\,u^{\,\mathrm{num}}\big(\,x^{\star}\,,\,t^{\star}_{\,j}\,\big)\moins u^{\,\mathrm{ref}}\big(\,x^{\star}\,,t^{\star}_{\,j}\,\big)\,\Bigg)^2\,} \,,
\end{align}
where $N_{\,t}$ is the number of temporal steps. 
In this work, the reference solution $u^{\mathrm{ref}}$ is given by the numerical \ainagul{solution} of the differential equation based on the \textsc{Chebyshev} polynomial and adaptive spectral methods. It is obtained using the function \texttt{pde23t} from the \texttt{Matlab\texttrademark} \citep{MATLAB_2018} open source package \texttt{Chebfun} \citep{Driscoll2014}. 

To validate the \textsc{Taylor} series expansion for each model output, another error $\varepsilon_{\mathrm{tay}}$ quantity is suggested. It evaluates the difference between two solutions. The first solution is obtained with the \textsc{Taylor} series formula for variables $k^{\,\star}_{\,1}\,\in\,[k^{\,\star\,\mathrm{min}}_{\,1}\,,k^{\,\star\,\mathrm{max}}_{\,1}]\,,c^{\star}_{\,1}\,\in\,[c^{\star\,\mathrm{min}}_{\,1}\,,c^{\star\,\mathrm{max}}_{\,1}]\,$. The second is a reference solution retrieved from the direct computation of the governing model equation with parameter values $k^{\,\star}_{\,1}$ and $c^{\,\star}_{\,1}$. For the model output $u\,$, the error is computed according to:
\begin{align}
\label{eq:eps2_taylor}
\varepsilon_{\,\mathrm{tay}}\,\bigl[\,u \,\bigr] \, \eqdef \, 
\sqrt{\, \nicefrac{1}{N} \, \nicefrac{1}{N_t} \ \cdot \ \sum_{i=1}^{N}\,\sum_{j=1}^{N_t}\, \Bigg(\,u^{\,\mathrm{tay}}\big(\,k^{\,\star}_{\,1}\,,\,c^{\,\star}_{\,1}\,\big)\moins u^{\,\mathrm{ref}}\big(\,k^{\,\star}_{\,1}\,,c^{\,\star}_{\,1}\,\big)\,\Bigg)^2\,} \,.
\end{align}
Similar formulas can be defined for other model outputs such as $j$ and $E\,$.    

Last, the efficiency of a numerical model can be measured by its computational (CPU) run time required to compute the solution. It is measured using the \texttt{Matlab\texttrademark} environment with a computer equipped with \texttt{Intel} i$7$ CPU and $16$ GB of RAM. 

\section{Validation of the continuous approach for the sensitivity analysis}
\label{sec:validation}

First, a simple case study is considered to validate the proposed continuous approach for the sensitivity analysis of the heat transfer process. The study considers a heat transfer through a two-layered wall with the following \textsc{Robin}-type boundary conditions:
\begin{align}
u^{\,\mathrm{L}}_{\,\infty}\,(\,t\,) & \egal 0.8\ \cdot \ \sin\,\biggl(\, \frac{\pi \cdot t}{3} \,\biggr) \,, 
&& g^{\,\mathrm{L}}_{\,\infty}\,(\,t\,)  \egal 0.6\ \cdot \ \sin^{\,2}\, \biggl(\, \frac{\pi \cdot t}{5} \,\biggr) \,, \\[4pt]
u^{\,\mathrm{R}}_{\,\infty}\,(\,t\,) & \egal 0.5\ \cdot \ \Biggl(\,1 \moins \cos \,\biggl(\, \frac{\pi \cdot t}{4} \,\biggr)\,\Biggr)\,. \nonumber
\end{align}
The \textsc{Biot} numbers equal $\mathrm{Bi}_{\,\mathrm{L}} \egal 0.1$ and $\mathrm{Bi}_{\,\mathrm{R}} \egal 0.2\,$. \ainagul{ A simple uniform initial condition is set as $u_{\,0} \egal 0\,$ to illustrate the approach.}  The \textsc{Fourier} number, the thermal conductivity, and the heat capacity are given as:
\begin{align*}
&&\mathrm{Fo} \egal 0.02\,,
&&k^{\,\star} \egal
\begin{cases}
\, k^{\star \,\circ}_{\,1} \,, & x^{\,\star} < 0.6  \\
\, k^{\star \,\circ}_{\,2} \,, & x^{\,\star} \geq 0.6 
\end{cases}\,, 
&&c^{\,\star} \egal
\begin{cases}
\, c^{\star \,\circ}_{\,1} \,, & x^{\,\star} < 0.6  \\
\, c^{\star \,\circ}_{\,2} \,, & x^{\,\star} \geq 0.6 
\end{cases} \,,
\end{align*} 
where $k^{\star \,\circ}_{\,1} \egal 0.1 \,$, $k^{\star \,\circ}_{\,2} \egal 0.3 \,$, $c^{\star \,\circ}_{\,1} \egal 0.2\,$, and $ c^{\star \,\circ}_{\,2} \egal 0.5 \,$. The time domain is defined as $t^{\,\star}\,\in\,[\,0\,,30\,]\,$. The problem is solved by implementing the \textsc{Dufort}--\textsc{Frankel} numerical scheme. The space discretization parameter is set to $\Delta\,x^{\,\star} \egal 10^{\,-2}$ and, time discretization parameter to $\Delta\,t^{\,\star} \egal 10^{\,-3}\,$.

\subsection{Validation of the numerical model}

\begin{figure}[h!]
\begin{center}
\subfigure[]{\includegraphics[width=0.45\textwidth]{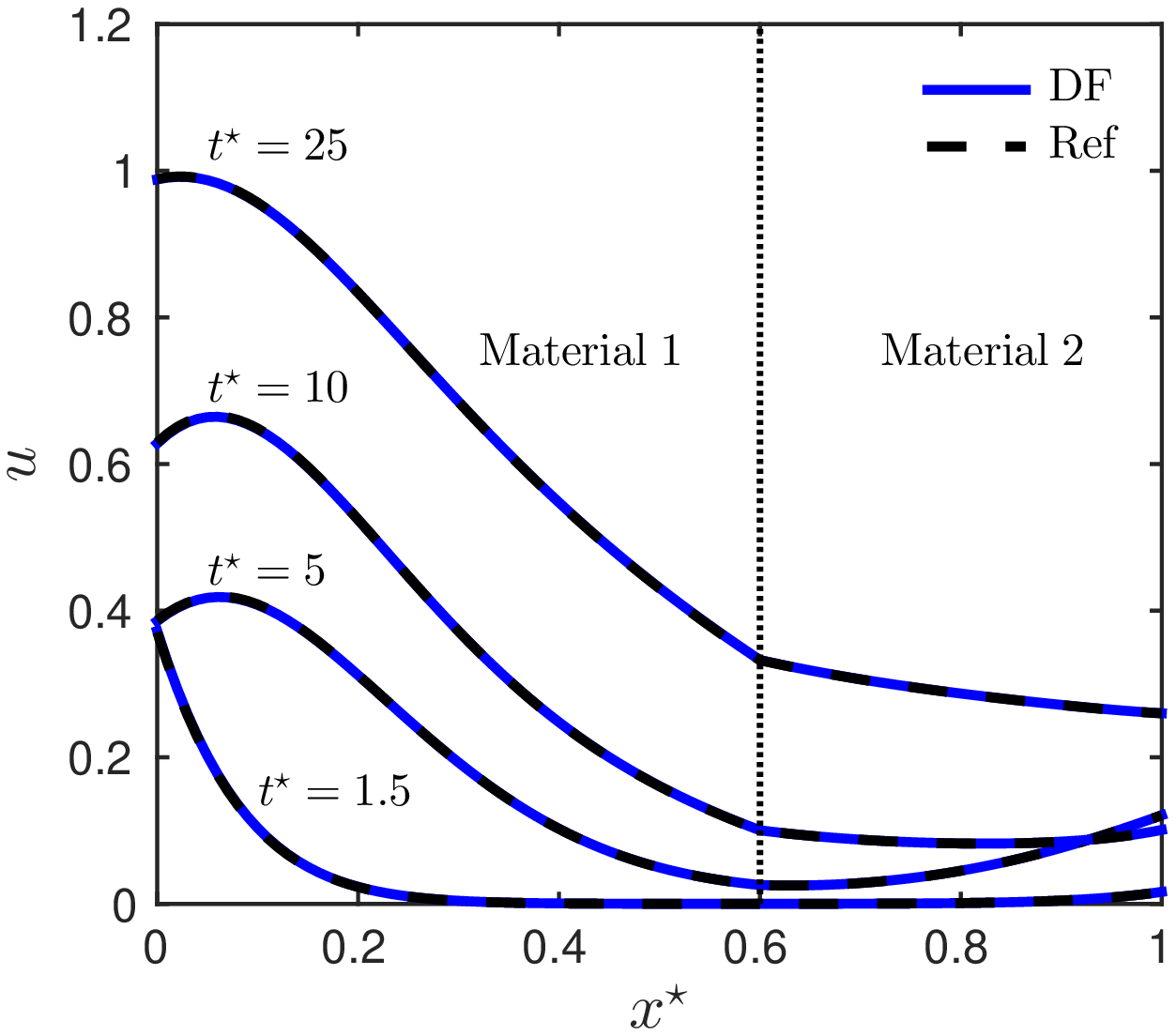}} \hspace{0.3cm}
\subfigure[]{\includegraphics[width=0.45\textwidth]{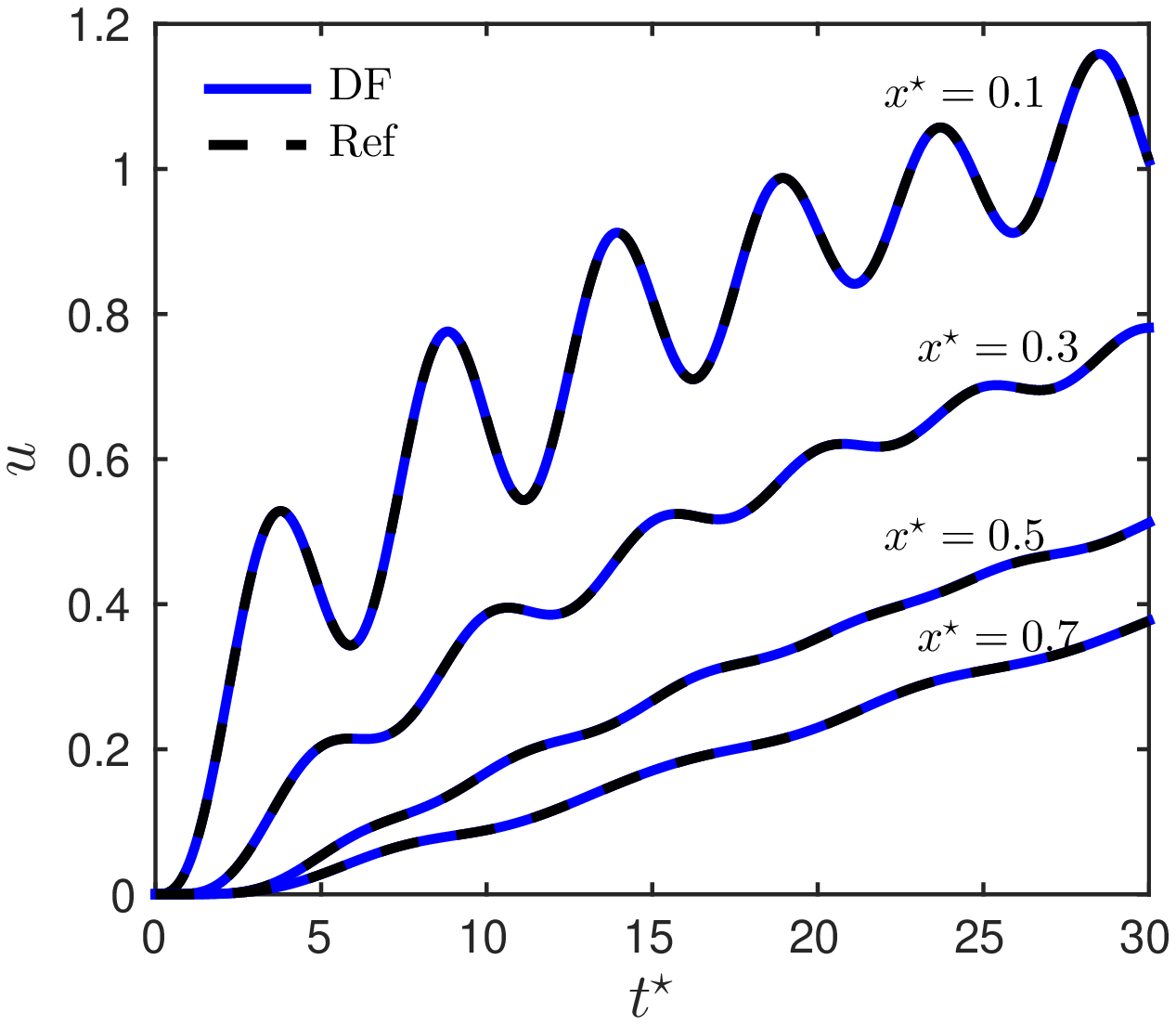}} 
\caption{Variation of the field $u$ as a function of \emph{(a)} space $x^{\,\star}$ and \emph{(b)} and time $t^{\,\star}\,$.}
\label{fig:solU}
\end{center}
\end{figure}

\begin{figure}[h!]
\begin{center}
\subfigure[]{\includegraphics[width=0.45\textwidth]{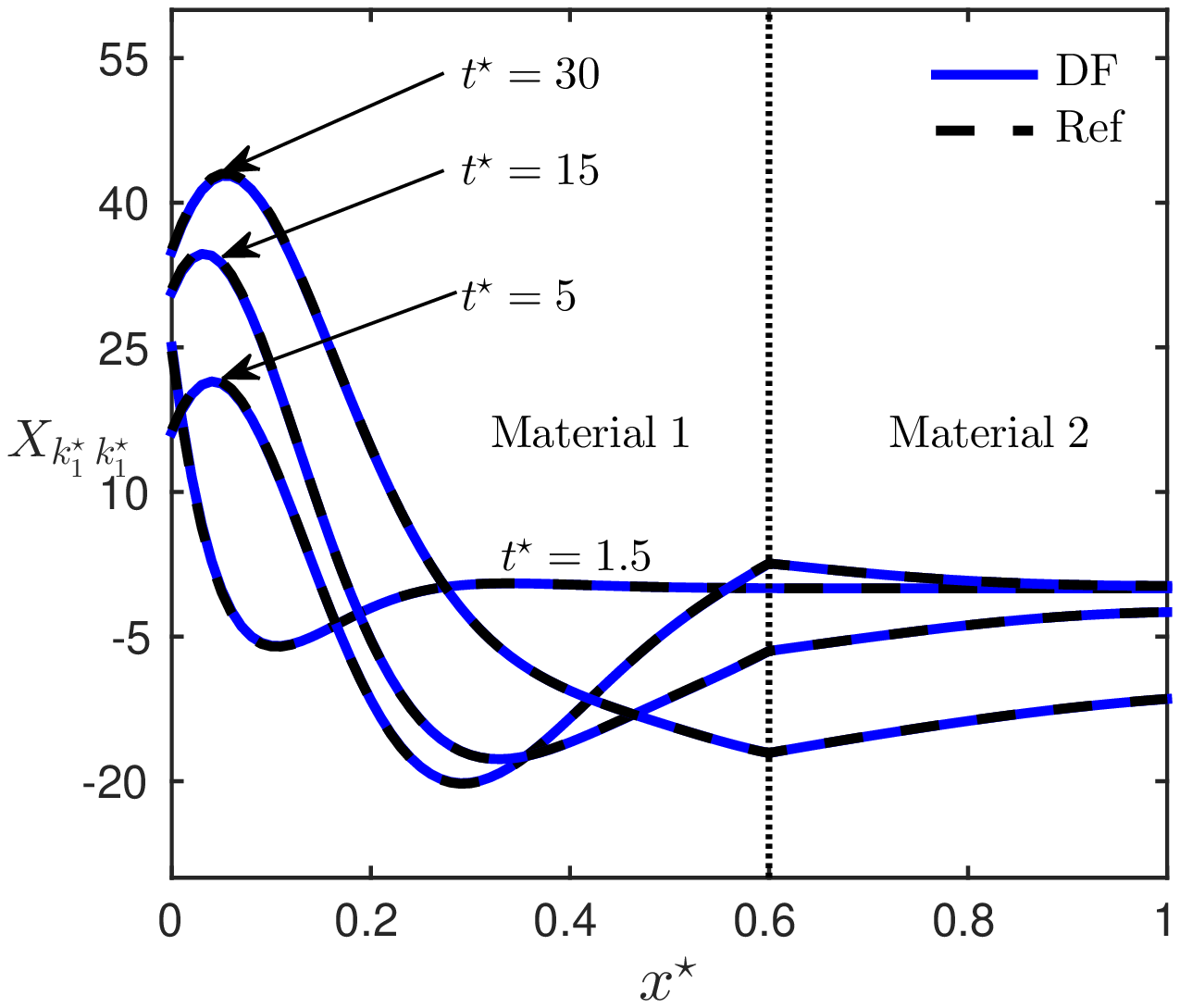}} \hspace{0.3cm}
\subfigure[]{\includegraphics[width=0.45\textwidth]{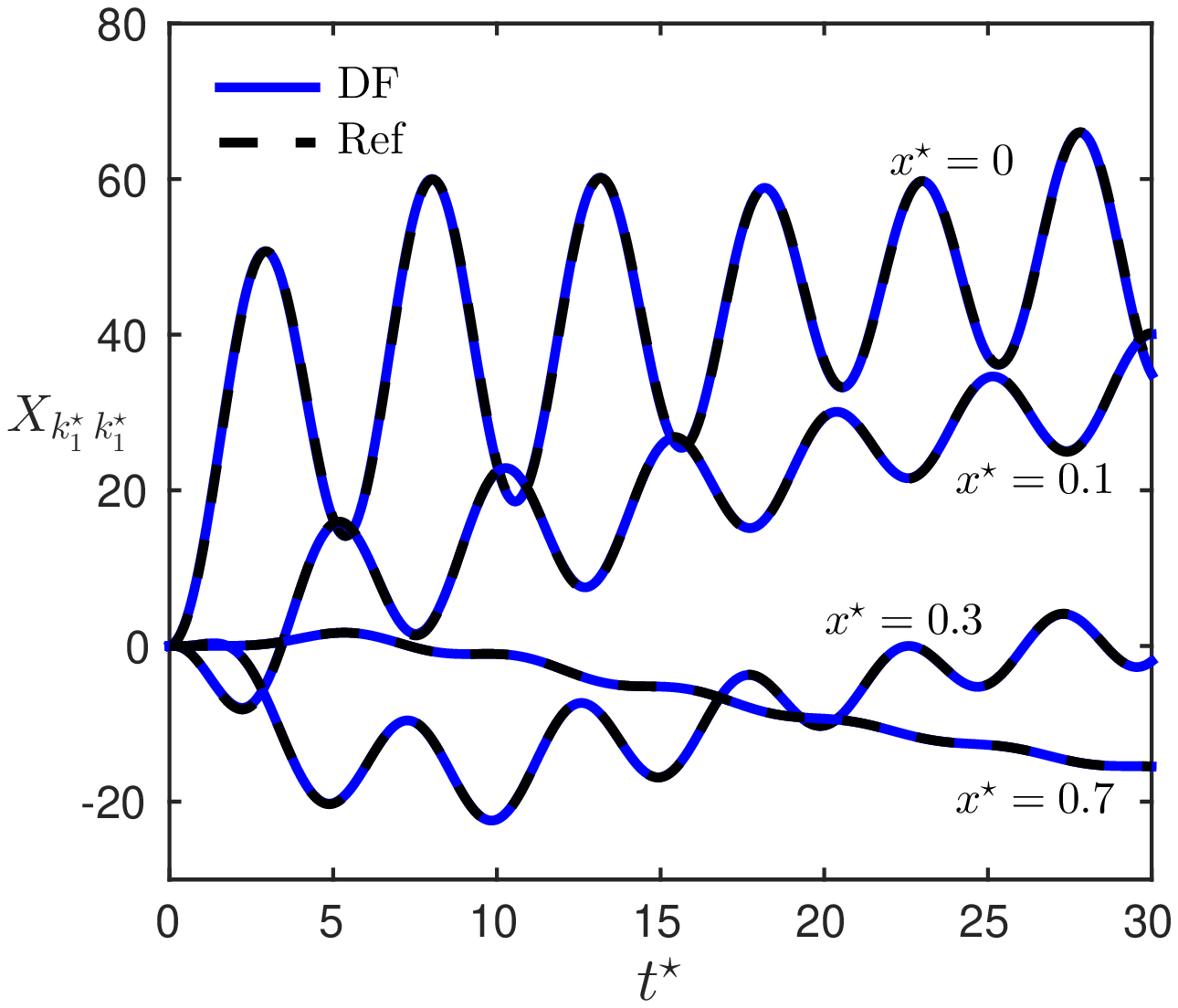}} 
\caption{Variation of the field $X_{\,k^{\,\star}_{\,1}\,k^{\,\star}_{\,1}}$ as a function of \emph{(a)} space $x^{\,\star}$ and \emph{(b)} and time $t^{\,\star}\,$.}
\label{fig:u1_KK}
\end{center}
\end{figure}

\begin{figure}[h!]
\begin{center}
\subfigure[]{\includegraphics[width=0.45\textwidth]{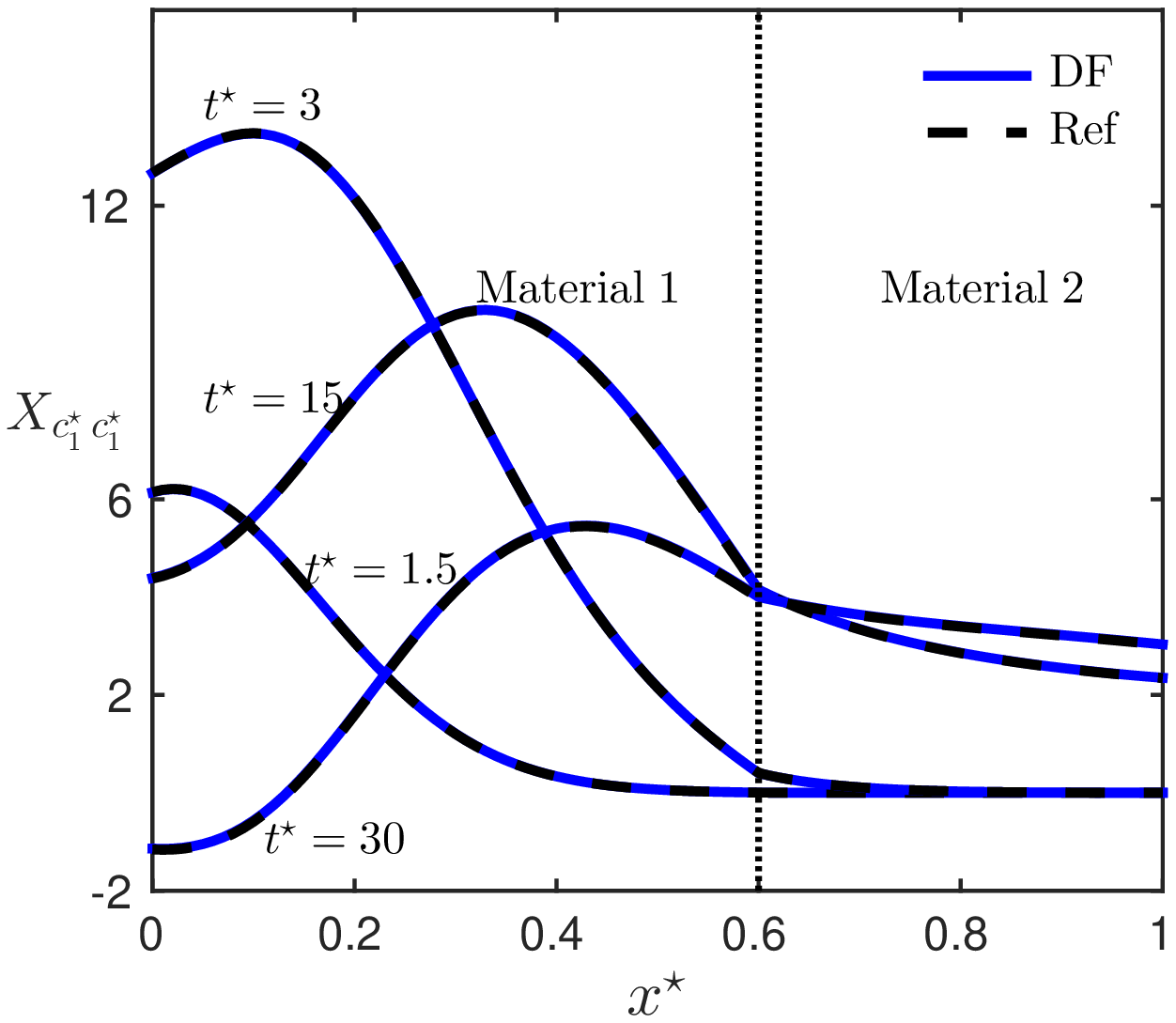}} \hspace{0.3cm}
\subfigure[]{\includegraphics[width=0.45\textwidth]{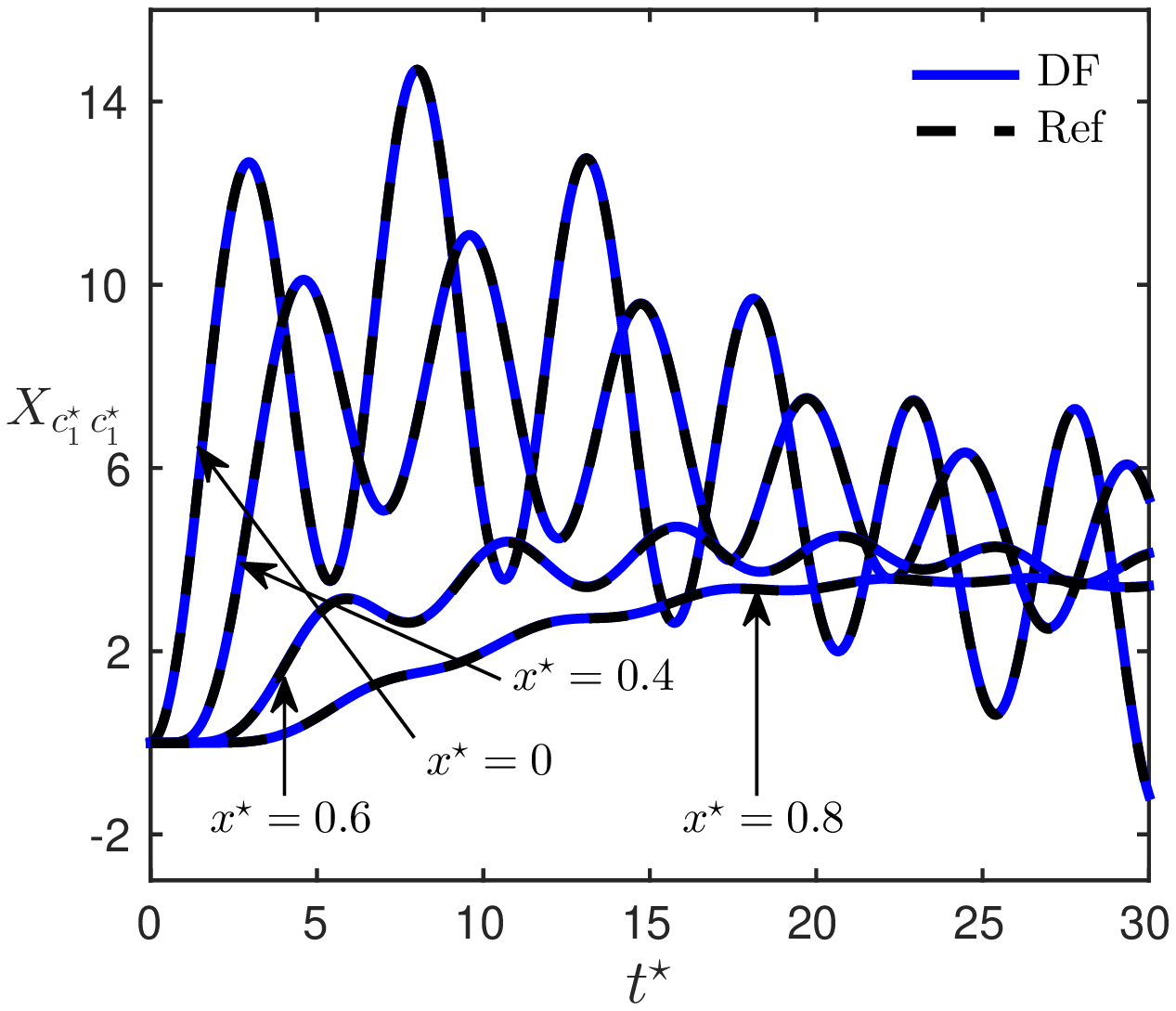}} 
\caption{Variation of the field $X_{\,c^{\,\star}_{\,1}\,c^{\,\star}_{\,1}}$ as a function of \emph{(a)} space $x^{\,\star}$ and \emph{(b)} and time $t^{\,\star}\,$.}
\label{fig:u1_CC}
\end{center}
\end{figure}

\begin{figure}[h!]
\begin{center}
\subfigure[]{\includegraphics[width=0.45\textwidth]{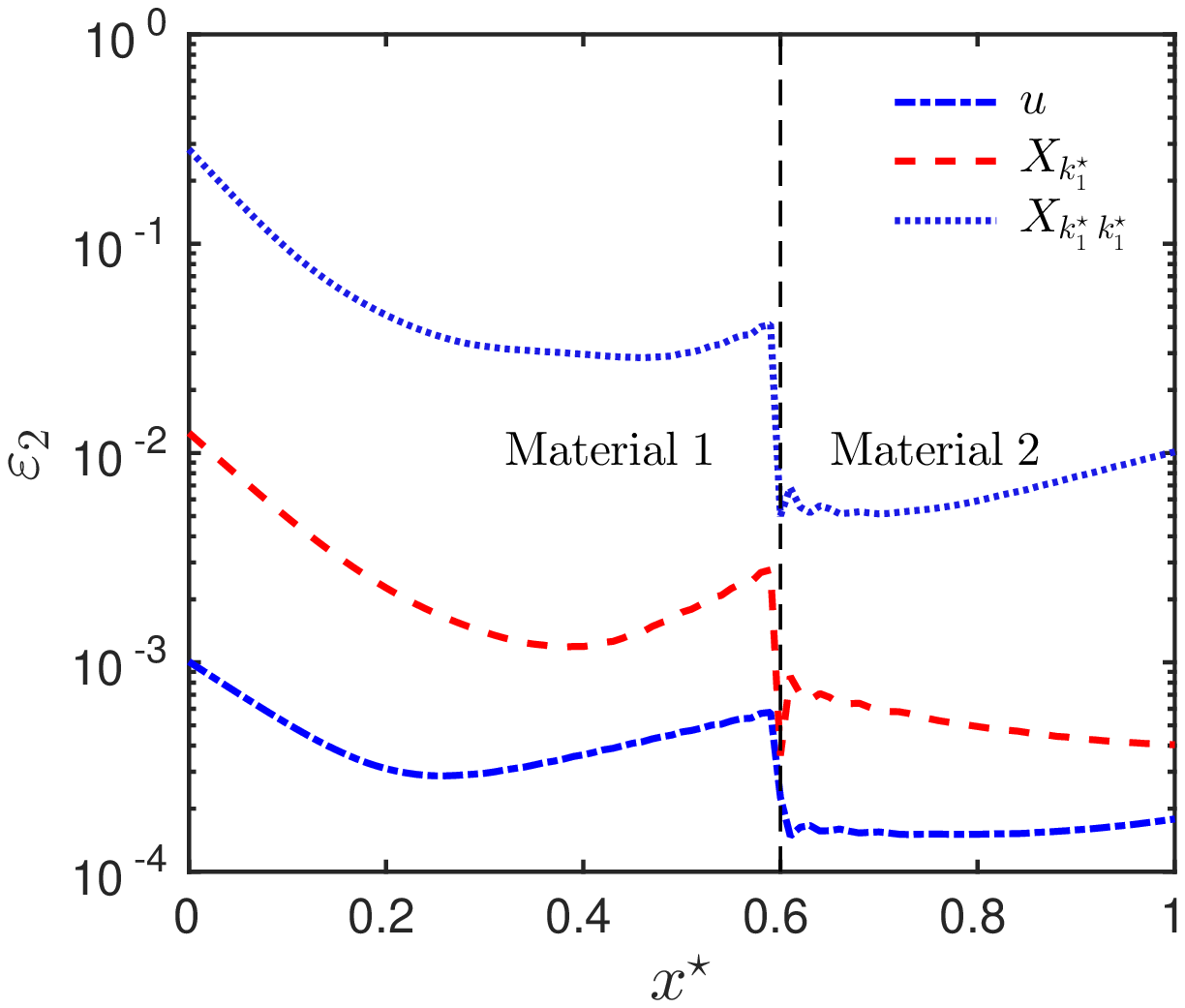}} \hspace{0.3cm}
\subfigure[]{\includegraphics[width=0.45\textwidth]{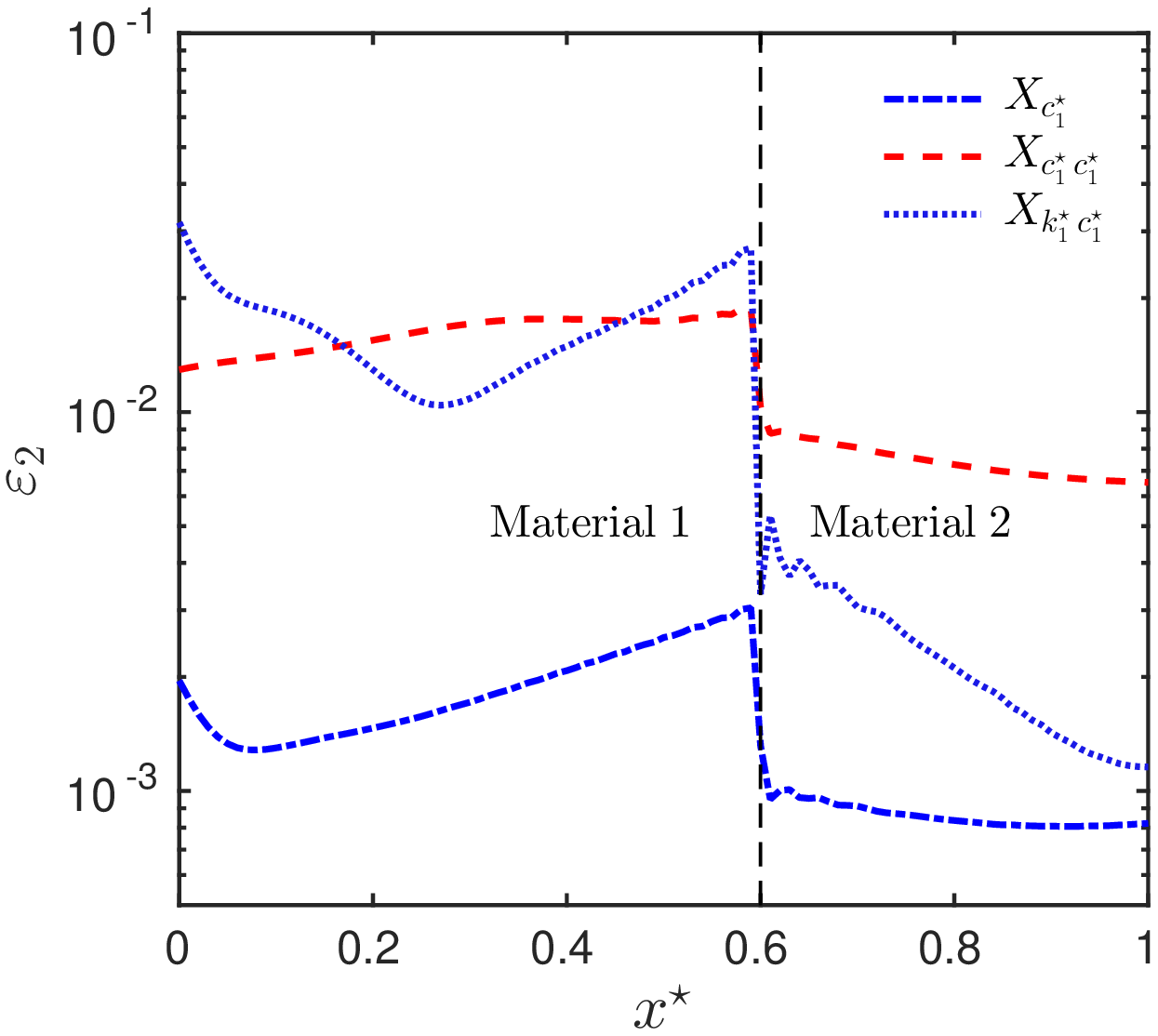}} 
\caption{Error $\varepsilon_{\,2}$ for sensitivity coefficients as a function of space.}
\label{fig:eps2_K_C}
\end{center}
\end{figure}

Figure~\ref{fig:solU} displays the variations of the solution $u$ as a function of space and time. It demonstrates a good agreement between the \textsc{Dufort}--\textsc{Frankel}  numerical solution and the reference solution. \ainagul{The reference solution is obtained through the \textsc{Chebyshev} polynomial interpolation, as mentioned in Section~\ref{sec:sec_ref}. Owing to time--dependent \textsc{Robin} boundary conditions on both wall surfaces combined with spatially dependent material properties, obtaining an analytical solution is difficult.} Figures~\ref{fig:u1_KK},~\ref{fig:u1_CC} show how second--order sensitivity coefficients of the first material vary over space and time. One may note that values of sensitivity coefficients for thermal conductivity are higher than for heat capacity, which means that the former has a greater influence on the temperature field. Figure~\ref{fig:eps2_K_C} shows the error for different quantities. The numerical solution for the temperature field has an error at the order of $10^{-3}$. Error orders for heat capacity sensitivity coefficients are below $10^{-2}$. A very satisfactory agreement is observed between the numerical and the reference solutions for variables $X_{\,k^{\,\star}_{\,1}}$ and $X_{\,k^{\,\star}_{\,1}\,c^{\,\star}_{\,1}}$. It can be noted that the error is higher, but still acceptable, for the coefficients $X_{\,k^{\,\star}_{\,1}\,k^{\,\star}_{\,1}}$ due to the important magnitudes of variation. These results validate the implementation of the numerical model to compute the solution and its sensitivity coefficients.


\begin{figure}[h!]
\begin{center}
\subfigure[\label{fig:eps2_localK2a}]{\includegraphics[width=0.45\textwidth]{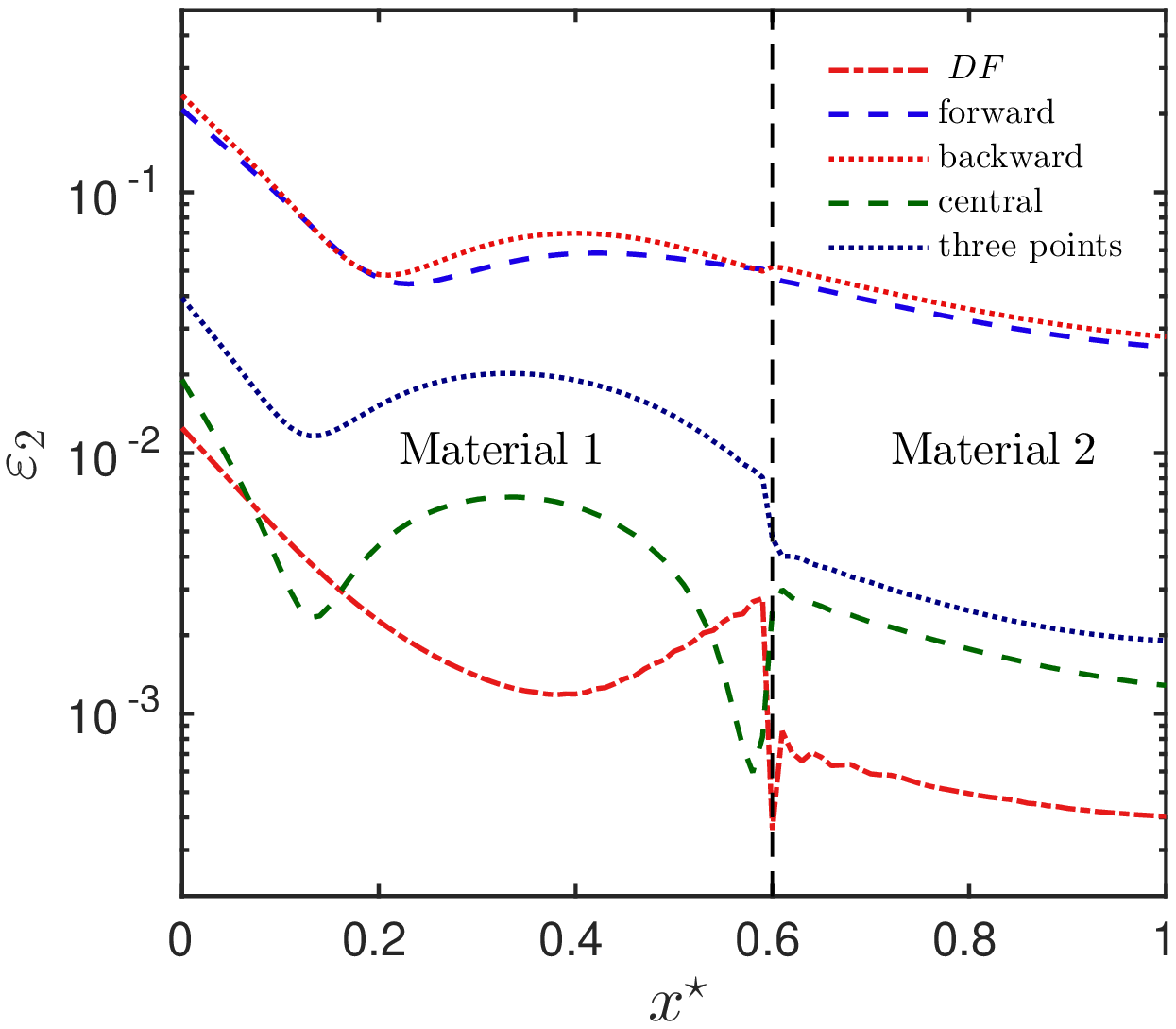}} \hspace{0.3cm}
\subfigure[\label{fig:eps2_localK2b}]{\includegraphics[width=0.45\textwidth]{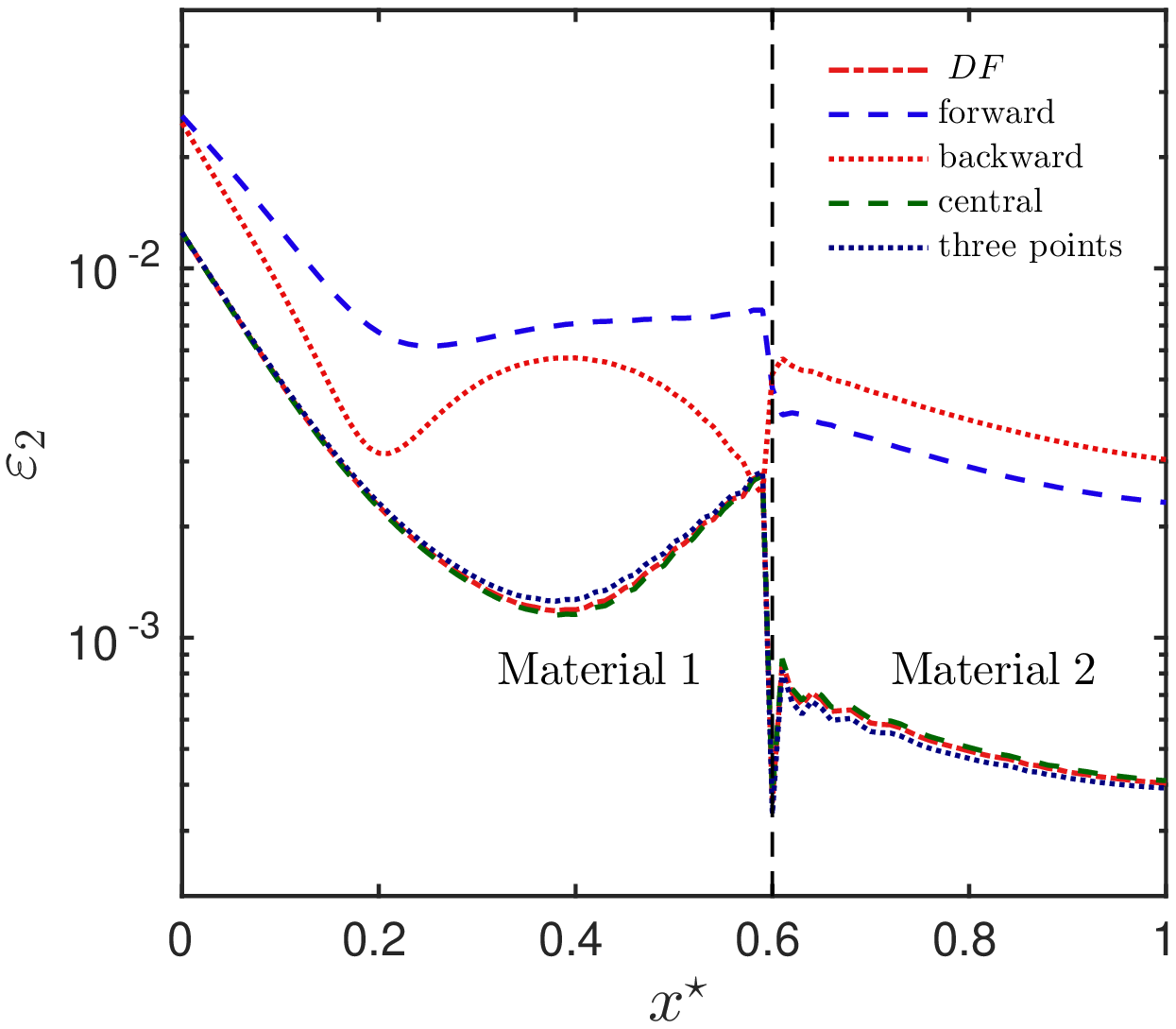}} 
\caption{Error $\varepsilon_{\,2}$ for sensitivity coefficient $X_{\,k^{\,\star}_{\,1}}$  as a function of space for \emph{(a)} $\Delta k^{\,\star}_{\,1} \egal 10^{-2}$ and \emph{(b)} $\Delta k^{\,\star}_{\,1} \egal 10^{-3}$.}
\label{fig:eps2_localK2}
\end{center}
\end{figure}

In addition, it is important to compare results given by the continuous approach with those obtained with discrete. Although discrete approximations are neither particularly accurate nor efficient, this approach is commonly used to approximate sensitivity coefficients thanks to its easy implementation. By solving only the governing problem equations one may obtain the sensitivity coefficient values. However, as mentioned before, its accuracy strongly depends on the choice of the parameter discretization step and the order of accuracy of the approximation.  Figures~\ref{fig:eps2_localK2a} and \ref{fig:eps2_localK2b} illustrate how the error between \ainagul{several} discrete approximations for sensitivity coefficient $X_{\,k^{\,\star}_{\,1}}$ and the reference solution changes according to the choice of discretiziation parameter $\Delta k^{\,\star}_{\,1}$. Space and time discretization steps remain the same and equal $\Delta \,x^{\,\star} \egal 10^{\,-2}\,$ and $\Delta\,t^{\,\star} \egal 10^{\,-3}\,,$ respectively. One may note that \emph{forward} and \emph{backward} approximations cannot achieve the accuracy of the continuous approach even when $\Delta k^{\,\star}_{\,1}$ is $10^{\,-3}\,$. Nevertheless, the \emph{central} and \emph{three--points} formulations have the same order of error as the continuous approach. To compute sensitivity values of $X_{\,k^{\,\star}_{\,1}}$ using the continuous approach $40\,\mathsf{s}$ is required, which is almost $1.5$ times greater than computation using \emph{central} approximation. The CPU time  to calculate \emph{central} and \emph{three--points} approximations corresponds to $27\,\mathsf{s}$ and $40\,\mathsf{s}\,$, respectively. 

\begin{figure}[h!]
\begin{center}
\subfigure[]{\includegraphics[width=0.45\textwidth]{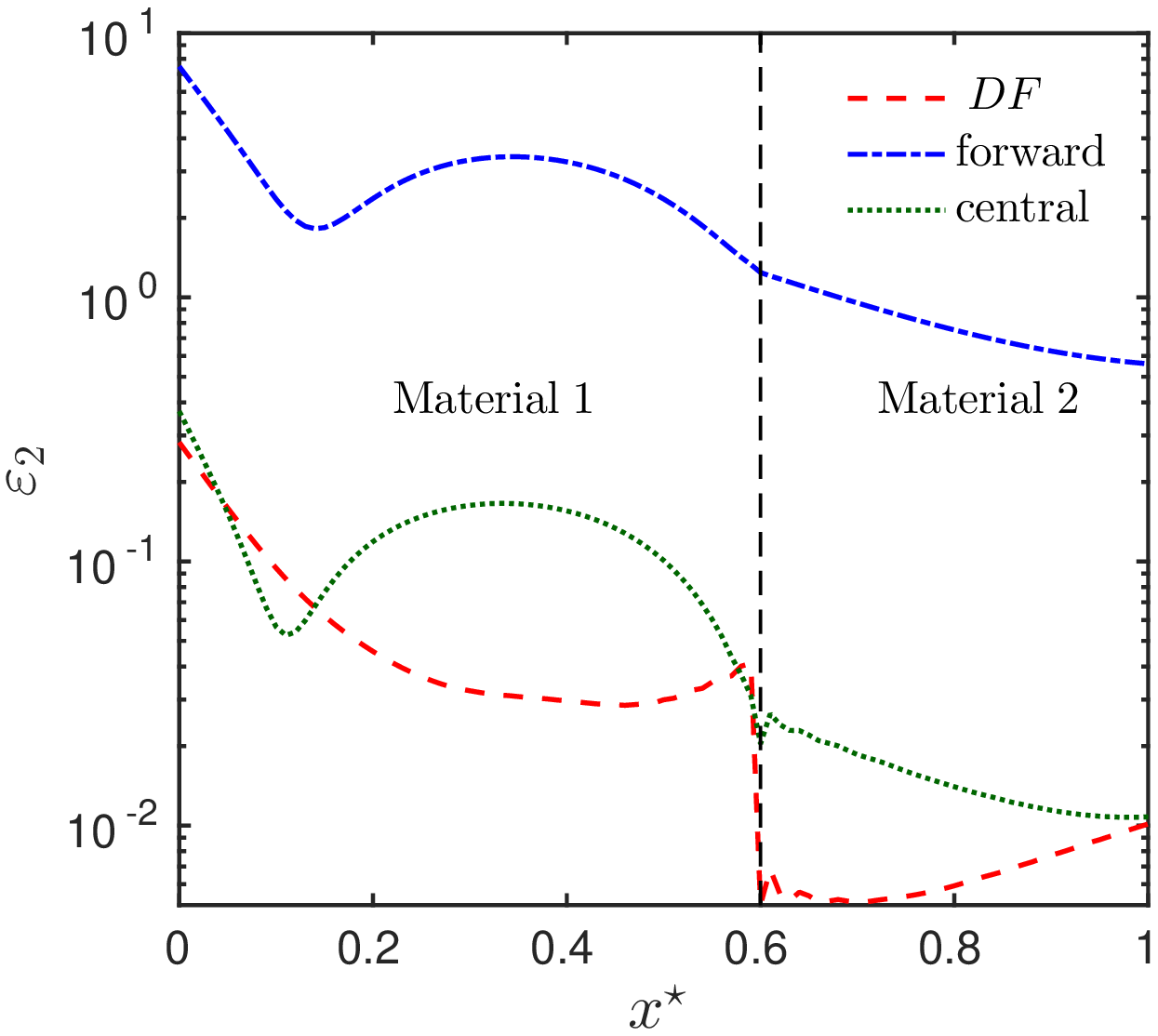}} \hspace{0.3cm}
\subfigure[]{\includegraphics[width=0.45\textwidth]{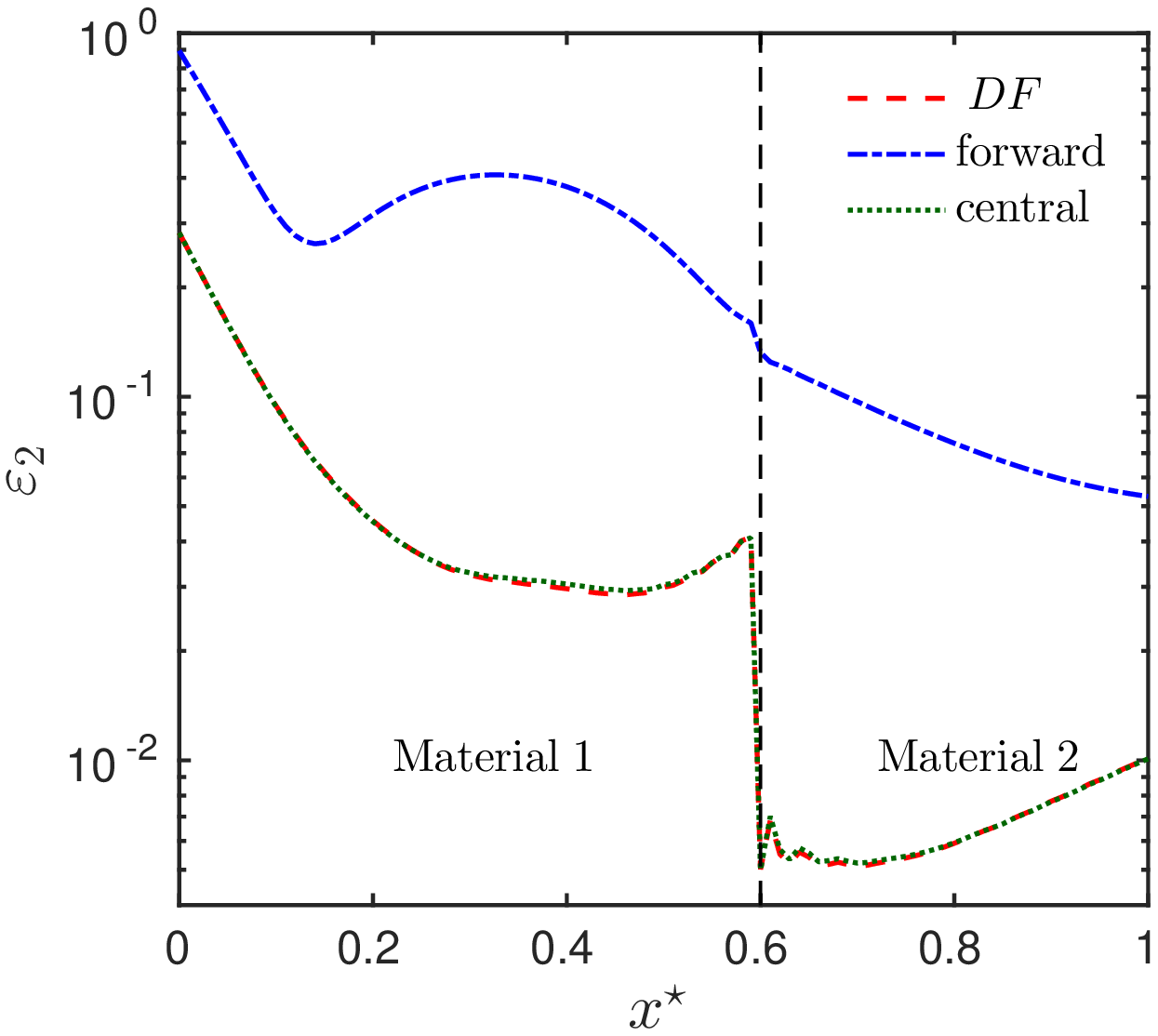}} 
\caption{Error $\varepsilon_{\,2}$ for sensitivity coefficient $X_{\,k^{\,\star}_{\,1}\,k^{\,\star}_{\,1}}$  as a function of space for \emph{(a)} $\Delta k^{\,\star}_{\,1} \egal 10^{-2}$ and \emph{(b)} $\Delta k^{\,\star}_{\,1} \egal 10^{-3}$.}
\label{fig:eps2_localK2K2}
\end{center}
\end{figure}

\begin{figure}[h!]
\begin{center}
\subfigure[]{\includegraphics[width=0.45\textwidth]{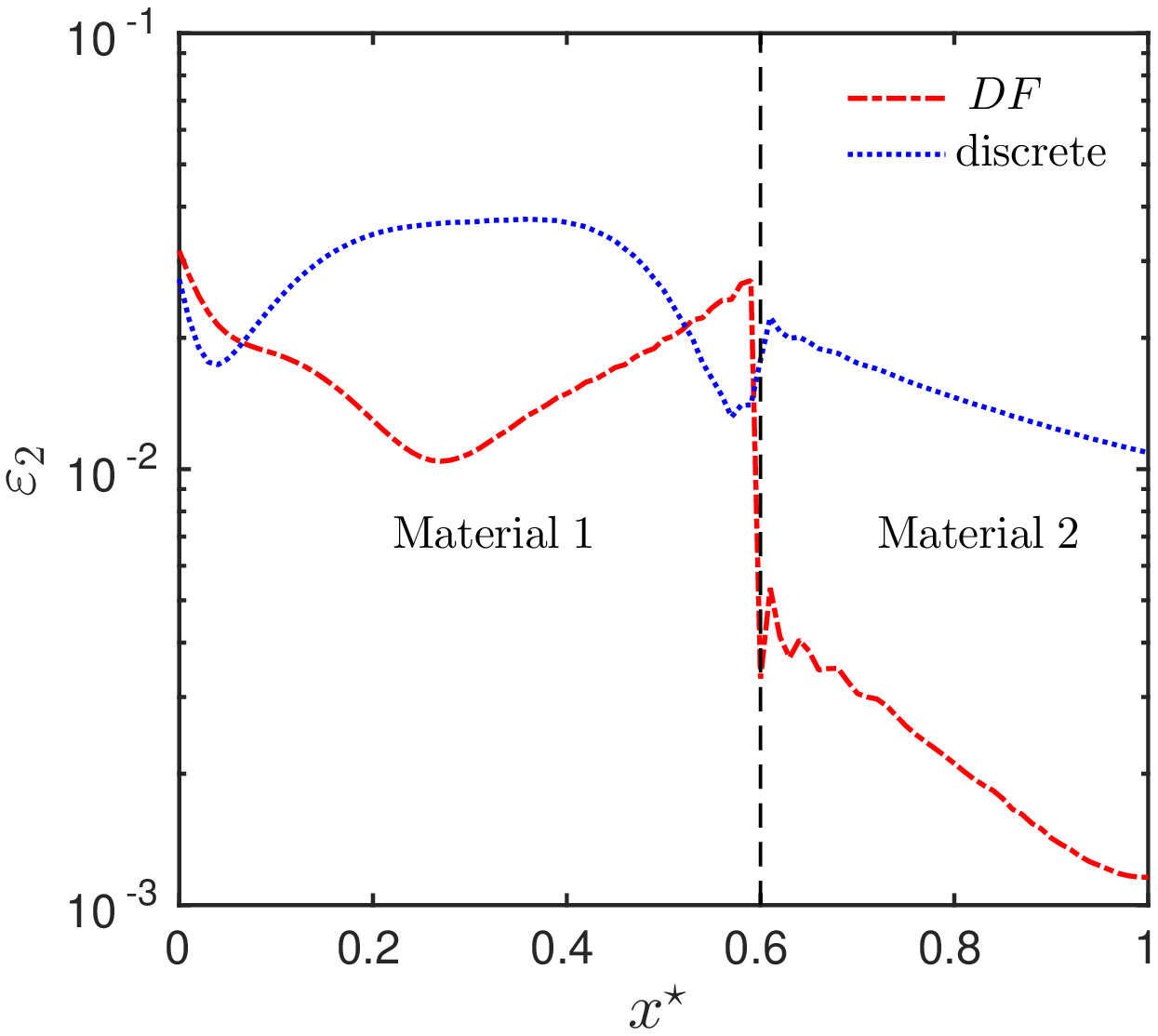}} \hspace{0.3cm}
\subfigure[]{\includegraphics[width=0.45\textwidth]{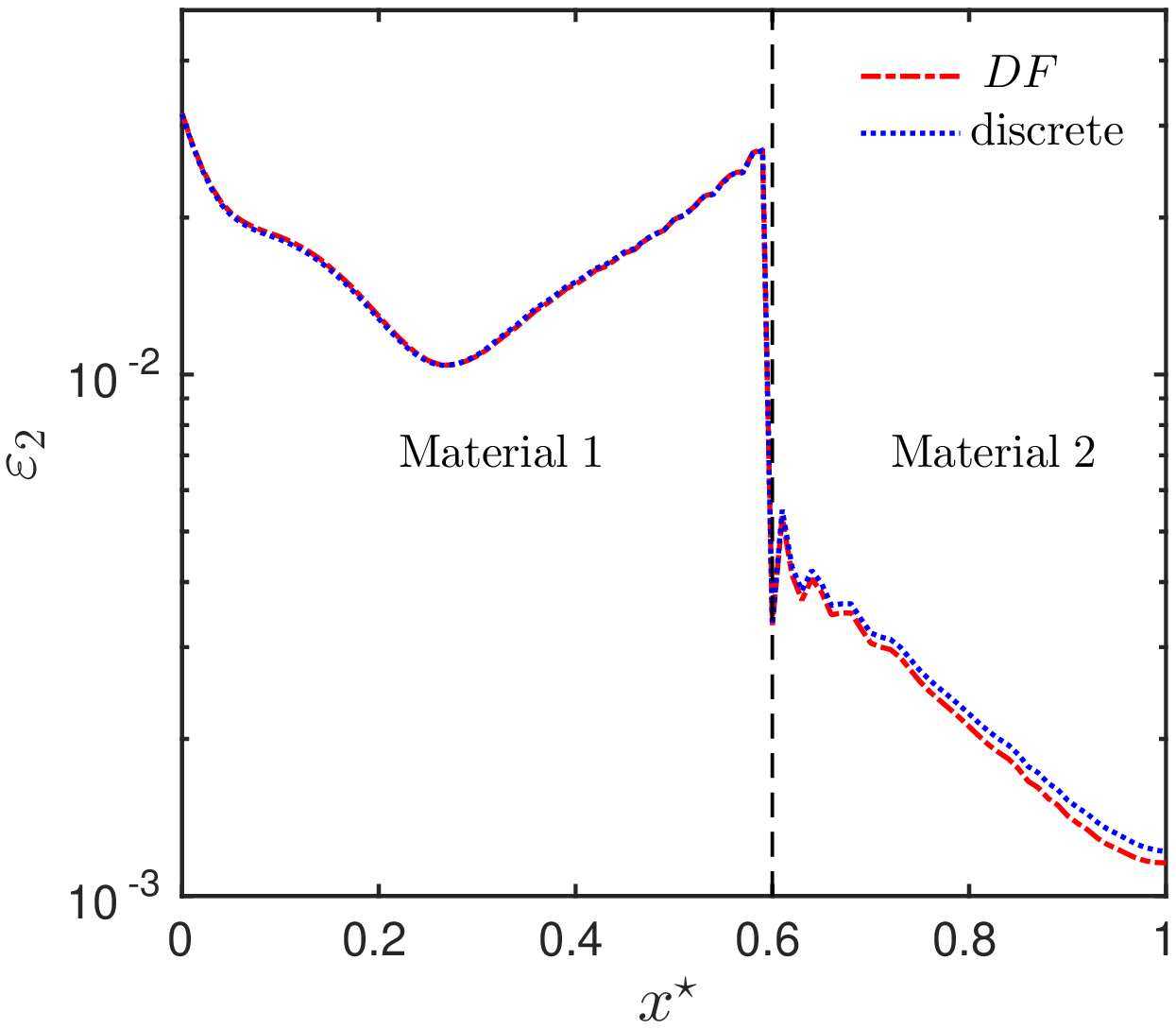}} 
\caption{Error $\varepsilon_{\,2}$ for sensitivity coefficient $X_{\,k^{\,\star}_{\,1}\,c^{\,\star}_{\,1}}$  as a function of space for \emph{(a)} $\Delta k^{\,\star}_{\,1} \egal 10^{-2}\,,\Delta c^{\,\star}_{\,1} \egal 10^{-2} $ and \emph{(b)} $\Delta k^{\,\star}_{\,1} \egal 10^{-3}\,,\Delta c^{\,\star}_{\,1} \egal 10^{-3}$.}
\label{fig:eps2_localK2C2}
\end{center}
\end{figure}

Computation of the second--order sensitivity coefficients  $X_{\,k^{\,\star}_{\,1}\,k^{\,\star}_{\,1}}$ and  $X_{\,k^{\,\star}_{\,1}\,c^{\,\star}_{\,1}}$ yields similar results. As shown in Figures~\ref{fig:eps2_localK2K2} and~\ref{fig:eps2_localK2C2} only second--order approximations give the same accuracy as a continuous approach when the discretization step for parameters $k^{\,\star}_{\,1}$ and $c^{\,\star}_{\,1}$ is $10^{\,-3}\,$. On the other hand, the efficiency of computation using the discrete approach is greater than with the continuous approach. The CPU time spent to obtain the discrete second--order sensitivity coefficients is $41\,\mathsf{s}$ , while the continuous approach requires $85\,\mathsf{s}$. However, the continuous approach allows us to obtain both first-- and second--order sensitivity coefficients.    

The results from the comparison between discrete and continuous approaches prove that the latter is more accurate and time--efficient for computing the sensitivity coefficients.

\subsection{Validation of \textsc{Taylor} series expansion}

The issue is now to evaluate the use of the coefficients, computed with the continuous approach, for the sensitivity analysis \ainagul{of the model outputs}. To this end, the \textsc{Taylor} expansion is used to analyze how the model outputs vary according to changes in the model input parameters. The study is carried out for the second material. A variation of the thermal conductivity and heat capacity by $\pm\,90\,\%$ of their \emph{a priori} values is made. Thus, the intervals of variation of the two parameters are $\Omega_{\,k_{\,2}} \egal k_{\,2}^{\star\,\circ} \cdot \bigl[\, 0.1 \,,\, 1.9 \,\bigr]$ and $\Omega_{\,c_{\,2}} \egal c_{\,2}^{\star\,\circ} \cdot \bigl[\, 0.1 \,,\, 1.9 \,\bigr] \,$, respectively. A total of $21$ discrete values of each parameter $k_{\,2}^{\,\star}$ and $c_{\,2}^{\,\star}$ in these intervals of variation are taken, or $N_{\,k2} \egal 21$ and $N_{\,c2} \egal 21$. Then, for each value of the parameters, the corresponding \textsc{Taylor} series value is computed. In parallel, the direct problem is solved with the corresponding parameter value, \ainagul{using \textsc{Chebyshev} polynomial interpolation}. This solution is used as a reference to compute the error $\varepsilon_{\,\mathrm{tay}}$ according to Eq.~\eqref{eq:eps2_taylor}. \ainagul{Calculated error values $\varepsilon_{\,\mathrm{tay}}$ validate the high--quality approximation of the model outputs. }

\begin{figure}[h!]
\begin{center}
\subfigure[\label{fig:eps2_T}]{\includegraphics[width=0.5\textwidth]{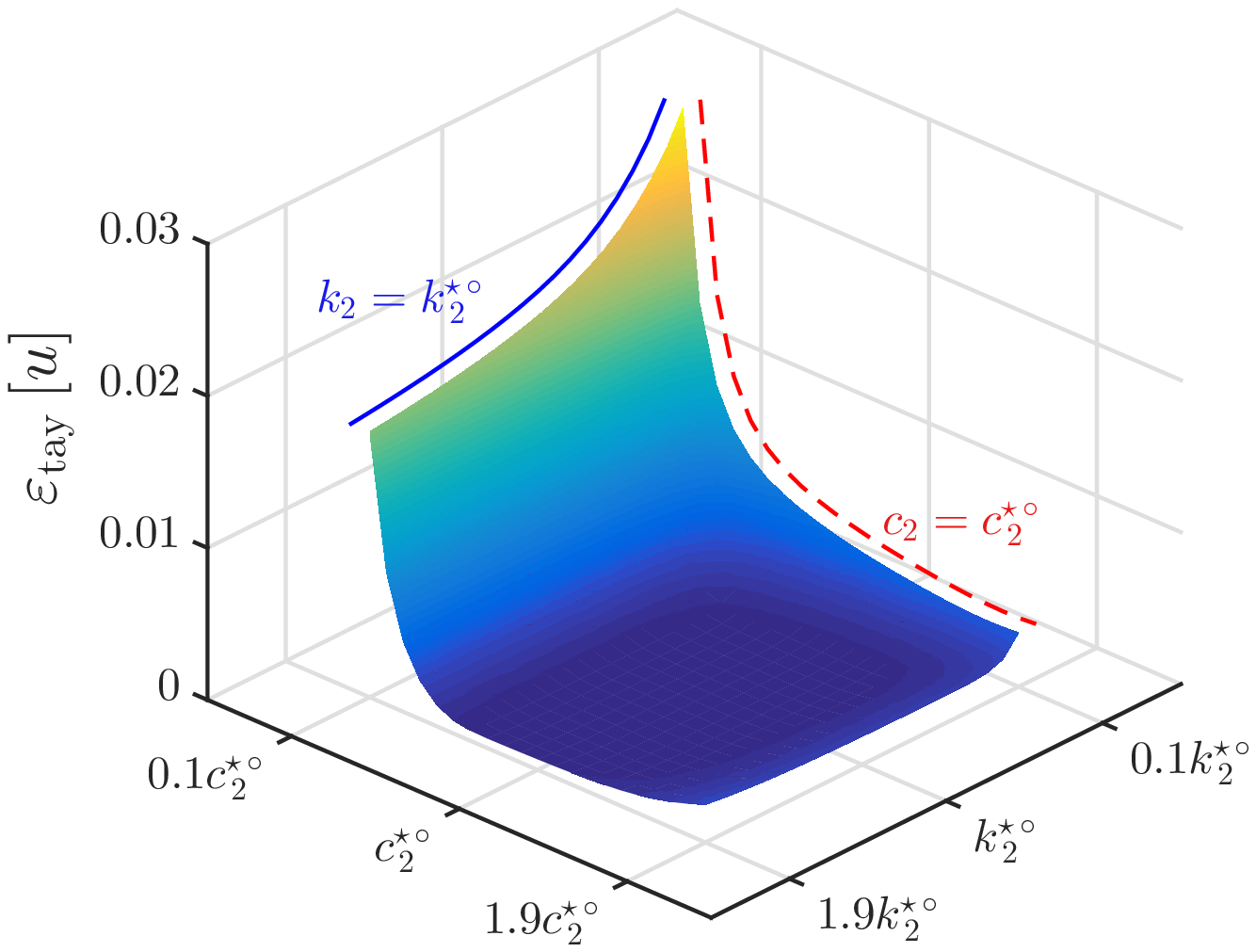}} \hspace{0.2cm}
\subfigure[\label{fig:u_K2C2}]{\includegraphics[width=0.42\textwidth]{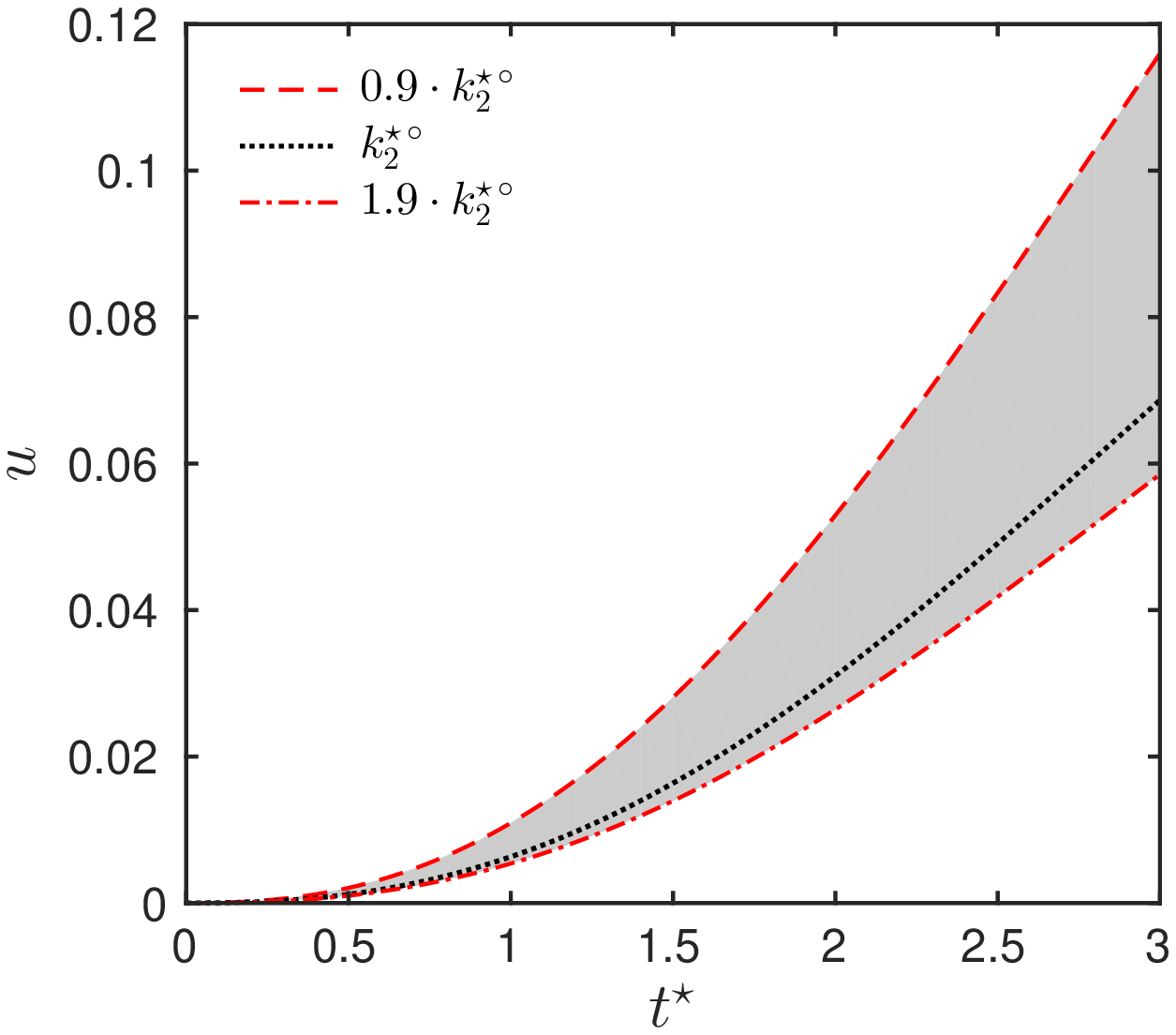}}
\caption{\emph{(a)} Error $\varepsilon_{\,\mathrm{tay}}$ of the temperature field $u$ as a function of variation $k^{\,\star}_2$ and $c^{\,\star}_2$, \emph{(b)} Variation of the temperature field $u$ according to the thermal conductivity $k^{\,\star}_2$.}
\end{center}
\end{figure}

\begin{figure}[h!]
\begin{center}
\subfigure[]{\includegraphics[width=0.45\textwidth]{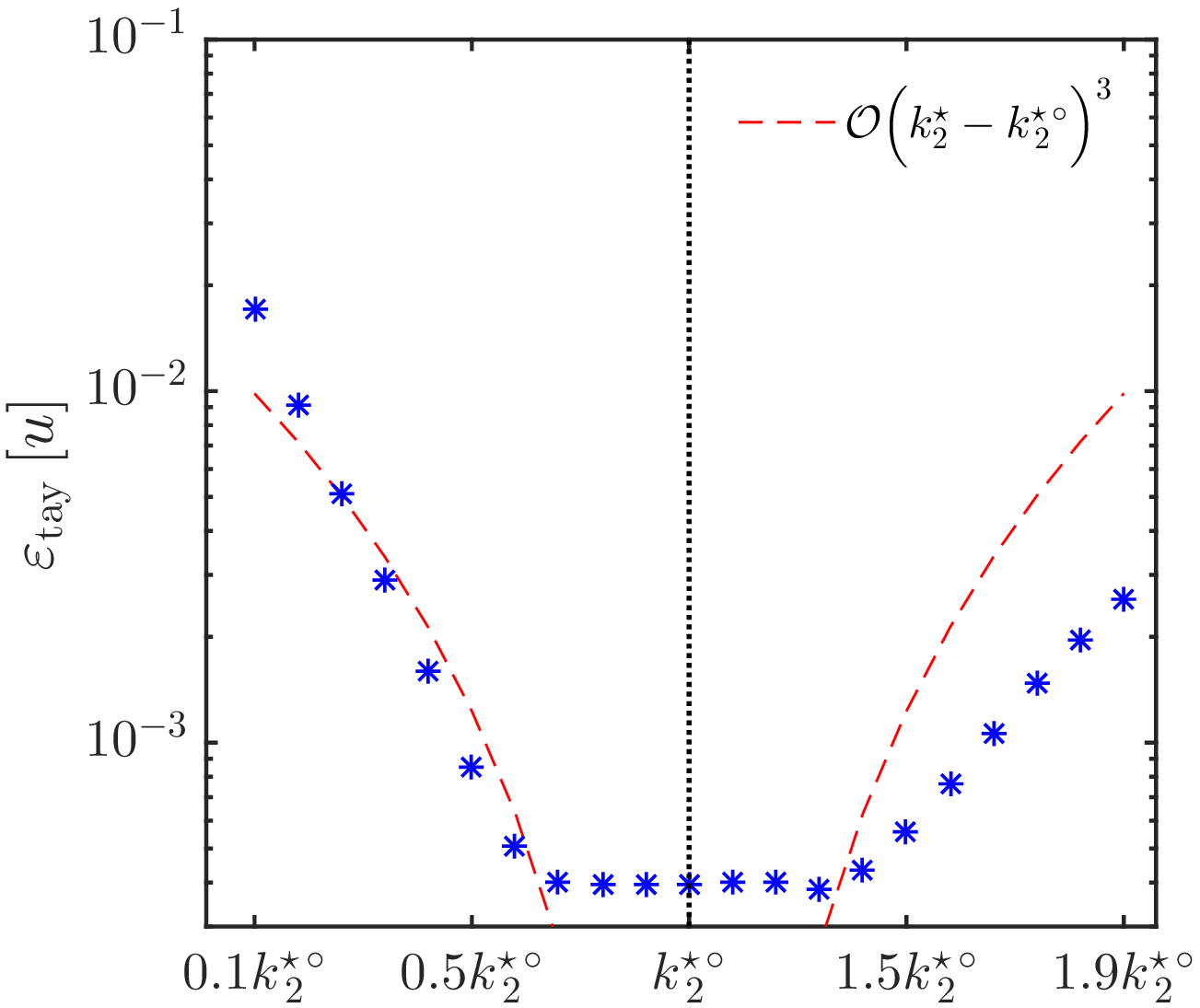}} \hspace{0.3cm}
\subfigure[]{\includegraphics[width=0.45\textwidth]{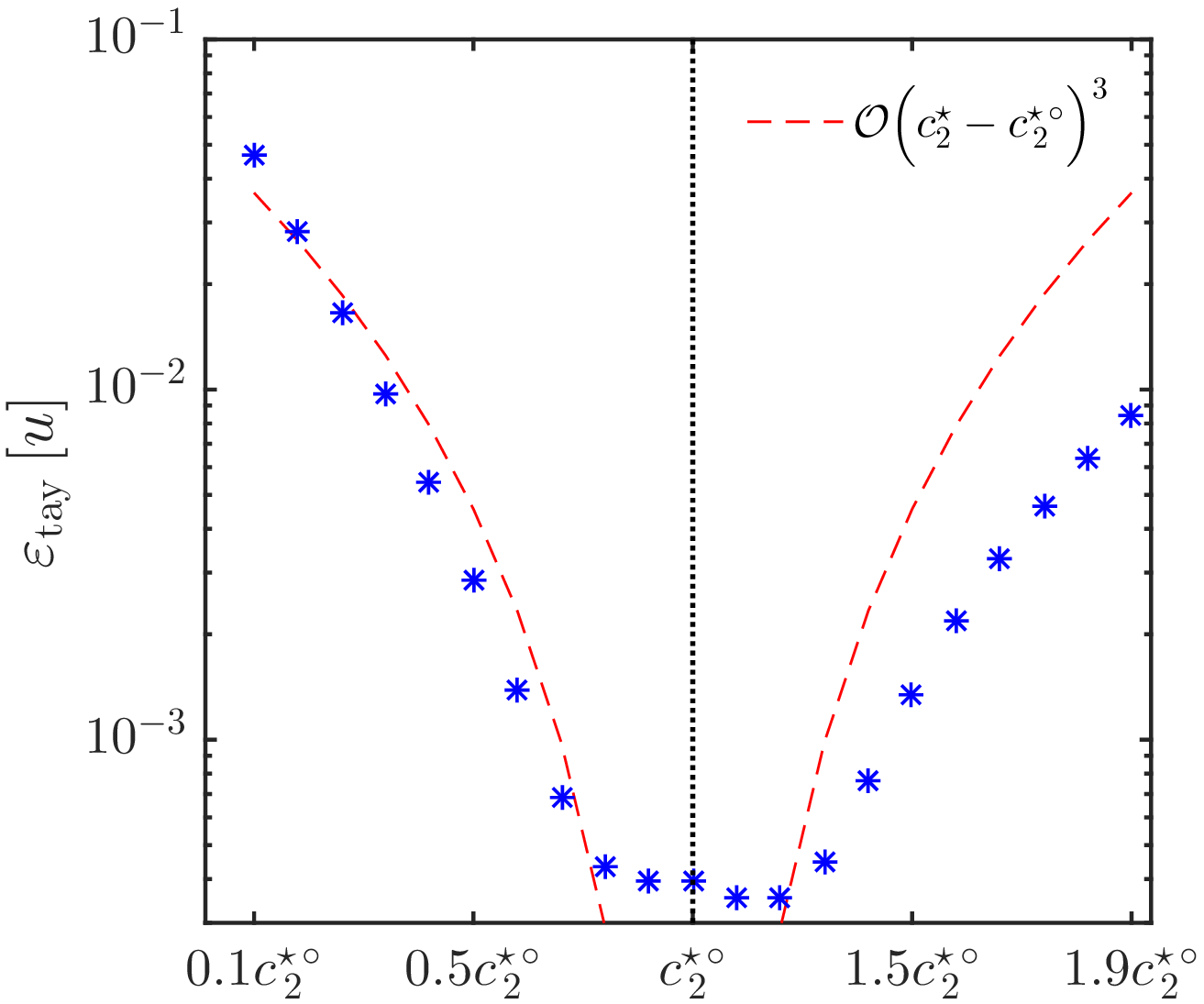}} 
\caption{Error $\varepsilon_{\,\mathrm{tay}}$ of the temperature field $u$ \emph{(a)} as a function of variation $k^{\,\star}_2$ and \emph{(b)} as a function of variation $c^{\,\star}_2$.}
\label{fig:eps_tay_u_k2c2}
\end{center}
\end{figure}

\begin{figure}[h!]
\begin{center}
\subfigure[\label{fig:eps_tay_jK2C2}]{\includegraphics[width=0.45\textwidth]{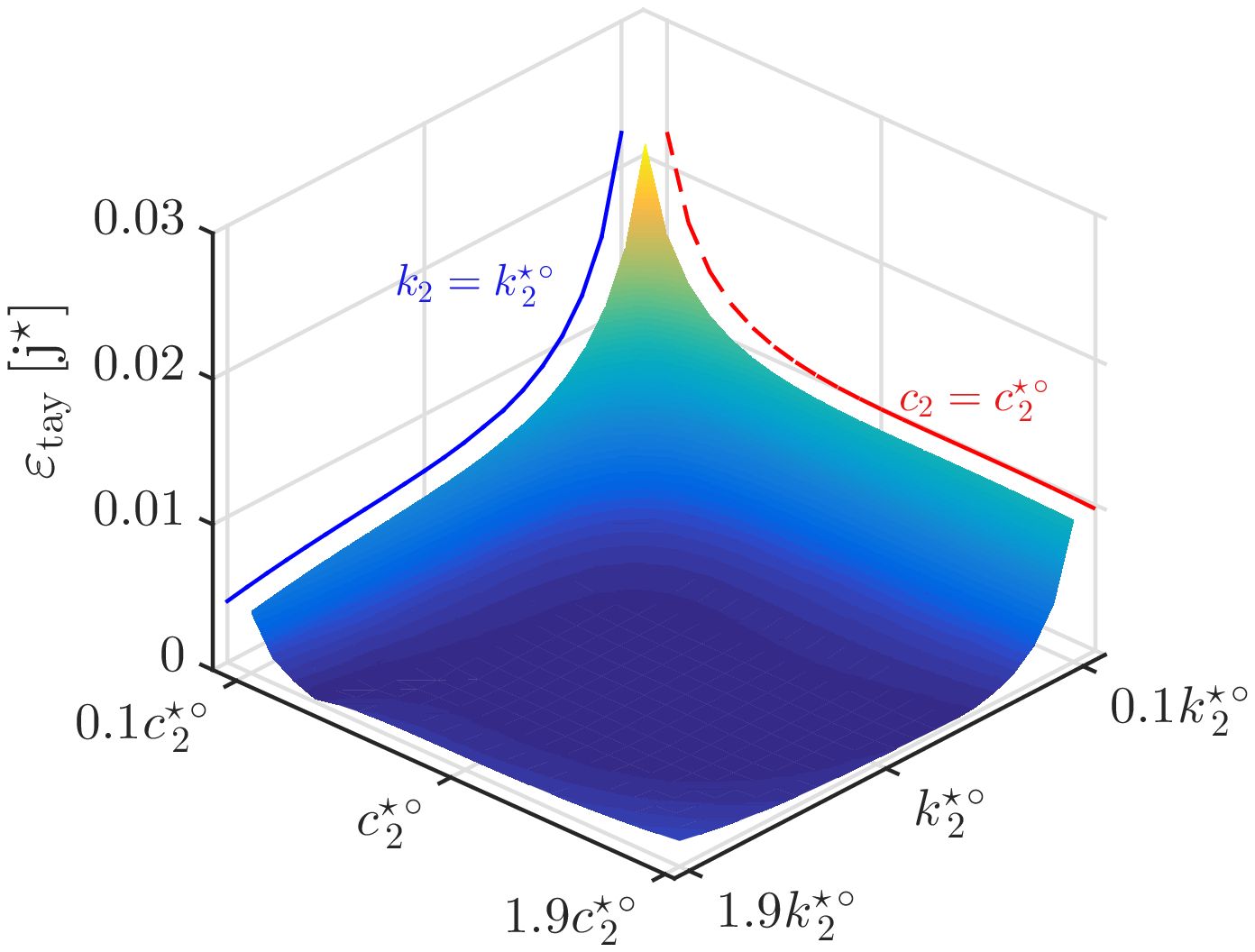}} \hspace{0.3cm}
\subfigure[\label{fig:eps_tay_EK2C2}]{\includegraphics[width=0.45\textwidth]{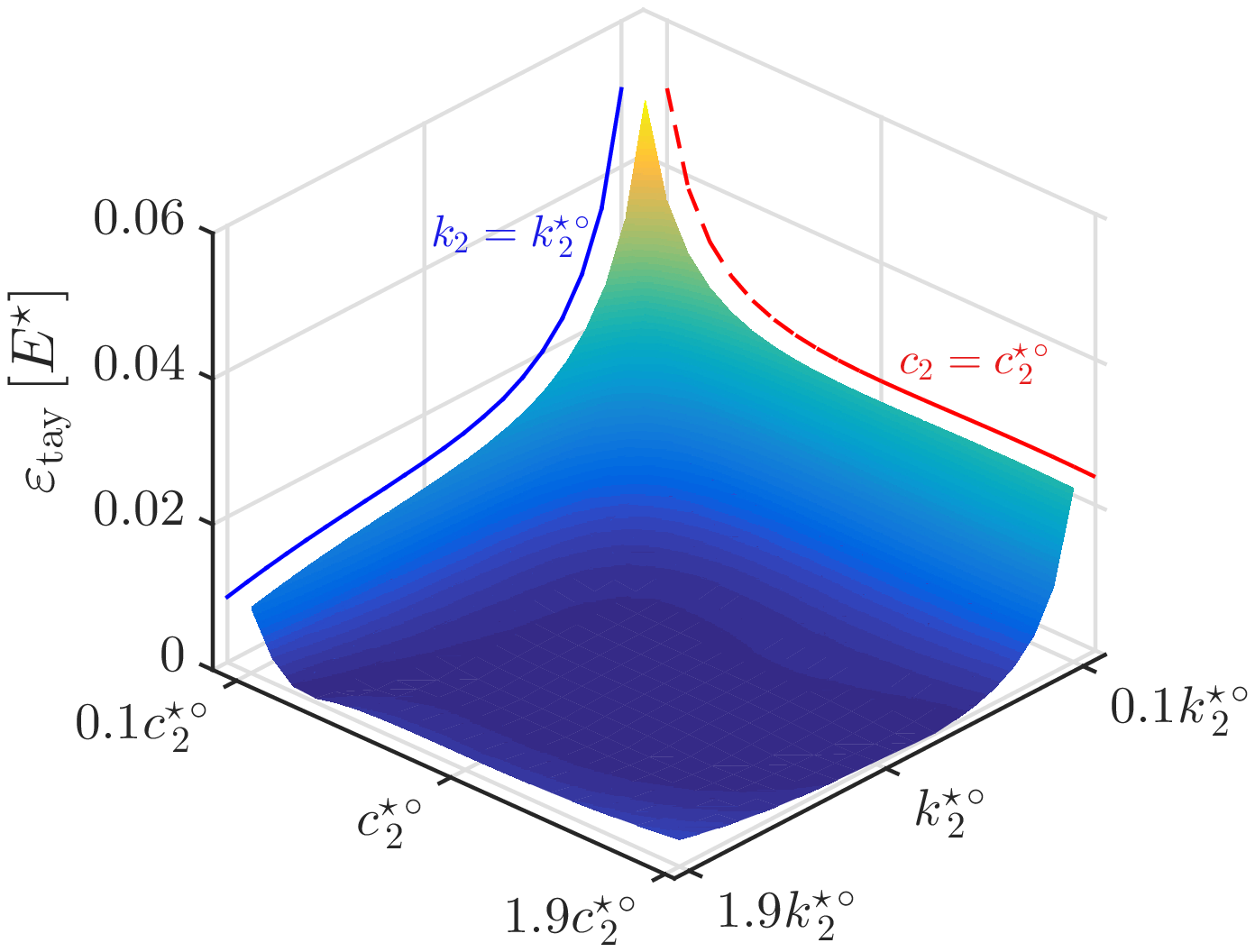}} 
\caption{\emph{(a)} Error $\varepsilon_{\,\mathrm{tay}}$ of the thermal flux $j^{\,\star}\,$ as a function of variation $k^{\,\star}_2$ and $c^{\,\star}_2$, \emph{(b)} Error $\varepsilon_{\,\mathrm{tay}}$ of the thermal loads $E^{\,\star}\,$ as a function of variation $k^{\,\star}_2$ and $c^{\,\star}_2$.}
\label{fig:eps_tay_j}
\end{center}
\end{figure}

Figure~\ref{fig:eps2_T} shows the error for the temperature field according to the variation of both parameters $k_{\,2}^{\,\star}$ and $c_{\,2}^{\,\star}$. \ainagul{Since the computed error is of the second order, the variation of the temperature field according to the variation of the parameters can be obtained through \textsc{Taylor} approximation. Figure~\ref{fig:u_K2C2} depicts how the interior temperature field varies, while the thermal conductivity value changes. One may note the temperature is strongly influenced as the value of the thermal conductivity decreases.} Figure~\ref{fig:eps_tay_u_k2c2} displays the variation in the error according to each parameter $k^{\,\star}_{\,2}$ or $c^{\,\star}_{\,2}\,$, while the other is fixed to the \emph{a priori} value. The error of the model expansion is acceptable until an order of $\mathcal{O}\bigl(\,10^{\,-2}\,\bigr)\,$. It corresponds to intervals of variation $k_{\,2}^{\star\,\circ} \cdot \bigl[\, 0.2 \,,\, 1.9\,\bigr]$ and $c_{\,2}^{\star\,\circ} \cdot \bigl[\, 0.35 \,,\, 1.9\,\bigr]$ for parameters $k^{\,\star}_{\,2}$ and $c^{\,\star}_{\,2}\,$, respectively. The error is proportional to $\mathcal{O}\Bigl(\, \bigr(\, k_{\,2}^{\,\star} \moins k_{\,2}^{\star\,\circ}\,\bigr)^{\,3} \,\Bigr)$ and $\mathcal{O}\Bigl(\, \bigr(\, c_{\,2}^{\,\star} \moins c_{\,2}^{\star\,\circ}\,\bigr)^{\,3} \,\Bigr)$ as expected from the theoretical results. Figure~\ref{fig:eps_tay_jK2C2} shows how thermal flux changes with the variation of both parameters. While parameters $k_{\,2}^{\,\star}$ and $c_{\,2}^{\,\star}$  vary in the range $k_{\,2}^{\star\,\circ} \cdot \bigl[\, 0.2 \,,\, 1.9\,\bigr]$ and  $c_{\,2}^{\star\,\circ} \cdot \bigl[\, 0.2 \,,\, 1.9\,\bigr]$, the \textsc{Taylor} expansion of the heat flux is satisfactory as the accuracy order equals $\mathcal{O}\bigl(\,10^{\,-2}\,\bigr)\,$. Similar conclusions are made for the thermal loads as a model output. When comparing the order of the error values between the heat flux and the thermal loads, we obtain higher error values for the thermal loads owing to the integration of the heat flux. 

The results of this study support the fact that the \textsc{Taylor} expansion can be used to predict how a model output varies according to changes in parameter values, using the continuous approach to compute the sensitivity coefficients. Information of the model output and its sensitivity coefficients provides an accurate approximation of the model output with parameter variation.

\subsection{Comparison with approaches from the literature}

In the two previous subsections, the use of the sensitivity coefficients and the \textsc{Taylor} expansion to approximate the variation of the output model were validated. Thus, the following section describes the comparison between the continuous approach and three other approaches widely used in the literature for sensitivity analysis: \emph{(i)} the Standardized Regression Coefficients (SRC) method~\citep{Saltelli_2000,Saltelli_2004}, \emph{(ii)} the \textsc{Sobol} method~\citep{Sobol_1990,HOMMA1996}, and the \textsc{Fourier} Amplitude Sensitivity Test (FAST) \citep{CUKIER1978}. The Latin Hypercube Sampling~\citep{Helton2003} {and the Sobol Sequence set~\citep{Sobol_1967} were used to calculate the \textsc{Sobol} indices} in order to cover the parameter ranges $\Omega_{\,k_{\,2}}$ and $\Omega_{\,c_{\,2}}$ effectively. The number of samples $N_{\,s}$ is mentioned below. The interval of variation of the parameters is still set to $\Omega_{\,k_{\,2}} \egal k_{\,2}^{\star\,\circ} \cdot \bigl[\, 0.1 \,,\, 1.9 \,\bigr]$ for $k_{\,2}^{\star}$ and $\Omega_{\,c_{\,2}} \egal c_{\,2}^{\star\,\circ} \cdot \bigl[\, 0.1 \,,\, 1.9 \,\bigr]$ for $c_{\,2}^{\star}\,$. 
The thermal loads are chosen as a model output. The time domain is decreased to $t^{\,\star} \, \in\, [\,0\,,5\,]$ to accelerate the calculation of the sensitivity indices. The space and time discretization steps are set to $\Delta\,x^{\,\star} \egal 10^{\,-2}\,$ and $\Delta\,t^{\,\star} \egal 10^{\,-3}\,$, respectively.

\begin{figure}[!ht] 
\centering
\includegraphics[width=0.5\linewidth]{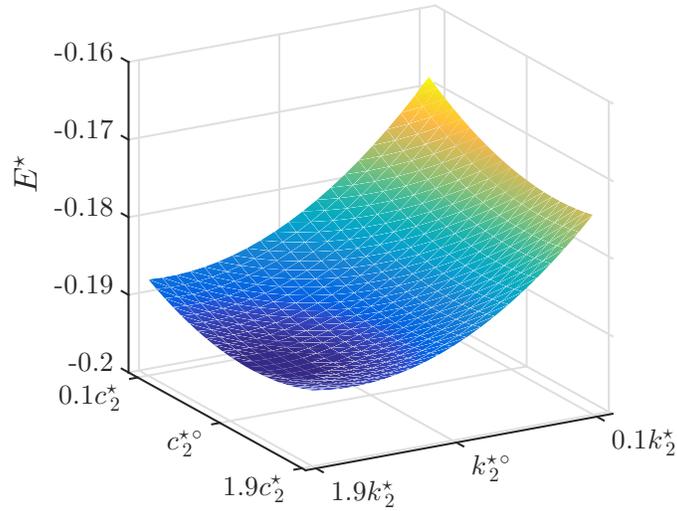}
\caption{Thermal loads variation over the changes in $k^{\,\star}_2$ and $c^{\,\star}_2$ parameters, obtained with the \textsc{Taylor} expansion and the sensitivity coefficients computed with the continuous approach.}
\label{fig:E_k2c2}
\end{figure}

Figure~\ref{fig:E_k2c2} displays how the thermal loads change over the parameter intervals $\Omega_{\,k_{\,2}}$ and $\Omega_{\,c_{\,2}}$. This variation is obtained through the \textsc{Taylor} expansion series. One may conclude that the thermal conductivity $k^{\,\star}_2$ has a greater influence on the output than the volumetric heat capacity $c^{\,\star}_2$. The subsequent results of the sensitivity indices reinforce this conclusion.

First, the SRC method is applied to measure the impact of the second--layer parameters on the model output. To determine the SRC coefficients, the Latin Hypercube Samplings on the domains $\Omega_{\,k_{\,2}}$ and $\Omega_{\,c_{\,2}}$  are taken with a different number of samplings $ N_{\,s}$. Then, the direct computation of the annual thermal loads is performed and through the Lasso optimization technique their corresponding standardized regression coefficients are evaluated. The regression coefficient of parameter $p_i$ is denoted as $\beta_{\,\mathrm{SRC}}\,[\,p_i\,]$. Since $\beta^{\,2}_{\,\mathrm{SRC}}\,[\,p_i\,]$ represents the share of variance \citep{Iooss2015}, results are normalized according to the formula:
\begin{align}
\label{eq:SRC_norm}
S_{\,\mathrm{SRC}}\,[\,p_{i}\,] \egal \displaystyle \dfrac{\beta_{\,\mathrm{SRC}}^{\,2}\,[\,p_i\,]}{\displaystyle \sum_{j\egal 1}^M\,\beta_{\,\mathrm{SRC}}^{\,2}\,[\,p_j\,]}\,,
\end{align}      
which allows us to compare SRC results with metrics, defined in Section~\ref{sec:metrics}.
\ajumabek{Similarly, the SRRC results were normalized.} The results are reported in Table~\ref{tab:SRC_coef} and Table~\ref{tab:SRC_valid}. 
\ajumabek{It can be noted that at least $150$ samples are required to obtain relevant results of the SRC and SRRC indices. The CPU time to calculate the thermal loads $150$ times is equal to $240\,\mathsf{s}$.  Additionally, it can be noted that SRRC coefficient values are greater than SRC indices, indicating that there is an interaction effect between parameters. The values of the regression coefficients metric demonstrate that parameter $k^{\,\star}_2$  has a greater impact on the thermal loads than parameter $c^{\,\star}_2$. }

Next, \textsc{Sobol} indices were calculated. The first--order sensitivity indices and total--effect indices were estimated for both second layer parameters. Results for a different number of model outputs are presented in Table~\ref{tab:Sobol_valid}, where $S_{\,\mathrm{SOB}}^{\,1}\,[\,p_{\,i}\,]$ is the first--order sensitivity index for parameter $p_{\,i}$, and $S_{\,\mathrm{SOB}}^{\,\mathrm{tot}}\,[\,p_{\,i}\,]$ is the total--effect sensitivity index for parameter $p_{\,i}$. It can be seen from the data in Table~\ref{tab:Sobol_valid} that acceptable index values start from the number of output samples equal to or greater than $1024$. Indeed, the total index values are greater than the first ones. The computation of \textsc{Sobol} indices is a very expensive procedure. The whole estimation process for $4096$ model outputs takes almost $3\,\mathsf{h}$. One may conclude that the thermal loads are more sensitive to the thermal conductivity parameter $k^{\,\star}_2$.  

The third step is to calculate the sensitivity indices using the FAST approach with a Random Balance Design \citep{TARANTOLA_2006}. The RBD--FAST approach calculates the first--order sensitivity indices with a reduced computational cost compared to the \textsc{Sobol} approach. As shown in Table~\ref{tab:FAST_valid}, samples of $500$ model outputs provide satisfactory results for this method, for which the computational time amounts to $17\,\mathsf{min}$. {One may note that sample sizes of $100$ and $200$ \citep{TARANTOLA_2006, EASI_2010} are not sufficient to analyze the sensitivity indices.} The results are in agreement with the \textsc{Sobol} indices, supporting the fact that the volumetric heat capacity parameter $c^{\,\star}_2$ has a smaller impact on the thermal loads.

Finally, the sensitivity coefficient metrics, introduced in Section~\ref{sec:metrics}, were computed. The local metric $\eta$ of a parameter $p$ considers the variation of output computed with fixed value of the parameter $p$, in this case, \emph{a priori} parameter value, while the global metric $\nu$ additionally explores the variation of the parameter $p$ and its influence on the output. The local metric values for thermal conductivity and volumetric heat capacity are $\eta_{\,k^{\,\star}_2} \egal 0.86$ and $\eta_{\,c^{\,\star}_2} \egal 0.14$, respectively. The duration of the calculation is $12\,\mathsf{s}$. 

On the other hand, the accuracy of the global metrics depends on the discretization of the parameter domains $\Omega_{\,k_{\,2}}$ and $\Omega_{\,c_{\,2}}$. Using a first--order \textsc{Taylor} expansion~\eqref{eq:FTaylor_E} of the thermal loads, the order of accuracy are $\mathcal{O}\Bigl(\, \max \bigl(\, \Delta^{\,2}_{\,k2^{\,\star}} \,,\, \Delta^{\,2}_{\,c2^{\,\star}}\,\bigr)\,\Bigr)\,$. Thus, if the discretization steps $\Delta_{\,k2^{\,\star}}$ and $\Delta_{\,c2^{\,\star}}$ are of the order $\mathcal{O}\,(\,10^{\,-2}\,)$, then the order of accuracy is equal to $\mathcal{O}\,(\,10^{\,-4}\,)$. Using these assumptions $N_{\,k2}\egal 20 $ and $N_{\,c2} \egal 20$ discrete parameter values ensure $\mathcal{O}\,(\,10^{\,-4}\,)$ order of accuracy. And values of $N_{\,k2}\egal 5 $ and $N_{\,c2} \egal 5$ maintain error at the  $\mathcal{O}\,(\,10^{\,-2}\,)$ order. The global sensitivity metrics for both discretization cases are given in Table~\ref{tab:metrics_valid}. Both metrics show that thermal conductivity has a greater influence on the thermal loads. 

\ajumabek{Additionally, the interaction between parameters was computed, it corresponds to the variable $\displaystyle \frac{\partial^{\,2}{E}}{\partial{k^{\,\star}_2}\,\partial{c^{\,\star}_2}}$, if the thermal loads are selected as model output. The crossed-based sensitivity measure is defined as follows:
\begin{align*}
    \nu_{\,k2\,,\,c2} \egal \displaystyle \int_{\,\Omega_{\,k_{\,2}}}\,\int_{\,\Omega_{\,c_{\,2}}} \, \frac{\partial^{\,2}{E}}{\partial{k^{\,\star}_2}\,\partial{c^{\,\star}_2}}\,\mathrm{d}\,k^{\,\star}_2\,\mathrm{d}\,c^{\,\star}_2\,.
\end{align*}
As demonstrated in~\citep{ROUSTANT_2014} $\nu_{\,k2\,,\,c2}$ provides maximal bound for the total \textsc{Sobol} indices of an interaction between two inputs, and if it is equal to $0$, there is no interaction between parameters. In this validation case study, value of $\nu_{\,k2\,,\,c2}$ is at order $\mathcal{O}\,(\,10^{\,-2}\,)$. It corresponds to the fact, that there is a small interaction between parameters. It is also consistent with the fact that there is almost no differences between the first and total-order \textsc{Sobol} indices. However, the computational cost increases, since this metric requires variation of both parameters simultaneously. In this particular case, it requires $N_{\,k2}\, \times \, N_{\,c2}$ model evaluations, or $400$ samples. This number is still smaller than the number of samples necessary for \textsc{Sobol} total-order indices. }

\ajumabek{ Additional verification is carried out by comparing the global metrics in Table~\ref{tab:metrics_valid} with the \textsc{Sobol} total variance $\mathrm{D}^{\,\mathrm{tot}}$. One may note that the results verify inequality~\eqref{eq:metrics_pi} for both cases. The global metrics provide a maximal bound for the \textsc{Sobol} total variance $\mathrm{D}^{\,\mathrm{tot}}$.  When  ${N_{\,p} \egal 20}$ for the parameter $c^{\,\star}_2$  even approximation~\eqref{eq:metrics_12} is true. One may note that global estimators provide an upper boundary for the total variance, thus, it can be used as quantitative measure to select significant parameters. This is consistent with the results from  \citep{Sobol2009}. }

\begin{table}[]
\centering
\caption{\ajumabek{Results of the SRC and SRRC indices}.}
\begin{tabular}{|c|c|c|c|c|c|}
\hline
\ajumabek{Number of model samples} $N_{\,s}$ & $20$ & $50$  & $100$ & $150$ & $\boldsymbol{200}$  \\ \hline
${\,\mathrm{SRC}}\,[\,k^{\,\star}_2\,]$       & $-0.65$    & $-0.77$  & $-0.75$  & $-0.7$ &  $\boldsymbol{-0.7}$ \\ \hline
${\,\mathrm{SRC}}\,[\,c^{\,\star}_2\,]$      & $-0.45$    & $-0.49$   & $-0.49$ & $-0.46$  & $\boldsymbol{-0.46}$ \\ \hline
${\,\mathrm{SRRC}}\,[\,k^{\,\star}_2\,]$       & $-0.66$    & -$0.85$  & $-0.82$  & $-0.78$ & $\boldsymbol{-0.75}$ \\ \hline
${\,\mathrm{SRRC}}\,[\,c^{\,\star}_2\,]$      & $-0.57$    & $-0.5$    & $-0.51$ & $-0.52$ &  $\boldsymbol{-0.54}$ \\ \hline
\end{tabular}
\label{tab:SRC_coef}
\end{table}

\begin{table}[]
\centering
\caption{\ajumabek{Results of the normalized SRC and SRRC indices}.}
\begin{tabular}{|c|c|c|c|c|c|c|}
\hline
\ajumabek{Number of model samples} $N_{\,s}$ & $20$ & $50$  & $100$ & $150$ &  $\boldsymbol{200}$  \\ \hline
$S_{\,\mathrm{SRC}}\,[\,k^{\,\star}_2\,]$       & $0.67$    & $0.71$  & $0.7$  & $0.69$ &  $\boldsymbol{0.69}$ \\ \hline
$S_{\,\mathrm{SRC}}\,[\,c^{\,\star}_2\,]$      & $0.33$    & $0.29$   & $0.3$ & $0.31$ &  $\boldsymbol{0.31}$ \\ \hline
$S_{\,\mathrm{SRRC}}\,[\,k^{\,\star}_2\,]$       & $0.58$    & $0.75$  & $0.72$  & $0.69$ &  $\boldsymbol{0.67}$ \\ \hline
$S_{\,\mathrm{SRRC}}\,[\,c^{\,\star}_2\,]$      & $0.42$    & $0.25$    & $0.28$ & $0.31$ & $\boldsymbol{0.33}$ \\ \hline
\end{tabular}
\label{tab:SRC_valid}
\end{table}

\begin{table}[]
\centering
\caption{First and total \textsc{Sobol} indices.}
\begin{tabular}{|c|c|c|c|c|c|c|c|}
\hline
Number of model samples $N_{\,s}$ & 
$128$ &
$256$ &
$512$ &
$\boldsymbol{1024}$ &
$2048$ &
$4096$ \\ \hline
$S_{\,\mathrm{SOB}}^{\,1}\,[\,k^{\,\star}_2\,]$  &
$1$ &
$0.69$ &
$0.68$ &
$\boldsymbol{0.66}$ &
$0.66$ &
${0.66}$ \\ \hline
$S_{\,\mathrm{SOB}}^{\,1}\,[\,c^{\,\star}_2\,]$  &
$0$ &
$0.37$ &
$0.34$ &
$\boldsymbol{0.33}$ &
$0.33$ &
${0.33}$ \\ \hline
$S_{\,\mathrm{SOB}}^{\,\mathrm{tot}}\,[\,k^{\,\star}_2\,]$  &
$1$ &
$0.63$ &
$0.65$ &
$\boldsymbol{0.67}$ &
$0.67$ &
${0.67}$ \\ \hline
$S_{\,\mathrm{SOB}}^{\,\mathrm{tot}}\,[\,c^{\,\star}_2\,]$&
$0$ &
$0.31$ &
$0.32$ &
$\boldsymbol{0.34}$ &
$0.34$ &
${0.34}$ \\ 
\hline 
$\mathrm{D}^{\,\mathrm{tot}}_{\,k_{\,2}}$ &
$0.0015$ &
$4.9\,\cdot\,10^{\,-4}$ &
$4.45\,\cdot\,10^{\,-4}$ &
$\boldsymbol{4.22\,\cdot\,10^{\,-4}}$ &
$4.14\,\cdot\,10^{\,-4}$ &
${4.08\,\cdot\,10^{\,-4}}$ \\ 
\hline 
$\mathrm{D}^{\,\mathrm{tot}}_{\,c_{\,2}}$ &
$0.0015$ &
$9.1\,\cdot\,10^{\,-4}$ &
$8.69\,\cdot\,10^{\,-4}$ &
$\boldsymbol{8.46\,\cdot\,10^{\,-4}}$ &
$8.37\,\cdot\,10^{\,-4}$ &
${8.31\,\cdot\,10^{\,-4}}$ \\  \hline
\end{tabular}
\label{tab:Sobol_valid}
\end{table}

\begin{table}[]
\centering
\caption{First--order sensitivity indices using RBD--FAST.}
\label{tab:FAST_valid}
\begin{tabular}{|c|c|c|c|c|c|c|c|c|c|}
\hline
Number of model samples $N_{\,s}$ & $100^{\,\ast}$ & $200^{\,\ast}$ & $300$ & $400$ & $\boldsymbol{500}$  & $1000$ & $1500$ & $2000$ & ${3000}$ \\ \hline
$S_{\,\mathrm{FAST}}^{\,1}\,[\,k^{\,\star}_2\,]$      & $\emph{0.76}$ &     $\emph{0.76}$ & $0.69$ & $0.69$ & $\boldsymbol{0.66}$ & $0.68$ & $0.65$ & $0.66$ & ${0.66}$ \\ \hline
$S_{\,\mathrm{FAST}}^{\,1}\,[\,c^{\,\star}_2\,]$       & $\emph{0.5}$ &     $\emph{0.49}$ & $0.34$ & $0.32$  & $\boldsymbol{0.32}$ & $0.37$ & $0.30$ & $0.33$ & ${0.32}$ \\ \hline
\end{tabular}
\end{table}

\begin{table}[]
\centering
\caption{\ajumabek{Global sensitivity coefficient metrics.}}
\label{tab:metrics_valid}
\begin{tabular}{|c|c|c|c|}
\hline
Cardinal of the intervals $\Omega_{\,k_{\,2}}$ and $\Omega_{\,c_{\,2}}$  
& $N_{\,k2} \egal N_{\,c2} \egal 5$
& $N_{\,k2} \egal N_{\,c2} \egal 20$ 
& $\mathrm{D}^{\,\mathrm{tot}}_{\,p_{\,i}}$ \\ \hline
Global metric $\gamma_{\,k_{\,2}}$ 
& $0.8 $
& $\boldsymbol{0.77}$ & \\ \hline
Global metric $\gamma_{\,c_{\,2}}$ 
& $0.2$
& $\boldsymbol{0.23}$ & \\ \hline
Global estimator $\nu_{\,k_{\,2}} \, / \, \pi^{\,2}$ 
& $101\,\cdot\,10^{\,-4}$   
& $\boldsymbol{25\,\cdot\,10^{\,-4}}$ 
& $4.22\,\cdot\,10^{\,-4}$ \\ \hline
Global estimator $\nu_{\,c_{\,2}} \, / \, \pi^{\,2}$ 
& $34\,\cdot\,10^{\,-4}$ 
& $\boldsymbol{10\,\cdot\,10^{\,-4}}$ 
& $8.46\,\cdot\,10^{\,-4}$ \\ \hline
Global estimator $\nu_{\,k_{\,2}} \, / \, 12$ 
& $83\,\cdot\,10^{\,-4}$   
& $21\,\cdot\,10^{\,-4}$
& $4.22\,\cdot\,10^{\,-4}$ \\ \hline
Global estimator $\nu_{\,c_{\,2}} \, / \, 12$ 
& $28\,\cdot\,10^{\,-4}$ 
& $8.5\,\cdot\,10^{\,-4}$ 
& $8.46\,\cdot\,10^{\,-4}$ \\ \hline
\end{tabular}
\end{table} 

Table~\ref{tab:all_valid} lists the sensitivity index values for each approach and compares the computational time. One may conclude that the results are similar to each other in a qualitative way. Specifically, the thermal conductivity parameter $k^{\,\star}_2$  has a greater effect on the thermal loads. Quantitatively, the metrics of the continuous approach are validated thanks to the accuracy of the \textsc{Taylor} expansion and the consistency with the theoretical results from \citep{Sobol2009}. This approach also provides sensitivity information faster than the RBD--FAST or \textsc{Sobol} methods. Indeed, these two methods require a high number of samples to provide accurate results, increasing the computational time $85$- and $900$-fold, respectively. 
Additionally, the computational time of the computation of the global metric can be decreased. The value presented in Table~\ref{tab:all_valid} refers to the number of parameter samples equal to  ${N_{\,p} \egal 20}$, while a smaller number of parameter samples give results faster {at the cost of losing accuracy.}

\ajumabek{From a theoretical point of view, the computational time is proportional to the number of output evaluations. For derivative based sensitivity analysis the computational cost is evaluated at order $\mathcal{O}\,(\,N_{\,p}\,\times\,N_{\,\Delta_{\,p}}\,)$, where $N_{\,p}$ is the number of parameters, and $N_{\,\Delta_{\,p}}$ is the number of discretizations in the parameter space. The appropriate number of discretizations can be calculated using first-order \textsc{Taylor} series expansion. If the discretization step $\Delta_{\,p}$ of the parameter space is at order $\mathcal{O}\,(\,10^{\,-2}\,)$, then the approximation has error at order $\mathcal{O}\,(\,10^{\,-4}\,)$. \\
In practice, an acceptable error's order is $\mathcal{O}\,(\,10^{\,-2}\,)$, thus, $\Delta_{\,p}$ is assumed to be at order $\mathcal{O}\,(\,10^{\,-1}\,)$. This corresponds to $N_{\,\Delta\,p}$ between $10$ and $20$, mostly, for all dimensionless parameters. Thus, the number of output evaluations is proportional to $\mathcal{O}\,(\,N_{\,p}\,\times\,10\,)$. \\
This evaluation can be compared to the other approaches from literature~\citep{Iooss2015}. For example, the regression-based methods have the same order of output computation. However, \textsc{FAST} methods require at least $\mathcal{O}\,(\,N_{\,p}\,\times\,10^{\,2}\,)$, which is $10$ times greater than derivative based approach. Additionally, it has at least $100$ times smaller computational cost compare to the variance-based methods, since \textsc{Sobol} approach requires $\mathcal{O}\,(\,N_{\,p}\,\times\,10^{\,3}\,)$ number of model evaluations.}

\begin{table}
\centering
\caption{Summary of the sensitivity coefficients and CPU time for each approach.}
\label{tab:all_valid}
\resizebox{\textwidth}{!}{
\begin{tabular}{|c|c|c|c|c|c|}
\hline
Method  & Sensitivity for $k^{\,\star}_2$   & Sensitivity for $c^{\,\star}_2$   & CPU time $ [\,\mathsf{min}\,]$   & CPU time $[\,\o\,]$ & Number of model samples $N_{\,s}$   \\ \hline
Continuous, local metric & $0.86$ & $0.14$ & $0.2$   & $t_{\,0}$ & 1   \\ \hline
Continuous, global metric     & $0.77$ & $0.23$ & $5$ & $25 \cdot t_{\,0}$  & 20 \\ \hline
\ajumabek{SRRC} & $0.67$ & $0.33$ & $4$   & $20 \cdot t_{\,0}$ & $150$    \\ \hline
\textsc{Sobol}     & $0.66$ & $0.31$ & $180$     & $900 \cdot t_{\,0}$ & 4096 \\ \hline
RBD--FAST & $0.66$ & $0.32$ & $17$  & $85 \cdot t_{\,0}$ & 500  \\ \hline
\end{tabular}}
\end{table}

\begin{answer}
\subsection{Summary of the results for the validation case}
On the basis of the present results, one may conclude the following:
\begin{itemize}
\item the comparison between the numerical and reference solution validates the implementation of the numerical model and its sensitivity coefficients;
\item the discrete partial derivatives lack accuracy and efficiency to compute sensitivity metrics; however, the continuous partial derivatives enable us to overcome these drawbacks;
\item the \textsc{Taylor} series expansion is a useful tool to evaluate continuously how the model output varies with the parameter variation using the sensitivity coefficients;
\item the partial derivatives can be used as global sensitivity indices, as demonstrated in previous works~\citep{Sobol2009};
\item the derivative based approach has a similar evaluation of the parameter influence to the results, obtained using the SRC, the \textsc{Sobol}, and the RBD--FAST methods;
\item the derivative based approach has the lowest computational cost among the sensitivity methods presented.
\end{itemize}
\end{answer}
\section{Real--world case study}
\label{sec:case_study}
\begin{figure}[h!]
\begin{center}
\subfigure[\label{fig:wall_real}]{\includegraphics[width=0.45\textwidth]{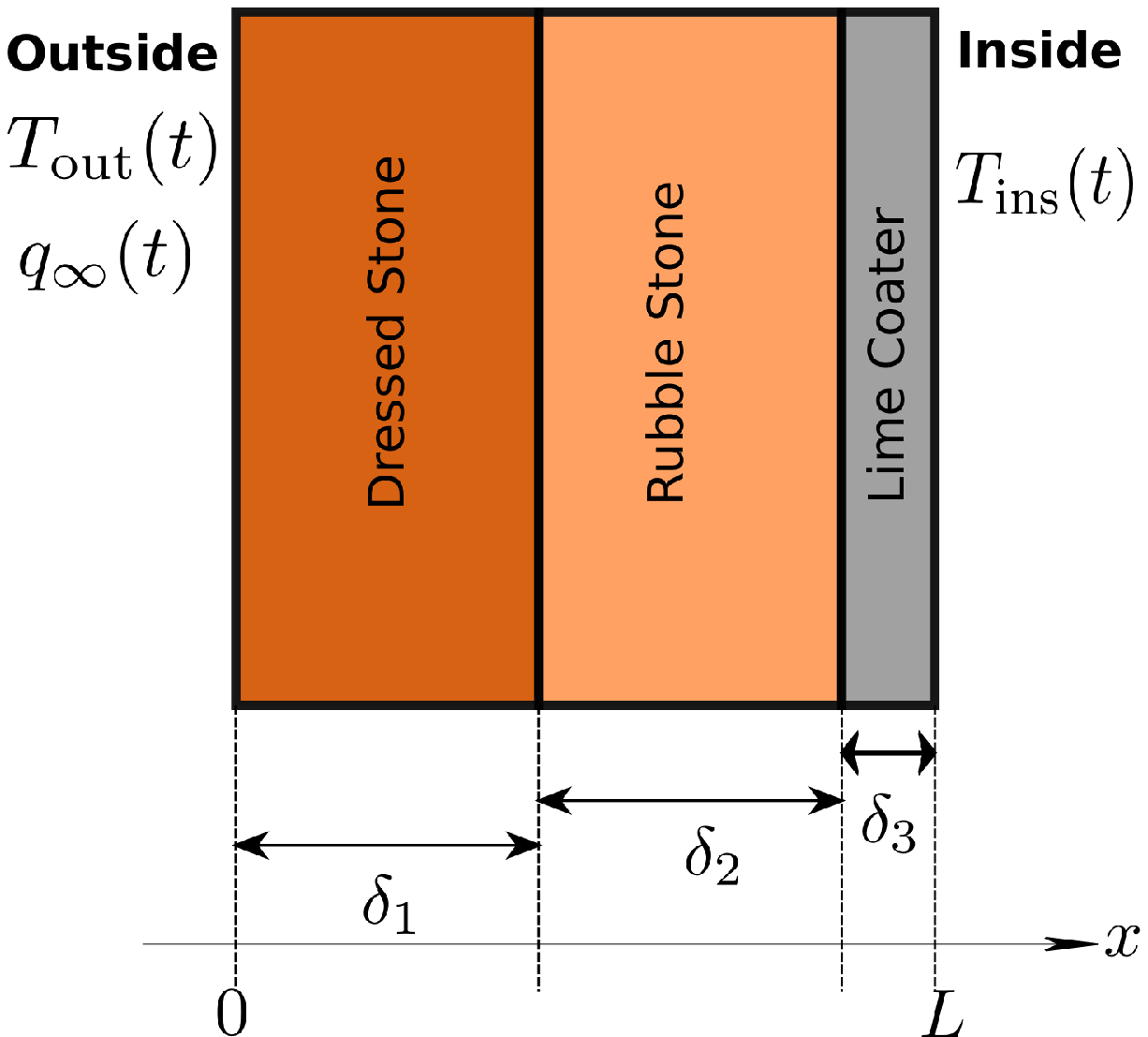}} \hspace{0.25cm}
\subfigure[\label{fig:T_bound}]{\includegraphics[width=0.45\textwidth]{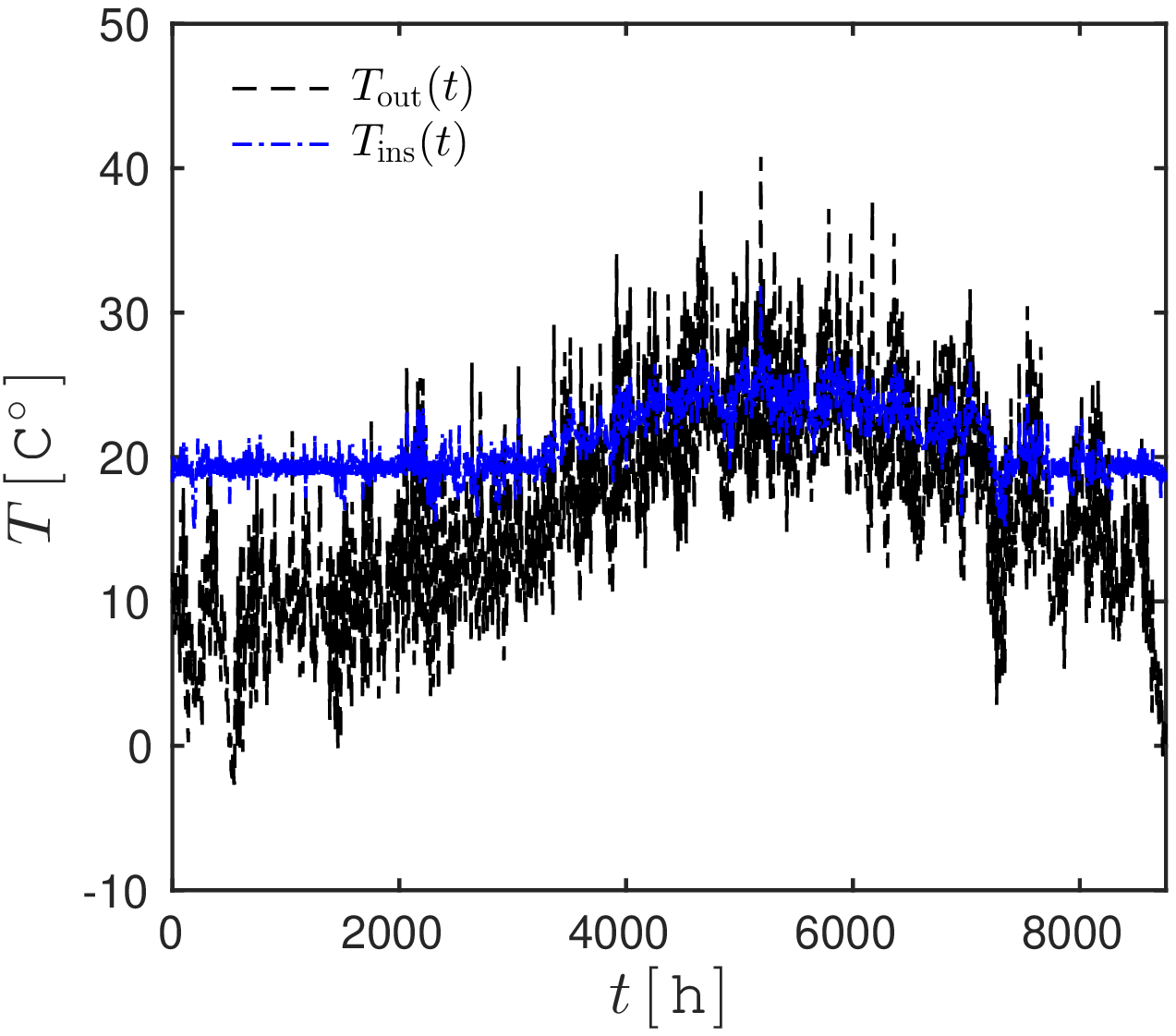}}
\caption{Illustration of the real--world case study \emph{(a)} with the boundary conditions \emph{(b).}}
\end{center}
\end{figure}

\ainagul{\subsection{Presentation of the Case Study}}
\label{sec:wall}
To illustrate the application of the continuous approach, a real--world case study of a  historical building is considered. The house, built in the nineteenth century, is located in Bayonne, France. The west--facing wall of the living room is considered for the study. The wall is composed of three materials: lime coater, rubble stone, and dressed stone, as illustrated in Figure~\ref{fig:wall_real}. To facilitate the thermal comfort of the inhabitants and improve the energy efficiency, it is important to provide an optimal retrofitting solution. For this, the designers and engineers need to perform sensitivity analysis, at a reduced computational cost, to identify the influential parameters on the thermal loads. 

The thermal properties of the wall are given in Table~\ref{tab:mat_properties},  obtained from the French standards \citep{fr_std}. The wall is monitored by sensors, which are placed on both sides of the wall surface and three are installed through the wall. Additionally, the \ainagul{exterior and} interior conditions are observed. Data acquisition took almost 1 year with a time step of $1 \ \mathsf{h}$. Supplementary information on the experimental design can be found in~\citep{BERGER_2016}. In this article, only measurements of the inside and outside \ainagul{air} temperature are used. Their variation over time is shown in Figure~\ref{fig:T_bound}.

\begin{figure}[h!]
\begin{center}
\subfigure[\label{fig:q_bound}]{\includegraphics[width=0.45\textwidth, height=7.5cm]{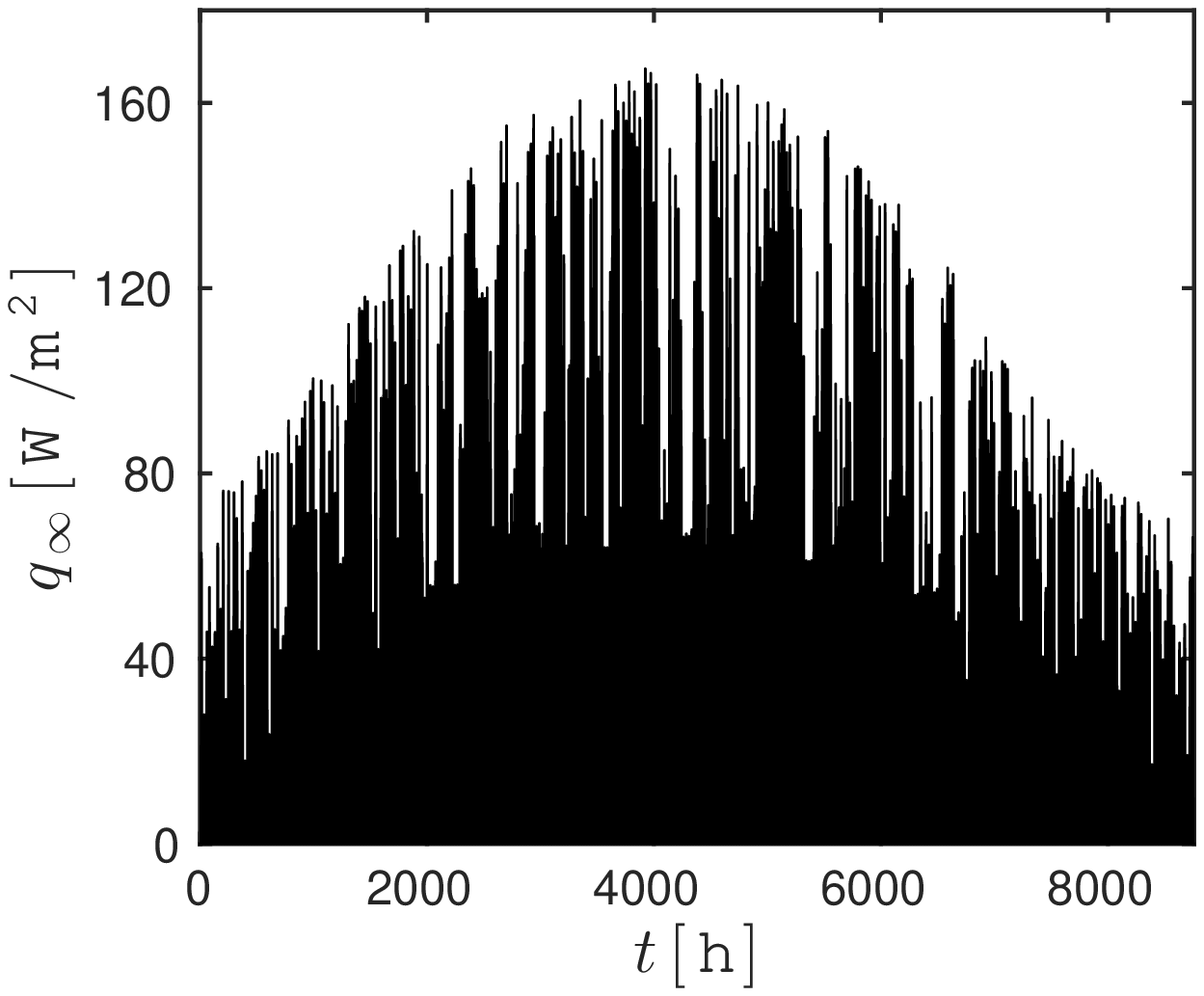}} \hspace{0.25cm}
\subfigure[\label{fig:T_0}]{\includegraphics[width=0.45\textwidth, height=7.6cm]{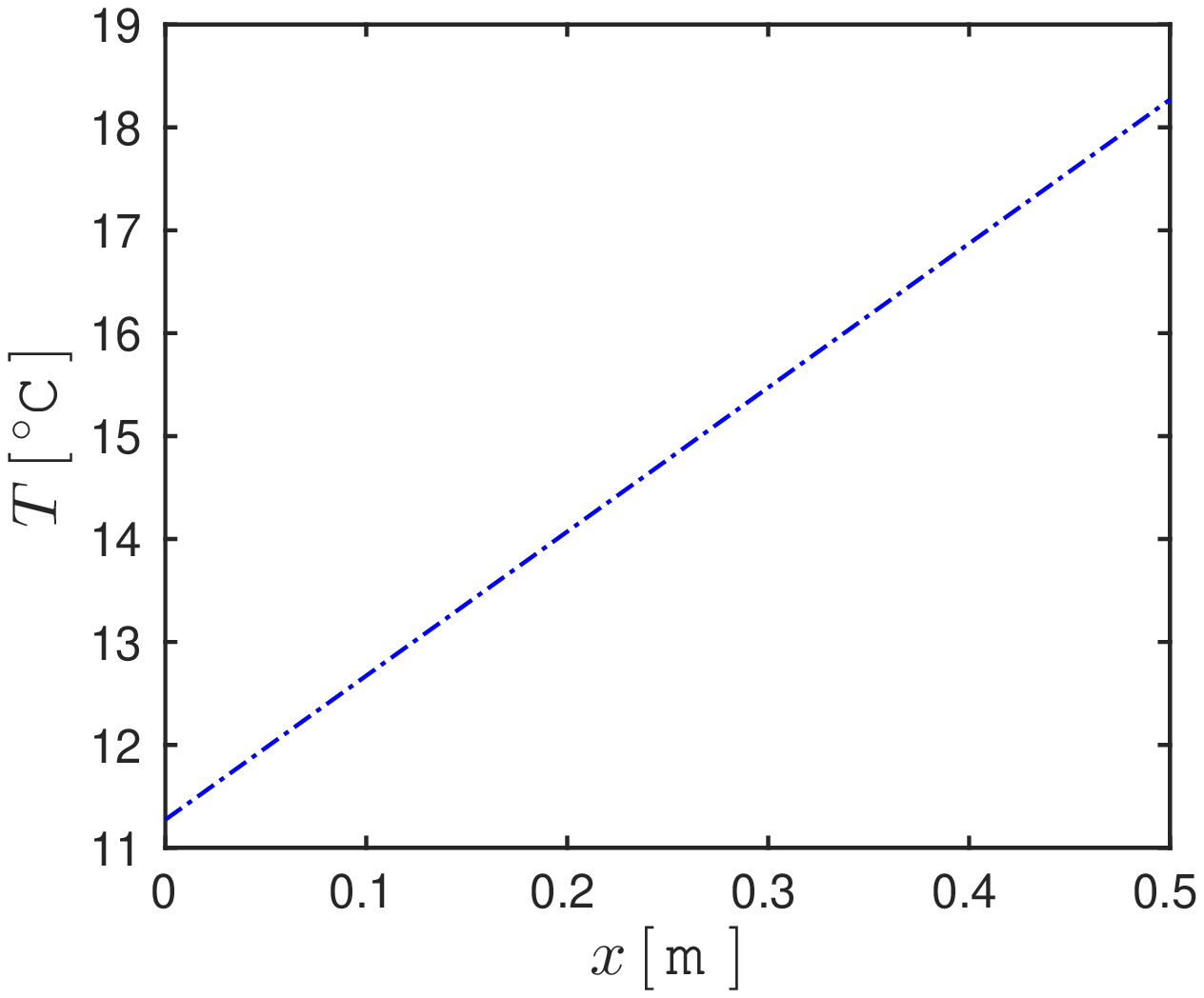}}
\caption{Illustration of the heat flux radiation on the exterior wall surface \emph{(a)} and the initial condition \emph{(b).}}
\end{center}
\end{figure}

The convective heat transfer coefficients are set to $h_{\,\mathrm{L}} \egal 15\, \mathsf{W/m^2\cdot K}$ and $h_{\,\mathrm{R}} \egal 8\, \mathsf{W/m^2\cdot K}$ for the external and internal surfaces, respectively. The heat flux radiation on the external wall is acquired using weather data. \ainagul{It is the total incident radiation, which includes the direct, diffuse, and reflective radiations}. Its variation over time is presented in Figure~\ref{fig:q_bound}. The initial condition of the problem is calculated as \ainagul{a linear interpolation between the temperatures on the wall surfaces. Figure~\ref{fig:T_0} displays a variation of the initial temperature}. 
The whole simulation is performed for 1 year. The temperature and its partial derivatives are calculated using the \DF~explicit scheme. The following reference values are used for the computations: $L \egal 0.5\, \mathsf{m} \,$, $t_{\,\mathrm{ref}} \egal 3600 \,  \mathsf{s}\,$, and $T_{\,\mathrm{ref}} \egal 293.15 \, \mathsf{K}\,$. The values of the thermal conductivity and the heat capacity of the first layer (dressed stone) are applied as reference values $k_{\,\mathrm{ref}}$ and $c_{\,\mathrm{ref}}\,$, respectively.  The space and time discretization are $\Delta x^{\,\star}\egal 10^{-2}$ and $\Delta t^{\,\star}\egal 10^{-2}$, corresponding, from a physical point of view, to $\Delta x \egal 5\,\cdot\,10^{-3}\, \mathsf{m}$ and $\Delta t \egal 3.6\,\mathsf{s}\,$,respectively. 
  
\begin{figure}[h!]
\begin{center}
\subfigure[\label{fig:eps_L}]{\includegraphics[width=0.45\textwidth, height=7.5cm]{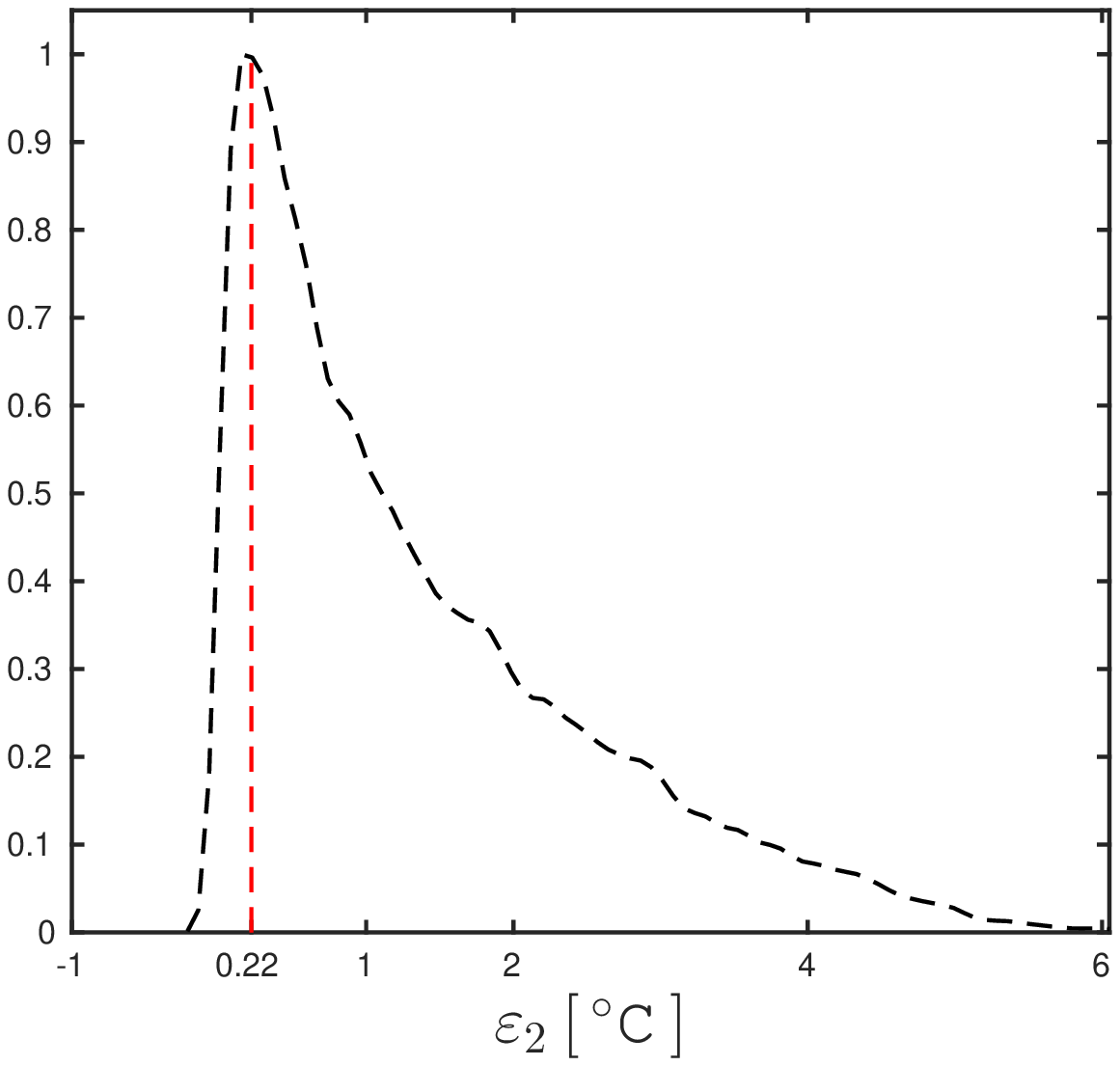}} \hspace{0.25cm}
\subfigure[\label{fig:eps_R}]{\includegraphics[width=0.45\textwidth, height=7.5cm]{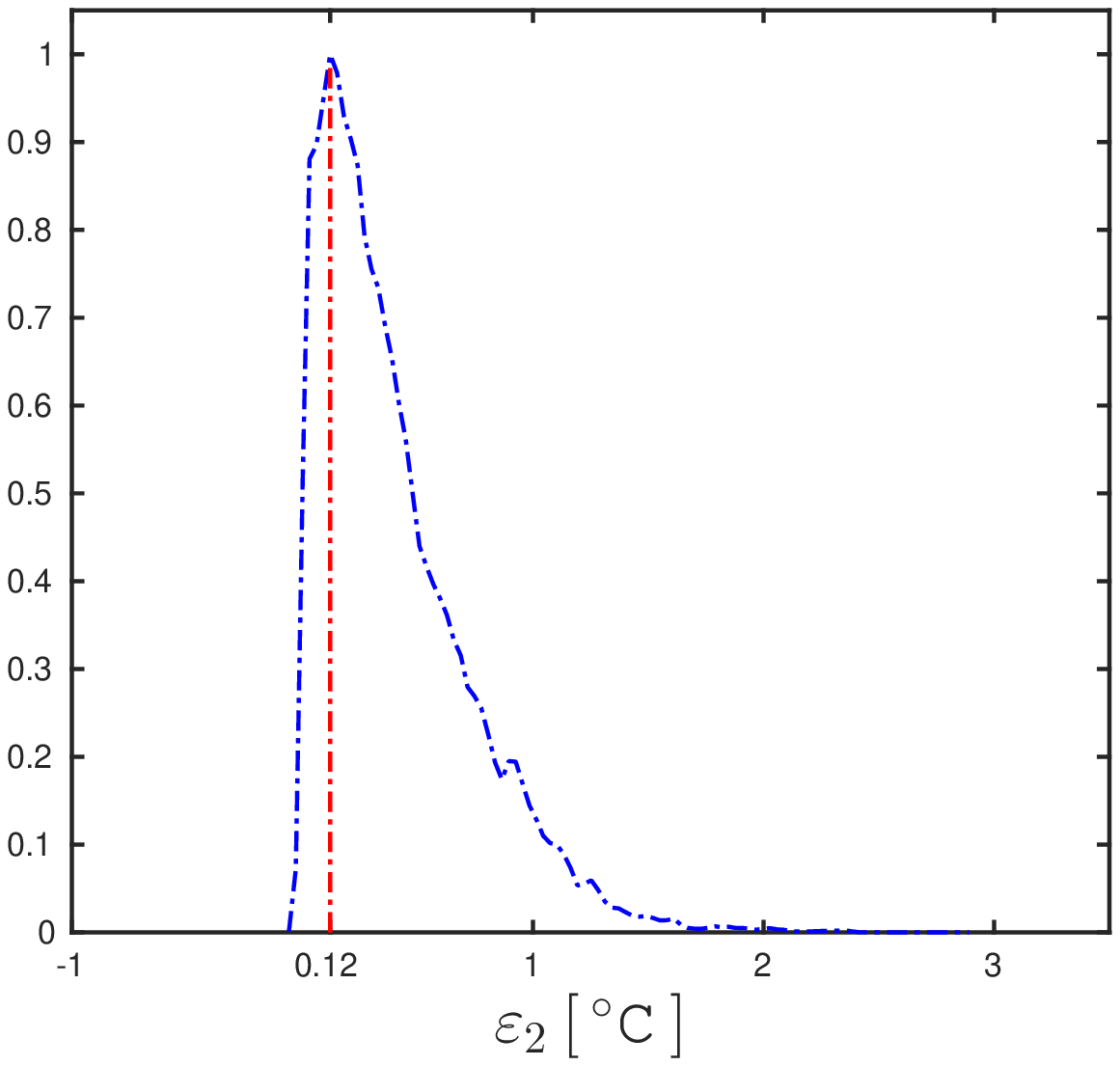}}
\caption{Normalized density function of the error between measurements and the calculated data on the exterior \emph{(a)} and the interior wall surfaces \emph{(b).}}
\label{fig:eps_solU}
\end{center}
\end{figure}


\begin{table}[]
\centering
\caption{\emph{A priori} thermal properties of each layer.}
\begin{tabular}{|c|c|c|c|}
\hline
Layer material     & Thermal conductivity $k_{\,i}^{\,\circ}$ $[\,\mathsf{W/m/K}\,]$ & Heat capacity $c_{\,i}^{\,\circ}$ $[\,\mathsf{J/m^3/K}\,]$ & Thickness $[\,\mathsf{m}\,]$ \\ \hline
Dressed Stone ($i \egal 1$)  &       $1.75$              &    $1.6 \,\times\,10^{\,6} $            &    $0.2$       \\ \hline
Rubble Stone ($i \egal 2$) &     $2.3$                 &       $2.8 \,\times\,10^{\,6} $         &     $0.28$      \\ \hline
 Lime Coater ($i \egal 3$) &        $0.8$              &       $2.2 \,\times\,10^{\,6} $        &       $0.02$    \\ \hline
\end{tabular}
\label{tab:mat_properties}
\end{table}

\subsection{Results and discussion}
\ainagul{The aim of this section is to determine a thermophysical property that has greater impact on the model outputs, and to predict how the outputs vary according to this significant parameter.
First, the error between the computed surface temperature values and the measurements is calculated. Figure~\ref{fig:eps_solU} shows the normalized density function of this error on the exterior and the interior wall surfaces. One may note that the mean values of the error are comparatively small. Therefore, the numerical model can be used for further investigation.}
The impact of the thermal conductivity and heat capacity of each layer on the model outputs is then discussed. Table~\ref{tab:sens_u} presents the values of importance local metric $\eta_{\,k_{\,i}}$  and $\eta_{\,c_{\,i}}$ for sensitivity of the temperature field and the annual thermal loads $E$ computed at the inner surface. It can be noted that thermal conductivity has more impact on these two outputs. Therefore, the heat capacity can be excluded from the scope of the thermal properties to estimate in the framework of a building assessment.    

The global metrics $\nu_{\,k_{\,i}}$  and $\nu_{\,c_{\,i}}$ can now be computed. These values highlight the sensitivity of the annual thermal loads $E$ computed at the inner surface by taking into account uncertainty in the parameters $k_{\,i}$ and $c_{\,i}\,$. A variation of $\pm 50\%$ of the \emph{a priori} values of both thermal parameters for each layer is assumed. \ainagul{The $\pm 50\%$ variation value is chosen to impose larger variations of the outputs. The same results are obtained with a $\pm 20\%$ variation value.} Thus, $\Omega_{\,k_{\,1}} \egal \bigl[\, 0.875 \,,\, 2.625 \,\bigr] \ \mathsf{W/(m \cdot K)}$ and $\Omega_{\,c_{\,1}} \egal  \bigl[\, 0.8\,,\, 2.4 \,\bigr] \,\cdot \,10^{\,6} \ \mathsf{J/m^3/K}$. Next, $\Omega_{\,k_{\,2}} \egal \bigl[\, 1.15 \,,\, 3.45 \,\bigr] \ \mathsf{W/(m \cdot K)}$ and $\Omega_{\,c_{\,2}} \egal  \bigl[\, 1.4 \,,\, 4.2 \,\bigr] \,\cdot \,10^{\,6} \ \mathsf{J/m^3/K}$. Finally, $\Omega_{\,k_{\,3}} \egal \bigl[\, 0.4 \,,\, 1.2 \,\bigr] \ \mathsf{W/(m \cdot K)}$ and $\Omega_{\,c_{\,3}} \egal  \bigl[\, 1.1 \,,\, 3.3 \,\bigr]\,\cdot \,10^{\,6}  \ \mathsf{J/m^3/K}$. 
To ensure an accuracy equal to $\mathcal{O}\,(\,10^{\,-2}\,)$, a step corresponding to $\mathcal{O}\,(\,10^{\,-1}\,)$ is used to discretize the domains $\Omega_{\,k_{\,2}}$ and $\Omega_{\,c_{\,2}}\,$.  Thus, each layer has a different number of discrete parameters: \emph{(layer 1)} $N_{\,k1}\egal N_{\,c1} \egal 5$; \emph{(layer 2)} $N_{\,k2}\egal N_{\,c2} \egal 7$; \emph{(layer 3)} $N_{\,k3}\egal N_{\,c3} \egal 5$. According to the results from Table~\ref{tab:sens_u}, one may also conclude that the influence of the heat capacity on the thermal loads can be ignored.

\begin{table}[!ht]
\centering
\caption{Sensitivity coefficients on the model outputs.}
\begin{tabular}{| c | c  c  c |}
  \hline
  \bf{} & \bf{Lime coater}  & \bf{Rubble stone} & \bf{Dressed stone} \\
  layer $i \egal \bigl\{\,1 \,,\,2 \,,\,3 \,\bigr\}$ & $1$  & $2$ & $3$ \\
  \hline
  \hline
  & \multicolumn{3}{c|}{Temperature} \\
  \hline
Local metric $\eta_{\,k_{\,i}}$ & 0.9999 & 0.9992 & 1 \\
  \hline
Local metric  $\eta_{\,c_{\,i}}$ & $\mathcal{O}\,(\,10^{-5}\,)$ & $\mathcal{O}\,(\,10^{-4}\,)$ & $\mathcal{O}\,(\,10^{-6}\,)$ \\
	\hline
  & \multicolumn{3}{c|}{Thermal loads} \\
  \hline
Local metric   $\eta_{\,k_{\,i}}$ & 0.9999 & 0.9996 & 1 \\
  \hline
Local metric   $\eta_{\,c_{\,i}}$ & 5e-5 & 4.2e-4 & 2.18e-08\\
  \hline
Global metric $\gamma_{\,k_{\,i}}$ & 0.96 & 0.75 & 0.98 \\
  \hline
Global metric $\gamma_{\,c_{\,i}}$ & 0.04 & 0.25 & 0.02\\
  \hline
\end{tabular}
\vspace{-10pt}
\label{tab:sens_u}
\end{table}

\begin{figure}[h!]
\begin{center}
\subfigure[\label{fig:E_tay}]{\includegraphics[width=0.45\textwidth]{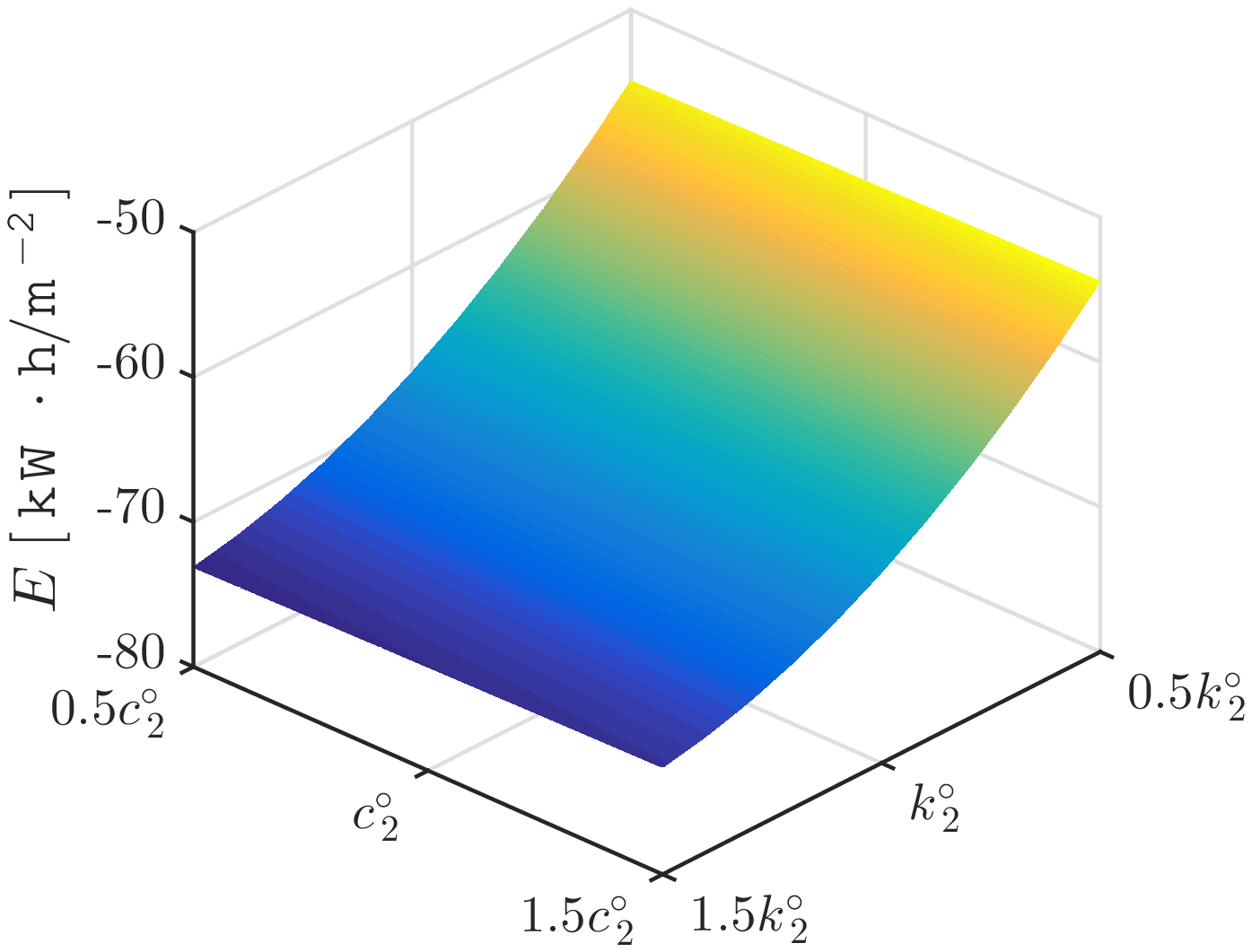}} \hspace{0.3cm}
\subfigure[\label{fig:E_SRC}]{\includegraphics[width=0.45\textwidth]{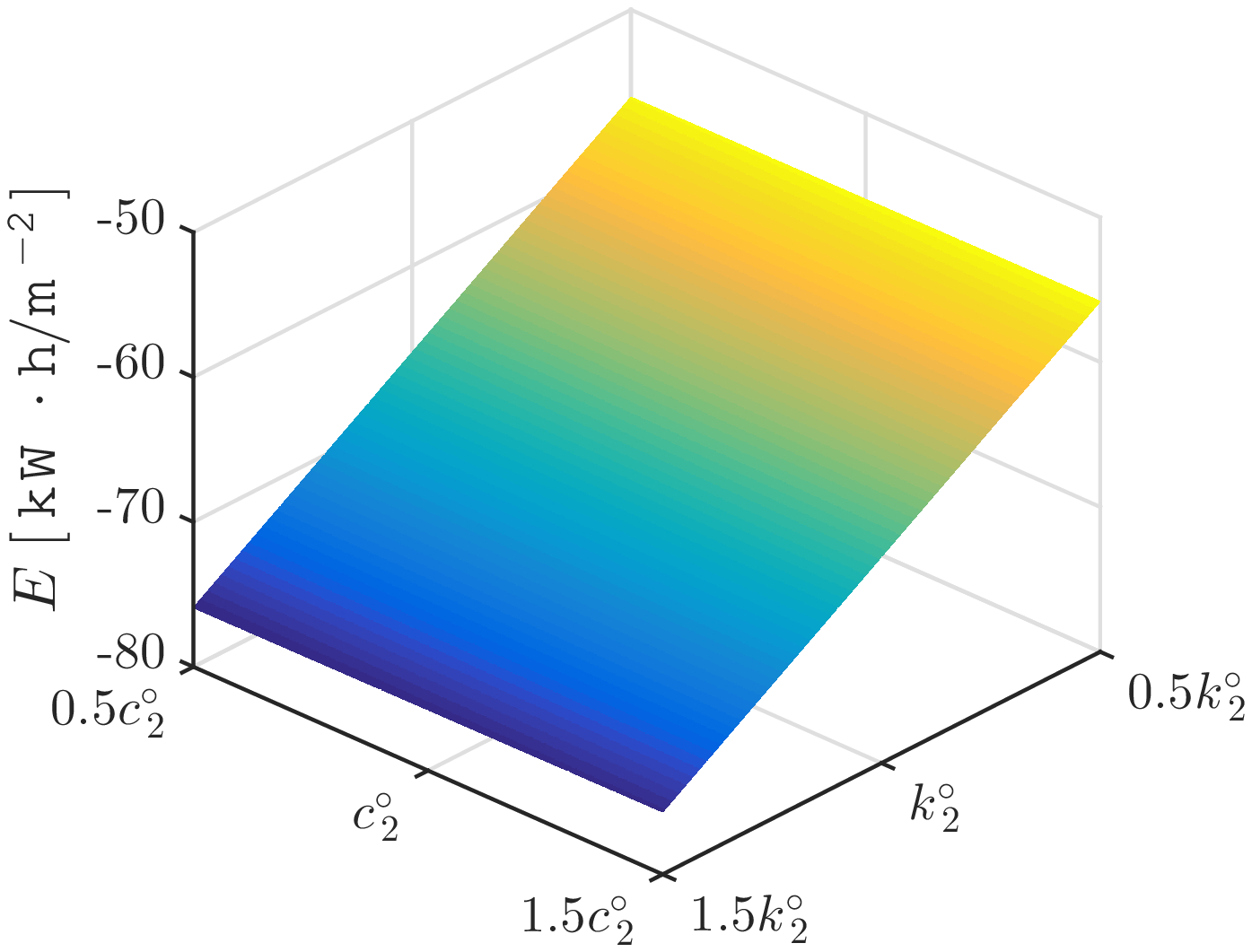}}
\caption{Variation of the annual thermal loads according to changes in the thermal properties of the second layer using \emph{(a)} \textsc{Taylor} expansion and  \emph{(b)} SRC approximation.}
\label{fig:SRC_Taylor}
\end{center}
\end{figure}

The next step presents how the annual thermal loads differ according to the variation of the thermal conductivity and the volumetric heat capacity. Without losing the generality, all further computations are performed within the second layer. Both thermal parameters still vary by $\pm\,50\%$ of their \emph{a priori} values. Subsequently, the annual thermal loads on the interior surface are approximated through the \textsc{Taylor} series expansion using the  sensitivity coefficients. The results of the approximation are displayed in Figure~\ref{fig:E_tay}. It can be seen that the variation of the annual thermal loads is not linear relative to the thermal conductivity $k_{\,2}\,$.

\begin{figure}[h!]
\begin{center}
\subfigure[\label{fig:Day_tay}]{\includegraphics[width=0.45\textwidth]{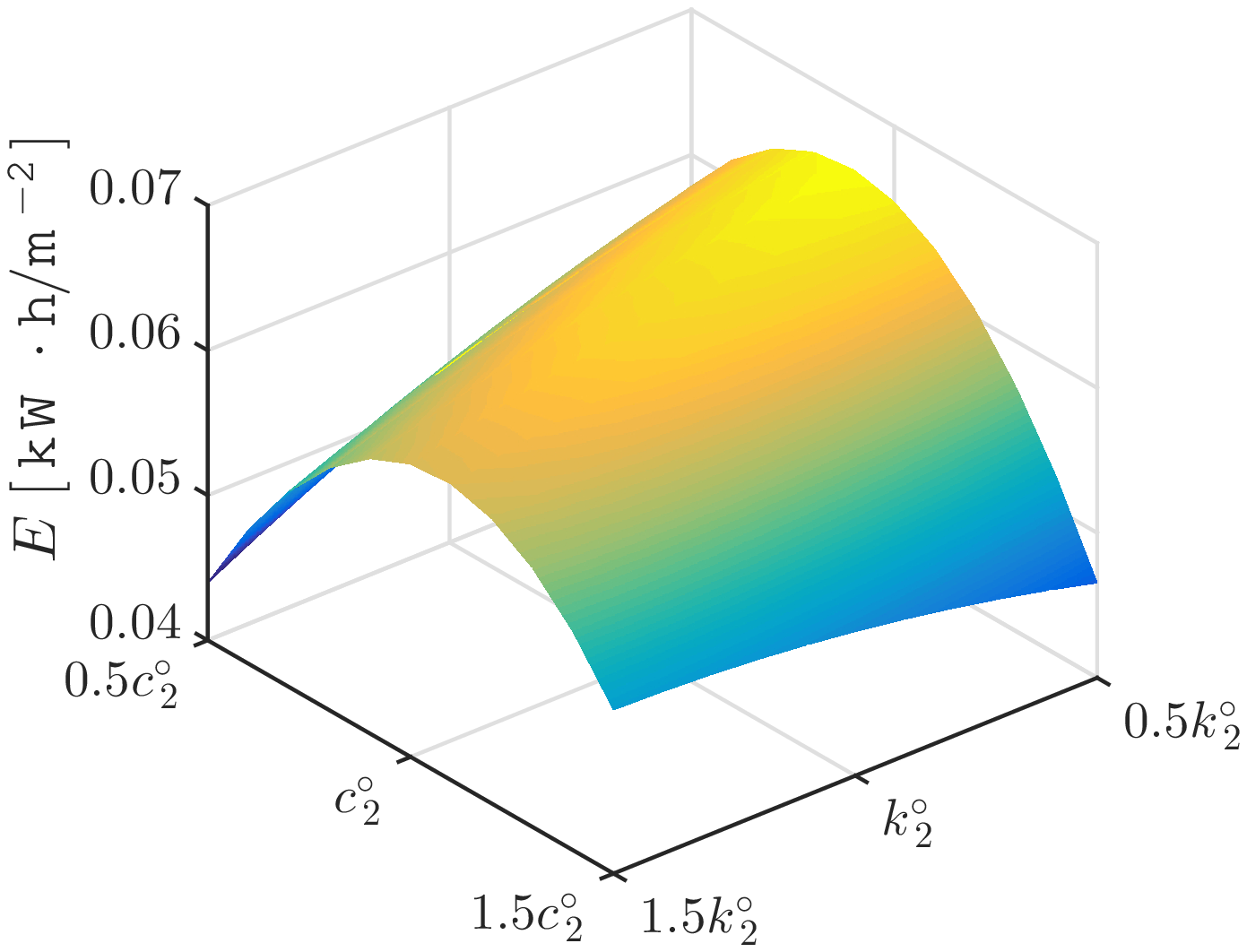}} \hspace{0.3cm}
\subfigure[\label{fig:Day_SRC}]{\includegraphics[width=0.45\textwidth]{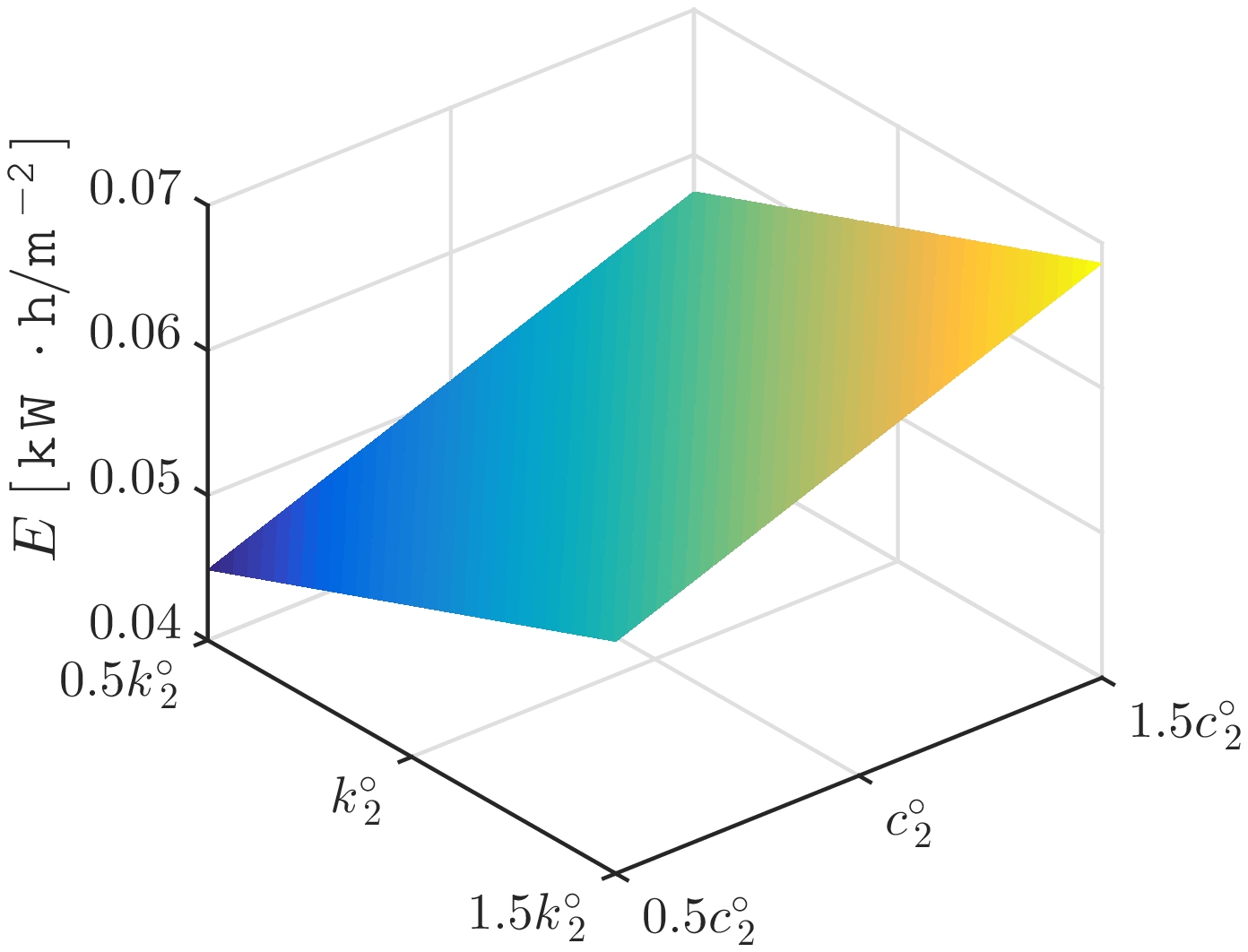}}
\caption{Variation of the daily thermal loads according to changes of the thermal properties of the second layer $k_{\,2}$ and $c_{\,2}$  using \emph{(a)} \textsc{Taylor} expansion and  \emph{(b)} SRC approximation.}
\label{fig:QQ_day}
\end{center}
\end{figure}

According to the previous results, the SRC requires less computational effort compared with \textsc{Sobol} or RBD--FAST. Thus, it is applied to compare its results with the sensitivity coefficients obtained through the continuous approach. To determine the SRC coefficients, the Latin Hypercube Samplings on the domains $\Omega_{\,k_{\,2}}$ and $\Omega_{\,c_{\,2}}$  are taken with a different number of samplings $ N_{\,s}$. Then, the normalized SRC indices using expression~\eqref{eq:SRC_norm} are calculated. Table~\ref{tab:SRC_E} summarizes the results for every value of sampling number and its respective sensitivity index for each parameter. The reported values are about of the same order, which validates the convergence of the approach. One may note that there are similarities between values from Table~\ref{tab:sens_u} and Table~\ref{tab:SRC_E}. Furthermore, the linear approximation of the annual thermal loads is shown in Figure~\ref{fig:E_SRC}. The SRC approach assumes a linear variation of $E\,$, which is not the case as when comparing Figures~\ref{fig:E_tay} and \ref{fig:E_SRC}. This assumption is no longer valid when the nonlinear behavior increases. For instance, the daily thermal loads could not be approximated using the SRC method owing to a nonlinear relationship to parameters $k_{\,2}$ and $c_{\,2}$, as seen in Figure~\ref{fig:QQ_day}.

\begin{table}[ht]
\vspace{-10pt}
\caption{The sensitivity indices, using SRC coefficients.}
\label{tab:SRC_E}
\centering
\begin{tabular}{| c | c | c | c | c |}
  \hline
  Number of model samples $N_{\mathrm{s}}$ & 5 & 10 & 20 & 50\\
  \hline
  $ S_{\,\mathrm{SRC}}\,[\,{k_{\,2}}\,]$ & 0.9995 & 0.9998 & 0.9999 & 0.9999\\
  \hline
  $ S_{\,\mathrm{SRC}}\,[\,{c_{\,2}}\,]$ & $5.0\,\cdot\,10^{-4}$ & $2.3\,\cdot\,10^{-4}$ & $9.9\,\cdot\,10^{-5}$  & $9.8\,\cdot\,10^{-5}$\\
  \hline
\end{tabular}
\vspace{-8pt}   
\end{table}

\begin{figure}[!ht] 
\centering
\includegraphics[width=0.5\linewidth]{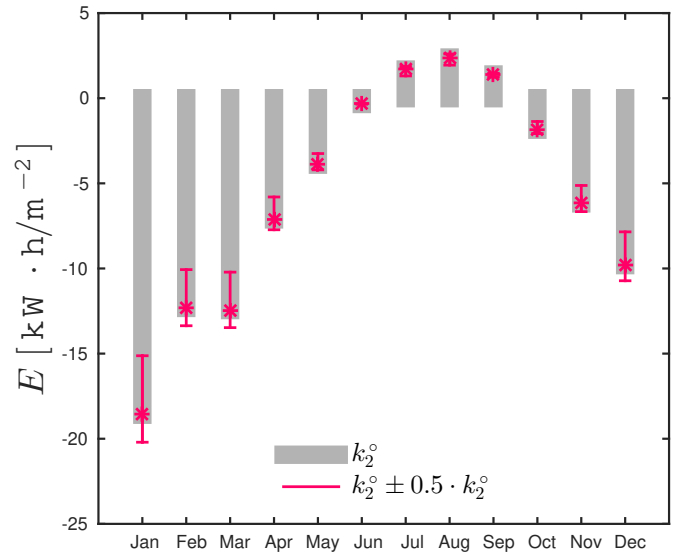}
\caption{The monthly thermal loads and its variation due to $k_{\,2}$ .}
\label{fig:QQ_year}
\end{figure}

Figure~\ref{fig:E_tay} demonstrates that the thermal loads do not vary while the volumetric heat capacity differs. For this reason, the influence of thermal conductivity represents the main interest. Furthermore, the monthly thermal loads values are computed according to the variation range of this parameter. The variations are computed using the \textsc{Taylor} expansion considering the same parameter domain $\Omega_{\,k_{\,2}} \egal \bigl[\, 1.15 \,,\, 3.45 \,\bigr] \ \mathsf{W/(m \cdot K)}$. The results are shown in Figure~\ref{fig:QQ_year}. It appears that the thermal loads are positive only for $3$ months, and, generally, the wall loses energy. One may note how the thermal conductivity strongly affects the thermal loads during the winter season. 

\begin{figure}[h!]
\begin{center}
\subfigure[]{\includegraphics[width=0.45\textwidth]{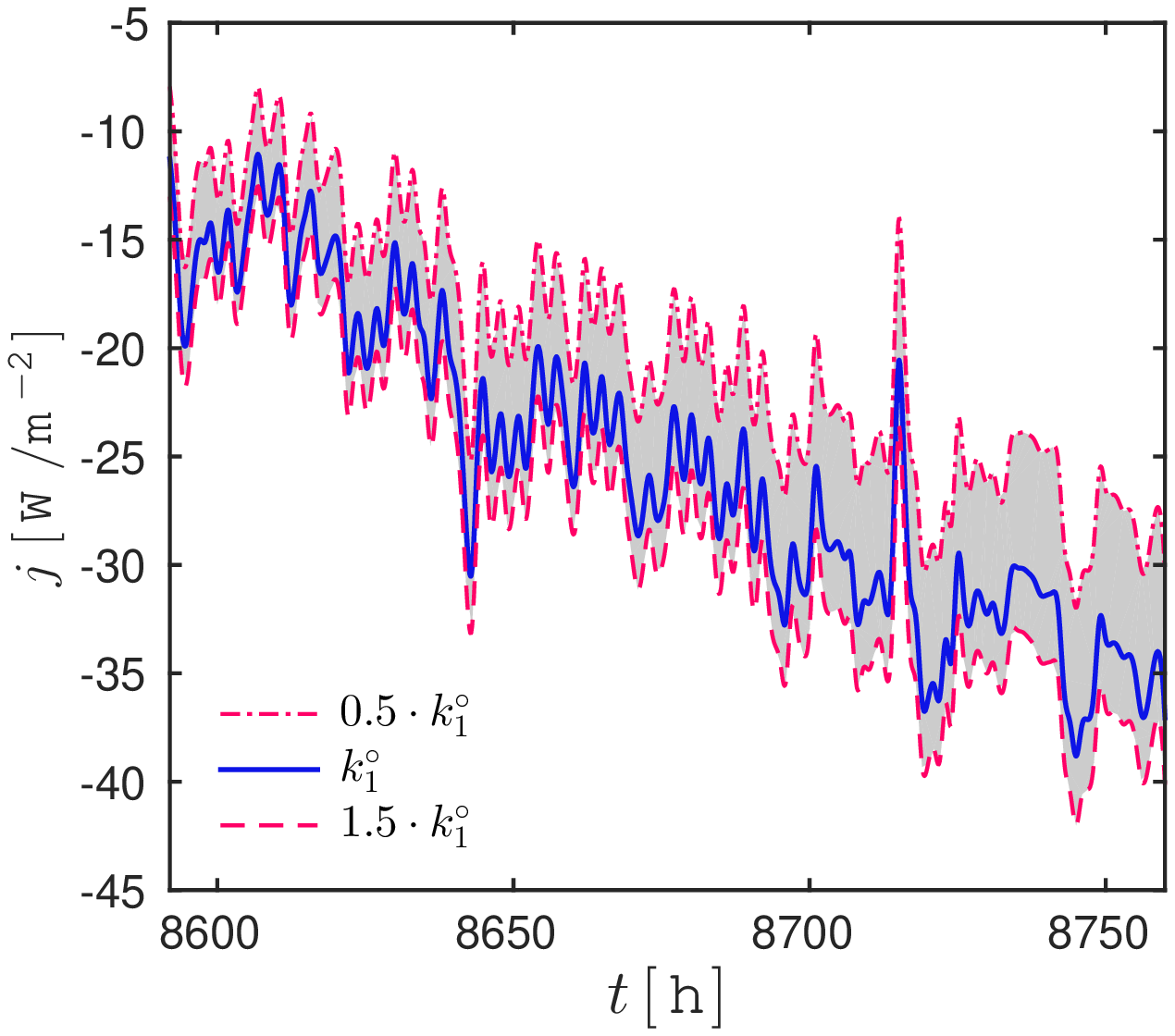}} \hspace{0.3cm}
\subfigure[]{\includegraphics[width=0.45\textwidth]{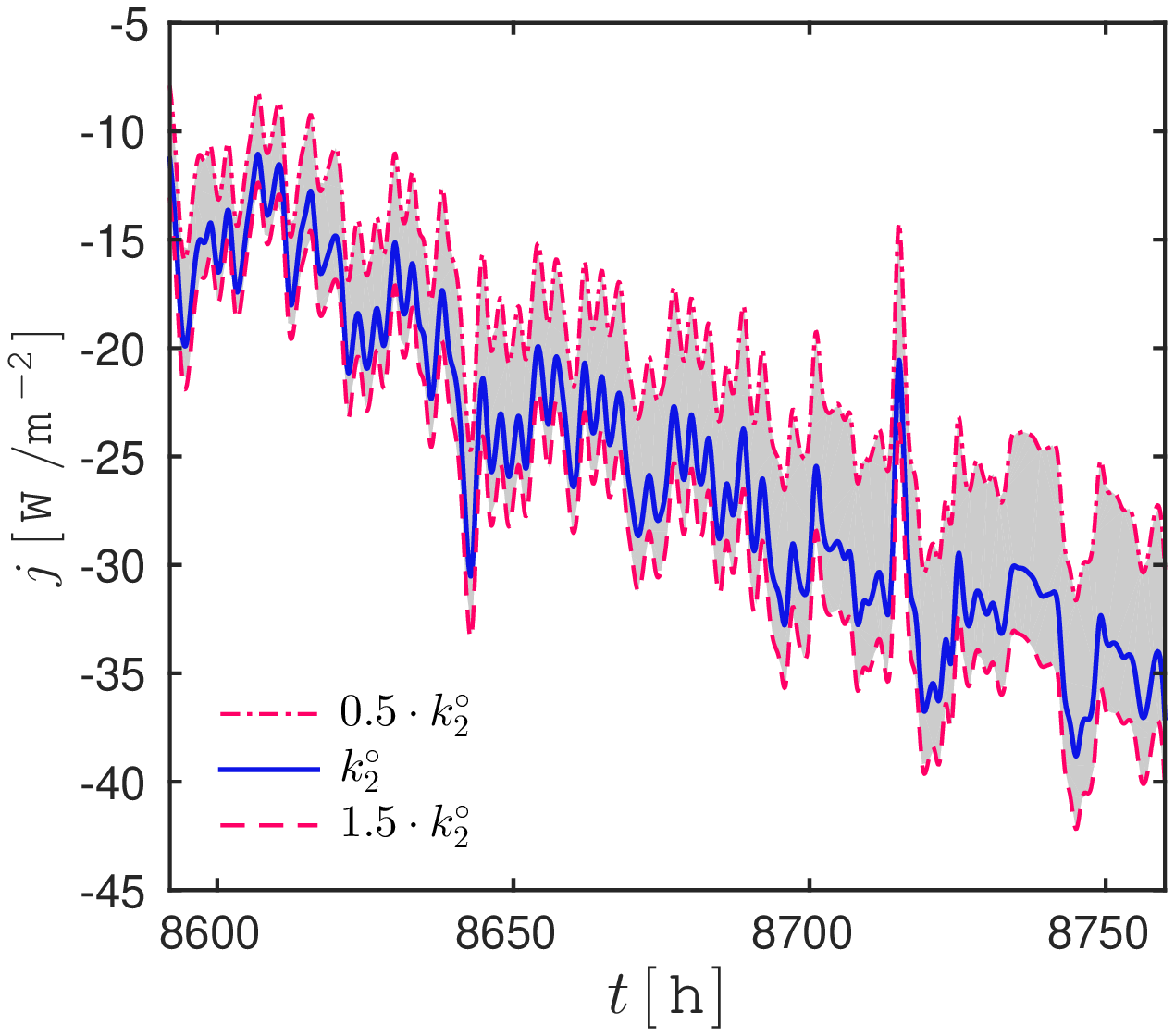}} 
\caption{Variation in heat flux according to the thermal conductivity of the first and second layers \ainagul{using \textsc{Taylor} series expansion}.}
\label{fig:J_k}
\end{center}
\end{figure}

One of the main merits of the continuous approach is that it can be used to explore the sensitivity of continuous model output due to changes in the parameters. Thus, it is possible to examine how the time variation in the heat flux differs according to the thermal conductivity of the dressed and rubble stone layers. As mentioned before, the influence of the volumetric heat capacity is negligible. Thus, only parameters $k_{\,1}$ and $k_{\,2}$ are altered by $\pm\,50\%$ of their \emph{a priori} values.  Figure~\ref{fig:J_k} shows the variation in the heat flux on the interior wall surface during the last week of the year. One may assess how the heat flux is affected by the thermal conductivity parameter. 
    
Now the remaining question is the computational effort of the continuous approach for the real--world case study. The results are summarized in Table~\ref{tab:cpu}. \ainagul{The computations are performed using the \texttt{Matlab\texttrademark} environment with a computer equipped with \texttt{Intel} i$7$ CPU and $16$ GB of RAM. }
According to the values in Table~\ref{tab:cpu} $1$ year simulation for the whole wall, composed of three layers, takes $2.5\,\mathsf{h}$ for the first--order sensitivities, and $7.5\,\mathsf{h}$ for the second--order sensitivities. One may note that the  computation of the local metric does not require the model output variation. However, calculation of the global metrics for all three layers involves at least $30$ output evaluations \ainagul{as noted in Table~\ref{tab:cpu}}. By contrast, standard global sensitivity analysis methods require a greater number of output samples.

\begin{table}[]
\centering
\caption{CPU time to perform sensitivity analysis with the continuous approach ($t_{\,0} \egal 12 \ \mathsf{min}$) for one layer.}
\label{tab:cpu}
\begin{tabular}{|c|c|c|}
\hline
Method  & CPU time $ [\,\o\,]$ & Number of model samples $N_{\mathrm{s}}$   \\ \hline
\DF~only one output & $t_{\,0}$ &  1  \\ \hline
Continuous, local metric first order & $4 \cdot t_{\,0}$  &   2\\ \hline
Continuous, local metric second order & $12.5 \cdot t_{\,0}$  & 1  \\ \hline
Continuous, global metric & $20 \cdot t_{\,0}$  & 10  \\ \hline
\end{tabular}
\end{table}

\begin{BS_Answer}
\subsection{Impact of window properties on the energy performance of a building envelope}
\label{sec:window_eq}
A building envelope composed of wall of historical building, described in Section~\ref{sec:wall}, and a single-glazed window is considered. The purpose is to investigate the sensitivity of the thermal loads of the envelope regarding glass reflectivity, transmissivity and wall thermal properties. To define the thermal loads of the glass, one may study heat transfer through the window. The physical problem considers one--dimensional heat conduction transfer through a building window. The window is a single glazed. The temperature in the window is defined on the domains  $\Omega_{\,x}\,:\, x \in [\,0\,,L_{\,w}\,]\,$ and $ \Omega_{\,t}\,:\,t \in [\,0\,,\tau_{\,\mathrm{max}}\,]\,$, where $L_{\,w}\,\mathsf{[\,m\,]}$ is the length of the window and $\tau_{\,\mathrm{max}} \,\mathsf{[\,s\,]}$ is the duration of the simulation.

In the window, the heat transfer is governed by diffusion and radiation mechanisms~\citep{howell2010,Gasparin2020}, which can be written as follows: 
\begin{align}
\label{eq:heat_eq_win}
\rho_{\,w}\,\cdot\,c_{\,p} \,\cdot\,\pd{T}{t} \egal \pd{}{x}\,\Biggl(\,k_{\,w}\,\cdot\,\pd{T}{x}\,\Biggr) \plus S\,,
\end{align}
where $\rho_{\,w}\,\mathsf{[\,kg/m^3\,]}$ is the material density, $c_{\,p}\,\mathsf{[\,J/(kg \cdot K)\,]}$ is the material specific heat, $k_{\,w}\,\mathsf{[\,W/(m \cdot K)\,]}$ is the material thermal conductivity, $S\,\mathsf{[\,W/m^3\,]}$ is an internal source term for the absorbed short-wave radiation part:
\begin{align*}
    S \egal A\,\cdot\,q^{\,{sw}}\,\cdot\,\frac{x}{L_{\,w}}\,,
\end{align*}
in which $A$ is the fraction of absorbed heat. This source term is valid only for glass materials.
The fraction of absorbed energy in a single glazing glass is written according to~\citep{howell2010,Gasparin2020}:
\begin{align*}
    A \egal \frac{(\,1\moins \tau\,)\,(\,1\moins \rho\,)}{(\,1\moins \rho\,\tau\,)}\,,
\end{align*}
where $\rho\,[\,-\,]$ is the glass reflectivity and $\tau\,[\,-\,]$ the glass transmissivity. 
Glass properties are defined in Table~\ref{tab:gl_properties}.
\FloatBarrier
\begin{table}[h]
\centering
\caption{ Glass properties.}
\begin{tabular}{|c c|}
\hline
Property & Value \\ \hline
Density $\rho_{\,w}\,\mathsf{[\,kg/m^3\,]}$   & $2200$ \\
Specific heat capacity $c_{\,p}$ $[\,\mathsf{J/m^3/K}\,]$ & $835$ \\
Thermal conductivity $k_{\,w}$ $[\,\mathsf{W/m/K}\,]$ & $1$\\
Reflectivity $\rho\,[\,-\,]$  & 0.15 \\
Transmissivity $\tau\,[\,-\,]$ & 0.26 \\
Glass thickness $L_{\,w}\,[\,\mathsf{m}\,]$ & 0.006 \\ \hline
\end{tabular}
\label{tab:gl_properties}
\end{table}
\FloatBarrier

The heat balance on the exterior window surface includes the convective exchange between the air outside and the window surface, as well as the absorbed radiation. The effect of radiation is included as a source term in the governing equation~\eqref{eq:heat_eq_win}, thus, it can be omitted. The exterior boundary condition can be written as:
\begin{align}
k_{\,w}\,\cdot\,\pd{T}{x} \egal & h_{\,L}\,\cdot\,\Bigl(\,T \moins T_{\,\infty}^{\,L}\,(\,t\,)\,\Bigr) \,,\qquad x\egal 0\,, 
\end{align}
where $T_{\,\infty}^{\,L}\,\mathsf{[\,K\,]}$ is the temperature of the outside air that varies over time, $h_{\,L}\,\mathsf{[\,W/(m^2 \cdot K)\,]}$ is the exterior convective heat transfer coefficient.
The interior heat balance consists of the convective exchange between the air inside $T_{\,\infty}^{\,R}\,\mathsf{[\,K\,]}$ and the window surface, and is given by the following expression:
\begin{align}
k_{\,w}\,\cdot\,\pd{T}{x}  \egal & \moins h_{\,R}\,\cdot\,\Bigl(\,T \moins T_{\,\infty}^{\,R}\,(\,t\,) \,\Bigr)\,,\qquad x\egal L_{\,w}\,,
\end{align}
where $h_{\,R}\,\mathsf{[\,W/(m^2 \cdot K)\,]}$ is the convective heat transfer coefficient on the inside boundary.

In addition, the thermal loads of the glass $\boldsymbol{\text{E}_{\,\text{glass}}}\;\mathsf{[\,W\cdot s/m^2 \,]}$  {are computed by integrating the heat flux at the inner surface over the chosen time interval~\citep{ALSANEA_2012,BERGER_2017_E}}:
\begin{align}
\text{E}_{\,\text{glass}} \egal  \int_{\,t_{}}^{\,t\plus \delta\,t} \, \Biggl(\, \moins k_{\,w}\,\cdot\,\pd{T}{x}\,\Biggr)\,\Bigg|_{\,x \egal L_{\,w}} \, \mathrm{d\tau} \,,
\end{align}
where $\delta\,t$ is a time interval such as day, week, or month.

Furthermore, the overall thermal loads is defined as sum of the thermal loads of the wall and the glass respectively, and is expressed as follows:
\begin{align}
    \widehat{\mathrm{E}} \egal \text{E}_{\,\text{glass}} \plus \text{E}_{\,\text{wall}}\,,
\end{align}
where $\widehat{\mathrm{E}}$ is overall energy performance of the envelope, $\text{E}_{\,\text{wall}}$ is the wall thermal loads, obtained using~\eqref{eq:E_wall}. 

According to the previous results the wall thermal loads mostly depend on the thermal conductivities of the first and second layers of the wall. Thus, the overall thermal loads can be written as:
\begin{align*}
    \widehat{\mathrm{E}} \egal \widehat{\mathrm{E}} \, \bigl(\,k_{\,1}^{\,\circ}\,, k_{\,2}^{\,\circ}\,,\tau\,, \rho\,\bigr)\,.
\end{align*}

Similarly, the sensitivity functions of the glass properties, such as reflectivity and transmissivity, were calculated as direct differentiation of the governing equation~\eqref{eq:heat_eq_win}. Next, their impact on the overall building energy performance compared with wall thermal properties was analyzed.  Results are shown in Table~\ref{tab:norm_E_glass}. Influence of wall thermal conductivities and glass properties on the thermal loads during one year was studied. It can be noted that glass properties do not have a large impact on the thermal loads, while the thermal conductivity of each wall's layer have a similar effect, which is several times greater compared to the glass properties. This effect is probably due to the bad efficiency of the wall materials from the thermal point of view.

\begin{table}[h!]
\centering
\caption{Sensitivity coefficients metric on the thermal loads $\widehat{\mathrm{E}}$  during one year.}
\begin{tabular}{|c|c|c|c|c|}
\hline
\bf{Layer} & \bf{Lime coater}  & \bf{Rubble stone} & \bf{Glass} & \bf{Glass} \\ \hline
Parameter & $k^{\,\circ}_{\,1}$ & $k^{\,\circ}_{\,2}$ & $\tau$ & $\rho$  \\  \hline
 & \multicolumn{4}{c|}{Thermal loads} \\  \hline
Global metric $\gamma_{\,p}$ & 0.53 & 0.46 & $\mathcal{O}\,(\,10^{\,-2}\,)$ &$\mathcal{O}\,(\,10^{\,-4}\,)$\\
\hline
\end{tabular}
\label{tab:norm_E_glass}
\end{table}

Using \textsc{Taylor} series expansion, one may calculate the impact on the monthly thermal loads according to variation in the selected parameter. The relative difference in the thermal loads is expressed as follows:
\begin{align*}
    \varepsilon_{\,r}\,(\,p\,)  \egal \displaystyle \Biggl|\, \dfrac{ \widehat{\mathrm{E}}\,(\,p \plus \delta\,p\,) \moins  \widehat{\mathrm{E}}\,(\,p\,)  }{ \widehat{\mathrm{E}}\,(\,p\,)}\,\Biggr|\,,
\end{align*}
where $\delta\,p$ is the variation of selected parameter.
Furthermore, the aforementioned measure is calculated for each property of the wall and glass, which was varied by $\pm\,50\,\%$ of the parameter's value. Results are presented in Figures~\ref{fig:e_glass_plus},~\ref{fig:e_glass_minus}. It can be noted that window properties have influence on the overall thermal loads around $1\,\%$ during summer period, and during other three seasons its effect can be neglected, since it is varies between  $0.1\,\%$ and  $1\,\%$. Results demonstrate that glass reflectivity has the smallest impact on the overall thermal loads. Wall thermal properties have a great effect on the energy performance of the envelope, and their impact are $10\,\%$ to the overall thermal loads.    

\begin{figure}[ht!] 
\centering
\subfigure[\label{fig:e_glass_plus}]{\includegraphics[width=0.85\linewidth]{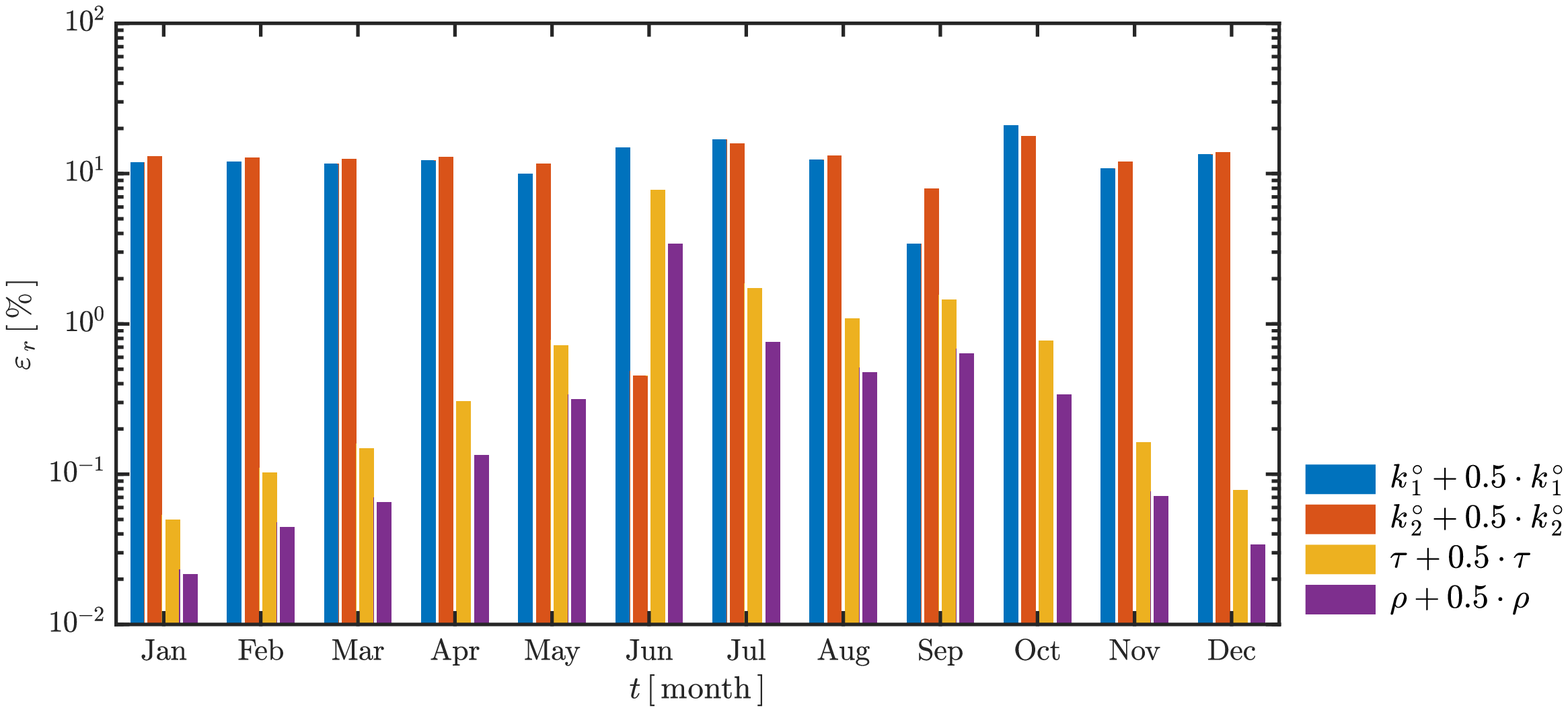}} \quad
\subfigure[\label{fig:e_glass_minus}]{\includegraphics[width=0.85\linewidth]{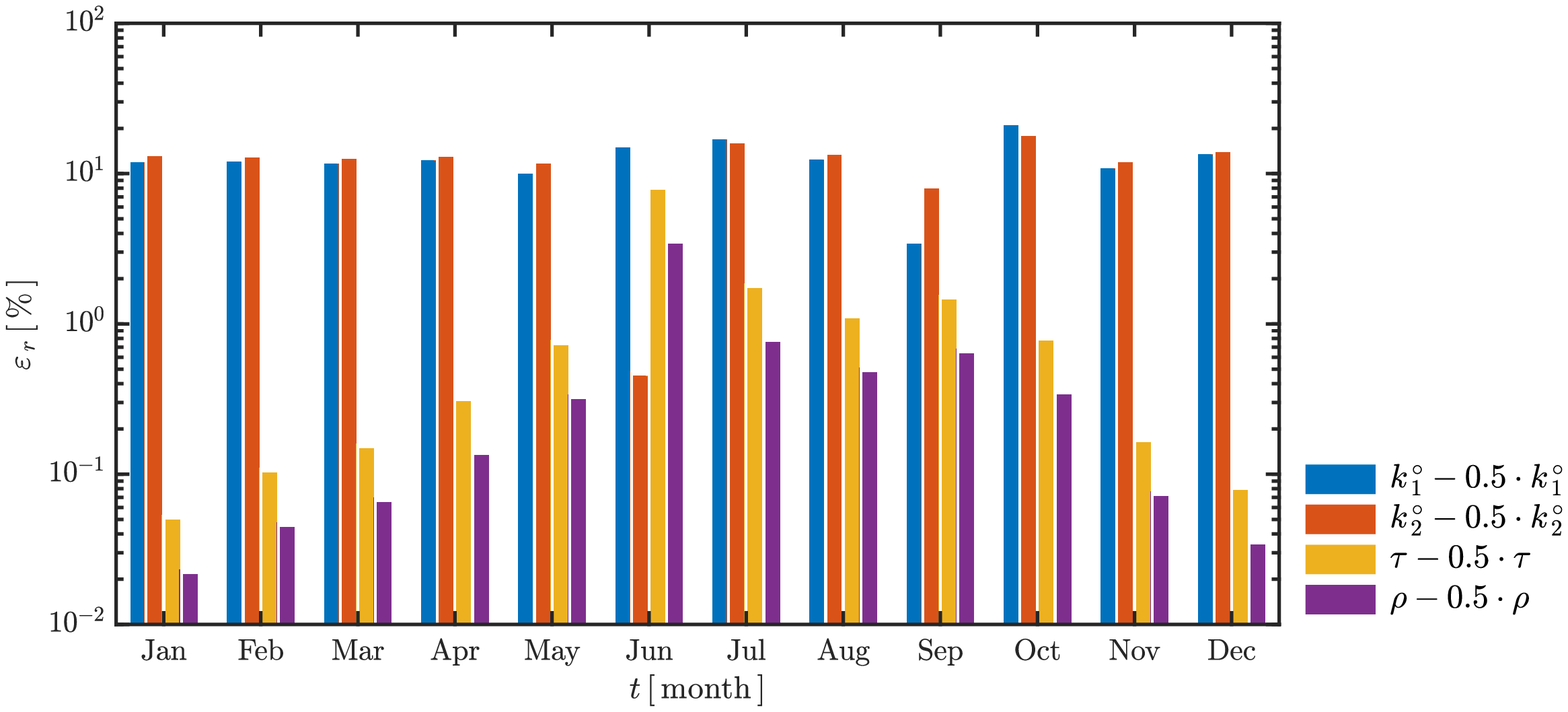}}
\caption{{Monthly variation of the relative difference in the thermal loads according to \emph{(a)} increased and \emph{(b)} decreased values of the selected parameters.}}
\end{figure}

\end{BS_Answer}

\subsection{Summary of results for the real--world case study}
The demonstration of the derivative -based sensitivity analysis is carried out on the wall of a historical building situated in France. The sensitivity coefficients are calculated using continuous partial derivatives. The influence of the wall thermal properties, such as thermal conductivity and volumetric heat capacity, on the wall temperature, the interior heat flux, and the thermal loads are computed. The results are summarized as follows: 
\begin{itemize}
\item in this case study, the thermal conductivity has a greater impact on the aforementioned model outputs;
\item as the heat capacity has a comparatively smaller impact on the temperature and the thermal loads, it can be neglected in a future parameter estimation problem;
\item the derivative -based approach has similar metric values of the parameter influence to the SRC method. However, the linear approximation of the thermal loads retrieved by the SRC method is not valid, for example, to simulate the daily thermal loads;
\item the time--varying model output variation according to the changes in parameters is assessed using the \textsc{Taylor} series expansion. As an example, the heat flux variation according to the variation of the thermal conductivity is presented; 
\item the derivative -based approach has the lowest computational cost among the sensitivity methods presented;
\item \ajumabek{the proposed approach can be extended to additional components of building envelopes, as demonstrated by carrying out a sensitivity analysis for the wall together with a single glazed window.}
\end{itemize}

\section{Conclusion}

Estimation of the properties of wall material plays an important role in assessment of building energy performance. To reduce computational cost, the concept of \emph{primary identifiability} is proposed to rank the unknown parameters and only estimate a few numbers. The parameters are selected by analyzing their impact on the model output. In other words, a sensitivity analysis is performed to eliminate the non-significant unknown parameters. However, the calculation of sensitivity coefficients requires a high computational effort since the methods usually employed in the literature involve a large number of model evaluations. Thus, the originality of this work lies in the computation of sensitivity coefficients through a continuous derivative -based approach. Incorporated in a reliable and efficient \DF~numerical model, the direct differentiation of the governing model equation provides accurate results with a reduced computational time.  

Initially, the results of first- and second-order sensitivity coefficients are compared with a reference solution and various discrete approximations. Although the discrete approach is faster, its accuracy depends greatly on the discretization step value. 
Then, \textsc{Taylor} series expansion is discussed. The sensitivity coefficients are used to explore how model outputs vary according to changes in parameters. Variations for the temperature field, the heat flux, and the thermal loads relative to modification of the thermal parameters are assessed with very satisfactory accuracy.
Finally, the comparison with variance and regression--based methods is conducted. Similar qualitative results highlight the reliability of the derivative -based approach. Additionally, the global metrics of the continuous approach comply with the \textsc{Sobol} total indices. Another advantage is CPU time. RBD--FAST and \textsc{Sobol} methods require at least $100$ times higher computational time. 

Since the validation case provides good results in terms of accuracy and computational time, the proposed method is used to determine the parameters that have the most influence on the energy performance of a real building wall, situated in France. The results of the local and global metrics show that the thermal conductivity has a greater impact on the thermal loads \ainagul{for this case study}. One may also analyze how the thermal loads vary during the day, month, or year or how the heat flux varies according to the changes in the thermal conductivity. In terms of computational time global metric values were obtained with fewer output evaluations using the derivative -based approach. \ajumabek{One may note that this approach can be employed together with the finite-difference approximations of sensitivity coefficients using the building simulation programs.}
\ajumabek{Additionally, an envelope, composed of a historical building wall and a single-glazed window, was studied. The influence of glass and wall thermal properties on the overall energy performance was calculated to illustrate that the methodology can be extended to the other components of building envelope.}

In brief, \ainagul{the derivative -based sensitivity analysis uses partial derivatives as global sensitivity coefficient estimators. Calculation of the partial derivatives through the direct differentiation of the model--governing equation provides continuous and time-varying sensitivity coefficients}. The \textsc{Taylor} approximation allows us to explore model output variation according to changes in the input parameters. The proposed approach accurately finds the most influential parameters on any chosen model output with small computational effort. Therefore, it can be used as a preliminary step for the parameter estimation problem. {However, these advantages, such as accuracy and computational cost, do not apply in general. \ainagul{They should be further investigated for other case studies and wall configurations}.} Further works should focus on extending the methodology for more complex mathematical models including, for instance, coupled heat and mass transfer in porous materials. \ainagul{Another possible extension is to widen the scope of the parameters scope by considering not only wall material properties but also the heat convection coefficients.}

\newpage
\section*{Nomenclature}

\begin{tabular}{|ccc|}
\hline
\multicolumn{3}{|c|}{\emph{Latin letters}} \\
      $c$   &  material volumetric heat capacity     & $\mathsf{[\,J/(m^3 \cdot K)\,]}$       \\
      $c_p$   &  \ainagul{material specific heat capacity}     & $\mathsf{[\,J/(kg \cdot K)\,]}$  \\
       $E$& thermal loads& $\mathsf{[\,W\cdot s/m^2 \,]}$      \\
      $h$ & convective heat transfer coefficient & $\mathsf{[\,W/(m^2 \cdot K)\,]}$   \\ 
      $j$ & heat flux & $\mathsf{[\,W/m^2 \,]}$ \\
      $k$ & thermal conductivity & $\mathsf{[\,W/(m \cdot K)\,]}$ \\ 
      $L$ & wall length & $\mathsf{[\,m\,]}$ \\
      $L_{\,w}$ & glass thickness & $\mathsf{[\,m\,]}$ \\
      $q_{\,\infty}$ & total incident radiation & $\mathsf{[\,W/m^2 \,]}$\\
      $T$ & temperature & $\mathsf{[\, K \,]}$\\
      $t$ & time & $\mathsf{[\,s\,]}$\\
      $x$ & thickness coordinate direction & $\mathsf{[\,m\,]}$
      \\ \hline
\end{tabular}

\begin{answer}
\begin{tabular}{|ccc|}
\hline
\multicolumn{3}{|c|}{\emph{Greek letters}} \\ 
	$\alpha$ & surface absorptivity &  \\
    $\rho_{\,w}$   &  material density     & $\mathsf{[\,kg/m^3\,]}$ \\
    $\Omega$ & variable domain & \\ 
    $\rho$ & glass reflectivity & $\mathsf{[\,-\,]}$ \\ 
    $\tau$ & glass transmissivity & $\mathsf{[\,-\,]}$ \\ \hline
\end{tabular}
\end{answer}

\begin{tabular}{|ccc|}
\hline
\multicolumn{3}{|c|}{\emph{Dimensionless values}} \\
   $A$ & fraction of absorbed heat & $\mathsf{[\,-\,]}$ \\ 
     $\mathrm{Bi}$  &  \textsc{Biot} number   & $\mathsf{[\,-\,]}$       \\
     $\mathrm{Fo}$  &  \textsc{Fourier} number   & $\mathsf{[\,-\,]}$      \\
     $g_{\,\infty}$  &  total incident radiation   & $\mathsf{[\,-\,]}$     \\ 
      $u$ & temperature field & $\mathsf{[\,-\,]}$ 
      
      \\ \hline
\end{tabular}

\begin{tabular}{|cc|}
\hline
\multicolumn{2}{|c|}{\emph{Subscripts and superscripts}} \\ 
     $L$  &  Left boundary $x\egal 0$        \\
     $R$  &  Right boundary $x\egal L$     \\
    $\star$  &  dimensionless parameter    \\ 
      $\circ$ & \emph{a priori} parameter value \\
      $\mathrm{tay}$ & \ainagul{\textsc{Taylor} series expansion} \\    
      $\mathrm{num}$ & \ainagul{numerical solution}
      \\
      $w$ & window \\ \hline
\end{tabular}

\begin{answer}
\begin{tabular}{|cc|}
\hline
\multicolumn{2}{|c|}{\emph{Abbreviations}} \\ 
    SRC  &  Standardized Regression Coefficients        \\
    SRRC  &  Standardized Rank Regression Coefficients        \\
    FAST &   \textsc{Fourier} Amplitude Sensitivity Test   \\
    RBD  &  Random Balance Design    
      \\ \hline
\end{tabular}
\end{answer}

\newpage
 
\bibliographystyle{unsrt}  
\bibliography{references}

\newpage
\appendix
\section{Dimensionless equation}
\label{sec:dim_eq}
This section introduces the dimensionless model equations. The dimensionless formulation has several advantages. First, once we operate the numbers of the same magnitude, we minimize the rounding numerical errors \citep{Kahan_1979}. Second, it enables a general investigation of the model behaviour regardless of the unit used to measure variable \citep{Berger2019,trabelsi2018response}. Finally, it simplifies equations by reducing the number of variables. 

Let us define the following dimensionless variables:
\begin{align*}
x^{\,\star} & \ \eqdef \ \dfrac{x}{L}\,, & u & \ \eqdef \   \dfrac{T}{T_{\,\mathrm{ref}}}\,,  & t^{\,\star} & \ \eqdef \  \dfrac{t}{t_{\,\mathrm{ref}}}\,,\\
 k^{\,\star} & \ \eqdef \  \dfrac{k}{k_{\,\mathrm{ref}}}\,,  & c^{\,\star} & \ \eqdef \ \dfrac{c}{c_{\,\mathrm{ref}}}\,,  & \mathrm{Fo} & \ \eqdef \  \dfrac{t_{\,\mathrm{ref}}\,\cdot\,k_{\,\mathrm{ref}}}{L^{\,2}\,\cdot\,c_{\,\mathrm{ref}}}\,,
\end{align*}
where subscripts $\boldsymbol{\mathrm{ref}}$ relate for a characteristic reference value, and superscript $\boldsymbol{\star}$ for dimensionless parameters. Thus, equation~\eqref{eq:heat_eq_ph} becomes:
\begin{align} \label{eq:dim_heat_eq}
c^{\,\star}\,\cdot\,\pd{u}{t^{\,\star}}  \egal \mathrm{Fo}\,\cdot\,\pd{}{x^{\,\star}}\,\Bigl(\,k^{\,\star}\,\cdot\,\pd{u}{x^{\,\star}}\,\Bigr)\,.
\end{align}

\textsc{Robin}--type boundary conditions are converted to:
\begin{align}
k^{\,\star}\,\cdot\,\pd{u}{x^{\,\star}}  \egal & \mathrm{Bi_{\,L}}\,\cdot\,\Bigl(\,u \moins u_{\,\infty}^\mathrm{\,L}\,\Bigr) \moins \alpha\,\cdot\,g_{\infty}^\mathrm{\,L}\,\,,\quad && x^{\,\star} \egal 0\\
k^{\,\star}\,\cdot\,\pd{u}{x^{\,\star}}  \egal & -\mathrm{Bi_{\,R}}\,\cdot\,\Bigl(\,u \moins u_{\,\infty}^\mathrm{\,R}\,\Bigr) \,,\quad && x^{\,\star} \egal 1\,,
\end{align}
with dimensionless quantities:
\begin{align*}
\mathrm{Bi_{\,L}} & \ \eqdef \  \dfrac{\mathrm{h_{\,L}}\,\cdot\,L}{k_{\,\mathrm{ref}}}\,, &  \mathrm{Bi_{\,R}} & \ \eqdef \  \dfrac{\mathrm{h_{\,R}}\,\cdot\,L}{k_{\,\mathrm{ref}}}\,,\\
 u_{\,\infty}^\mathrm{\,L} & \ \eqdef \  \dfrac{T_{\,\infty}^\mathrm{\,L}}{T_{\,\mathrm{ref}}}\,,& u_{\,\infty}^\mathrm{\,R} & \ \eqdef \  \dfrac{T_{\,\infty}^\mathrm{\,R}}{T_{\,\mathrm{ref}}}\,,& g_{\,\infty}^\mathrm{\,L} & \ \eqdef \  \dfrac{q_{\,\infty}^\mathrm{\,L}\,\cdot\,L}{T_{\,\mathrm{ref}}\,\cdot\,k_{\,\mathrm{ref}}}\,.
\end{align*}

The initial condition is transformed to
\begin{align*}
u \egal u_{\,0}\,,\qquad \mathrm{where}\qquad u_{\,0} \ \eqdef \ \dfrac{T_{\,0}}{T_{\,\mathrm{ref}}}\,.
\end{align*}

The dimensionless heat flux $\mathrm{j^{\,\star}}$  is expressed as:
\begin{align*}
    \mathrm{j^{\,\star}} \,:\, t^{\,\star} \, \longmapsto \, \dfrac{\mathrm{j}}{\mathrm{j_{\,ref}}}\,, \qquad \text{where} \qquad \mathrm{j_{\,ref}} \ \eqdef \  \dfrac{T_{\,\mathrm{ref}}\,\cdot\,k_{\,\mathrm{ref}}}{L}\,.
\end{align*}

The dimensionless thermal loads $\mathrm{E^{\,\star}}$ are calculated as:
\begin{align*}
    \mathrm{E^{\,\star}} \,:\, \delta\,t^{\,\star} \, \longmapsto \, \dfrac{\mathrm{E}}{\mathrm{E_{\,ref}}}\,, \qquad \text{where} \qquad \mathrm{E_{\,ref}} \ \eqdef \ \dfrac{T_{\,\mathrm{ref}}\,\cdot\,k_{\,\mathrm{ref}}\,\cdot\,t_{\,\mathrm{ref}}}{L}\,.
\end{align*}

Next section presents how to quantify the change of the model output value according to the change of one or set of input parameters. 

\newpage
\begin{answer}
\section{Dicrete function derivatives}
\label{sec:discr_diff}
The \emph{Central difference approximation} with the second--order accuracy can be defined as follows:
\begin{align}
    X_{\,p^{\,\star}}  \egal & \dfrac{u\,(\,p^{\,\star} \plus \Delta p^{\,\star} \,)\moins u\,(\,p^{\,\star} \,\moins \Delta p^{\,\star} ) }{2\,\cdot\,\Delta p^{\,\star}} \plus \mathcal{O}\,\Bigl(\,\Delta p^{\,\star\,2}\,\Bigr)\,.
\end{align}
By adding another term to the \emph{backward difference} approach, the second--order accuracy is achieved. The \emph{three points approximation} can be expressed by the following formulation: 
\begin{align}
    X_{\,p^{\,\star}}  \egal & \dfrac{3\,\cdot\,u\,(\,p^{\,\star} \,)\moins 4\,\cdot\,u\,(\,p^{\,\star} \,\moins \Delta p^{\,\star})\plus u\,(\,p^{\,\star}\,\moins 2\,\cdot\,\Delta p^{\,\star})}{2\,\cdot\,\Delta p^{\,\star}} \plus \mathcal{O}\,\Bigl(\,\Delta p^{\,\star\,2}\,\Bigr)\,.
\end{align}

Similarly, one may estimate the higher order sensitivity coefficients. The second--order derivatives can be approximated by the \emph{forward difference approximation}:
\begin{align}
    X_{\,p^{\,\star}\,p^{\,\star}}  \egal & \dfrac{u\,(\,p^{\,\star} \plus 2\,\cdot\,\Delta p^{\,\star}\,)\moins 2\,\cdot\,u\,(\,p^{\,\star}\,\plus \Delta p^{\,\star})\plus u\,(\,p^{\,\star}\,)}{\Delta p^{\,\star\,2}} \plus \mathcal{O}\,\Bigl(\,\Delta p^{\,\star}\,\Bigr)
\end{align}
or the \emph{central difference approximation}:
\begin{align}
    X_{\,p^{\star}\,p^{\,\star}}  \egal & \dfrac{u\,(\,p^{\,\star} \plus \Delta p^{\,\star}\,)\moins 2\,\cdot\,u\,(\,p^{\,\star}\,)\plus u\,(\,p^{\,\star} \moins \Delta p^{\,\star}\,)}{\Delta p^{\,\star\,2}} \plus \mathcal{O}\,\Bigl(\,\Delta p^{\,\star\,2}\,\Bigr)\,.
\end{align}

Joint effect of the two parameters $p_{\,i}^{\,\star}$ and $p_{\,j}^{\,\star}$ on the model output can be presented as mixed derivative with the following approximation:
\begin{align}
    X_{\,p^{\,\star}_{\,i}\,p^{\,\star}_{\,j}}  \egal & \dfrac{1}{4\,\cdot\,\Delta p^{\,\star}_{\,i}\,\cdot\,\Delta p^{\,\star}_{\,j}}\Biggl(\, u\,(\,p^{\,\star}_{\,i} \plus \Delta p^{\,\star}_{\,i}\,,p^{\,\star}_{\,j} \plus \Delta p^{\,\star}_{\,j}\,)\moins u\,(\,p^{\,\star}_{\,i} \plus \Delta p^{\,\star}_{\,i}\,,p^{\,\star}_{\,r} \moins \Delta p^{\,\star}_{\,j}\,) \\
    &\moins u\,(\,p^{\,\star}_{\,i} \moins \Delta p^{\,\star}_{\,i}\,,p^{\,\star}_{\,j} \plus \Delta p^{\,\star}_{\,j}\,) \plus  u\,(\,p^{\,\star}_{\,i} \moins \Delta p^{\,\star}_{\,i}\,,p^{\,\star}_{\,j} \moins \Delta p^{\,\star}_{\,j}\,) \, \Biggr) \plus \mathcal{O}\,\Bigl(\,\Delta p^{\,\star\,2}_{\,i}\,\cdot\,\Delta p^{\,\star\,2}_{\,j}\,\Bigr)\,.\nonumber
\end{align}
\end{answer}

\newpage
\begin{answer}
\section{Continuous function derivatives}
\label{sec:cont_diff}
he second order partial differentiation of Eq.~\eqref{eq:dim_heat_eq} gives us the second--order sensitivity coefficients:
\begin{align*}
X_{\,k^{\,\star}_{\,1}\,k^{\,\star}_{\,1}} \,:\, \bigl(\,x^{\,\star}\,,t^{\,\star}\,,k^{\,\star}_{\,1}\,\bigr) \,& \longmapsto \, \dfrac{\partial^{2}{u}}{\partial{k^{\,\star}_{\,1}}^2}\,, \\
X_{\,c^{\,\star}_{\,1}\,c^{\,\star}_{\,1}} \,:\, \bigl(\,x^{\,\star}\,,t^{\,\star}\,,c^{\,\star}_{\,1}\,\bigr) \, & \longmapsto \, \dfrac{\partial^{2}{u}}{\partial{c^{\,\star}_{\,1}}^2}\,, \\
X_{\,k^{\,\star}_{\,1}\,c^{\,\star}_{\,1}} \,:\, \bigl(\,x^{\,\star}\,,t^{\,\star}\,,k^{\,\star}_{\,1}\,,c^{\,\star}_{\,1}\,\bigr) \, & \longmapsto \, \dfrac{\partial^{2}{u}}{\partial{k^{\,\star}_{\,1}\,{\partial c^{\,\star}_{\,1}}}}\,.
\end{align*}
Thus, the sensitivity equations to calculate the quantities $X_{\,k^{\,\star}_{\,1}\,k^{\,\star}_{\,1}}$ and $X_{\,c^{\,\star}_{\,1}\,c^{\,\star}_{\,1}}$ are:
\begin{align}
\pd {X_{\,k^{\,\star}_{\,1}\,k^{\,\star}_{\,1}}}{t^{\,\star}}  \egal & \frac{\mathrm{Fo}}{c^{\,\star}}\,\cdot\,\pd{}{x^{\,\star}}\,\biggl(\,\frac{\partial^{\,2}\,k^{\,\star} }{\partial\,k^{\,\star\,2}_{\,1}}\,\cdot\,\pd{u}{x^{\,\star}} \plus 2\,\cdot\,\pd{k^{\,\star}}{k^{\,\star}_{\,1}}\,\cdot\,\pd{X_{\,k^{\,\star}_{\,1}}}{x^{\,\star}} \plus k^{\,\star}\,\cdot\,\pd{X_{\,k^{\,\star}_{\,1}\,k^{\,\star}_{\,1}}}{x^{\,\star}}\,\biggr)\,.
\end{align}
\begin{align}
\pd {X_{\,c^{\,\star}_{\,1}\,c^{\,\star}_{\,1}}}{t^{\,\star}}  \egal & \frac{2\,\cdot\,\mathrm{Fo}}{c^{\,\star\,3}}\,\cdot\,\Biggl(\,\pd{c^{\,\star}}{c^{\,\star}_{\,1}}\,\Biggr)^{\,2}\,\cdot\,\pd{}{x^{\,\star}}\,\Biggl(\,k^{\,\star}\,\cdot\,\pd{u}{x^{\,\star}}\,\Biggr) \moins \frac{\mathrm{Fo}}{c^{\,\star\,2}}\,\cdot\,\frac{\partial^{\,2}\,c^{\,\star}}{\partial\,c^{\,\star\,2}_{\,1}}\,\cdot\,\pd{}{x^{\,\star}}\,\Biggl(\,k^{\,\star}\,\cdot\,\pd{u}{x^{\,\star}}\,\Biggr) \moins \\ 
&{} \frac{2\,\cdot\,\mathrm{Fo}}{c^{\,\star\,2}}\,\cdot\,\pd{c^{\,\star}}{c^{\,\star}_{\,1}}\,\cdot\,\pd{}{x^{\,\star}}\,\Biggl(\,k^{\,\star}\,\cdot\,\pd{X_{\,c^{\,\star}_{\,1}}}{x^{\,\star}}\,\Biggr) \plus 
\frac{\mathrm{Fo}}{c^{\,\star}}\,\cdot\,\pd{}{x^{\,\star}}\,\Biggl(\,k^{\,\star}\,\cdot\,\pd{X_{\,c^{\,\star}_{\,1}\,c^{\,\star}_{\,1}}}{x^{\,\star}}\,\Biggr)\,.\nonumber
\end{align}
The coefficient $X_{\,k^{\,\star}_{\,1}\,c^{\,\star}_{\,1}}$ measures the impact of the two parameters $k^{\,\star}_{\,1}\;\text{and}\;c^{\,\star}_{\,1}$ on the model output, which is computed as a solution of the following equation:
\begin{align}
\pd {X_{\,k^{\,\star}_{\,1}\,c^{\,\star}_{\,1}}}{t}  = & -\frac{\mathrm{Fo}}{c^{\star\,2}}\,\pd{c^{\,\star}}{c^{\,\star}_{\,1}}\,\pd{}{x^{\,\star}}\,\biggl(\pd{k^{\,\star}}{k^{\,\star}_{\,1}}\,\pd{u}{x^{\,\star}} \plus
        k^{\,\star}\,\pd{X_{\,k^{\,\star}_{\,1}}}{x^{\,\star}}\biggr) \plus
&{} \frac{\mathrm{Fo}}{c^{\,\star}}\,\pd{}{x^{\,\star}}\,\biggl(\,\pd{k^{\,\star}}{k^{\,\star}_{\,1}}\,\pd{X_{\,c^{\,\star}_{\,1}}}{x^{\,\star}} \plus \,k^{\,\star}\,\pd{X_{\,k^{\,\star}_{\,1}\,c^{\,\star}_{\,1}}}{x^{\,\star}}\,\biggr)\,.
   \end{align}
\end{answer}

\newpage
\begin{answer}
\section{Taylor series expansion of model outputs}
\label{sec:taylor_pt2}
The following expression describes the estimation of the function $u\,(\,x^{\star}\,,t^{\star}\,)$ at every point of the parameters $\{\,k^{\star}_{\,1}\,,c^{\star}_{\,1}\,\}$ around the $\{\,k_{\,1}^{\star\,\circ}\,,c_{\,1}^{\star\,\circ}\,\}$, which are the \emph{a priori} parameter values.   
\begin{align*}
u\,(\,x^{\star}\,,t^{\star}\,,k^{\star}_{\,1}\,,c^{\star}_{\,1}\,) \egal & u\,(\,x^{\star}\,,t^{\star}\,,k_{\,1}^{\star\,\circ}\,,c_{\,1}^{\star\,\circ}\,) \plus  \pd{u}{k^{\star}_{\,1}}\,\Bigg|_{k^{\star}_{\,1}=k_{\,1}^{\star\,\circ}}\,\bigg(\,k^{\star}_{\,1}\moins k_{\,1}^{\star\,\circ}\,\bigg) \plus\\ &  \pd{u}{c^{\star}_{\,1}}\,\Bigg|_{c^{\star}_{\,1}=c_{\,1}^{\star\,\circ}}\,\bigg(\,c^{\star}_{\,1}\moins c_{\,1}^{\star\,\circ}\,\bigg) \plus  \frac{1}{2}\,\dfrac{\partial^{2}{u}}{\partial{k^{\star\,2}_{\,1}}}\,\Bigg|_{k^{\star}_{\,1}=k_{\,1}^{\star\,\circ}}\,\bigg(\,k^{\star}_{\,1}\moins k_{\,1}^{\star\,\circ}\,\bigg)^2 \plus \\ &
\frac{1}{2}\,\dfrac{\partial^{2}{u}}{\partial{c^{\star\,2}_{\,1}}}\,\Bigg|_{c^{\star}_{\,1}=c_{\,1}^{\star\,\circ}}\,\bigg(\,c^{\star}_{\,1}\moins c_{\,1}^{\star\,\circ}\,\bigg)^2 \plus \frac{1}{2}\,\dfrac{\partial^{2}{u}}{\partial{k^{\star}_{\,1}\,\partial{c^{\star}_{\,1}}}}\,\Bigg|_{k^{\star}_{\,1}=k_{\,1}^{\star\,\circ}\atop c^{\star}_{\,1}=c_{\,1}^{\star\,\circ}}\,\bigg(\,k^{\star}_{\,1}\moins k_{\,1}^{\star\,\circ}\,\bigg)\,\bigg(\,c^{\star}_{\,1}\moins c_{\,1}^{\star\,\circ}\,\bigg) \plus \\ &
\mathcal{O}\,\Biggl(\,\mathrm{max}\bigg\{\bigg(\,k^{\star}_{\,1}\moins k_{\,1}^{\star\,\circ}\,\bigg)^3\,,\bigg(\,c^{\star}_{\,1}\moins c_{\,1}^{\star\,\circ}\,\bigg)^3\,\bigg\}\,\Biggr)\,.
\end{align*}
The \textsc{Taylor} expansion for heat flux is given by:
\begin{align}
j^{\star}\,(\,t^{\star}\,,k^{\star}_{\,1}\,,c^{\star}_{\,1}\,) \egal & j^{\star}\,(\,t^{\star}\,,k_{\,1}^{\star\,\circ}\,,c_{\,1}^{\star\,\circ}\,) \plus \pd{j^{\star}}{k^{\star}_{\,1}}\,\Bigg|_{k^{\star}_{\,1}=k_{\,1}^{\star\,\circ}}\,\bigg(\,k^{\star}_{\,1}\moins k_{\,1}^{\star\,\circ}\,\bigg) \plus \\ & \pd{j^{\star}}{c^{\star}_{\,1}}\,\Bigg|_{c^{\star}_{\,1}=c_{\,1}^{\star\,\circ}}\,\bigg(\,c^{\star}_{\,1}\moins c_{\,1}^{\star\,\circ}\,\bigg) \plus 
\frac{1}{2}\,\dfrac{\partial^{2}{j^{\star}}}{\partial{k^{\star\,2}_{\,1}}}\,\Bigg|_{k^{\star}_{\,1}=k_{\,1}^{\star\,\circ}}\,\bigg(\,k^{\star}_{\,1}\moins k_{\,1}^{\star\,\circ}\,\bigg)^2 \plus \nonumber \\ &
\frac{1}{2}\,\dfrac{\partial^{2}{j^{\star}}}{\partial{c^{\star\,2}_{\,1}}}\,\Bigg|_{c^{\star}_{\,1}=c_{\,1}^{\star\,\circ}}\,\bigg(\,c^{\star}_{\,1}\moins c_{\,1}^{\star\,\circ}\,\bigg)^2 \plus
\frac{1}{2}\,\dfrac{\partial^{2}{j^{\star}}}{\partial{k^{\star}_{\,1}\,\partial{c^{\star}_{\,1}}}}\,\Bigg|_{k^{\star}_{\,1}=k_{\,1}^{\star\,\circ}\atop c^{\star}_{\,1}=c_{\,1}^{\star\,\circ}}\,\bigg(\,k^{\star}_{\,1}\moins k_{\,1}^{\star\,\circ}\,\bigg)\,\bigg(\,c^{\star}_{\,1}\moins c_{\,1}^{\star\,\circ}\,\bigg)\ \nonumber \\ &
\mathcal{O}\,\Biggl(\,\mathrm{max}\bigg\{\bigg(\,k^{\star}_{\,1}\moins k_{\,1}^{\star\,\circ}\,\bigg)^3\,,\bigg(\,c^{\star}_{\,1}\moins c_{\,1}^{\star\,\circ}\,\bigg)^3\,\bigg\}\,\Biggr)\,. \nonumber
\end{align}
Last, the thermal loads value can also be approximated through a second order \textsc{Taylor} series expansion as:
\begin{align}
\label{eq:FTaylor_E}
E^{\star}\,(\,k^{\star}_{\,1}\,,c^{\star}_{\,1}\,) \egal & E^{\star}\,(\,k_{\,1}^{\star\,\circ}\,,c_{\,1}^{\star\,\circ}\,) \plus \pd{E^{\star}}{k^{\star}_{\,1}}\,\Bigg|_{k^{\star}_{\,1}=k_{\,1}^{\star\,\circ}}\,\bigg(\,k^{\star}_{\,1}\moins k_{\,1}^{\star\,\circ}\,\bigg) \plus \\ & \pd{E^{\star}}{c^{\star}_{\,1}}\,\Bigg|_{c^{\star}_{\,1}=c_{\,1}^{\star\,\circ}}\,\bigg(\,c^{\star}_{\,1}\moins c_{\,1}^{\star\,\circ}\,\bigg) \nonumber \plus 
\frac{1}{2}\,\dfrac{\partial^{2}{E^{\star}}}{\partial{k^{\star\,2}_{\,1}}}\,\Bigg|_{k^{\star}_{\,1}=k_{\,1}^{\star\,\circ}}\,\bigg(\,k^{\star}_{\,1}\moins k_{\,1}^{\star\,\circ}\,\bigg)^2 \plus \\ \nonumber &
\frac{1}{2}\,\dfrac{\partial^{2}{E^{\star}}}{\partial{c^{\star\,2}_{\,1}}}\,\Bigg|_{c^{\star}_{\,1}=c_{\,1}^{\star\,\circ}}\,\bigg(\,c^{\star}_{\,1}\moins c_{\,1}^{\star\,\circ}\,\bigg)^2 \plus
\frac{1}{2}\,\dfrac{\partial^{2}{E^{\star}}}{\partial{k^{\star}_{\,1}\,\partial{c^{\star}_{\,1}}}}\,\Bigg|_{k^{\star}_{\,1}=k_{\,1}^{\star\,\circ}\atop c^{\star}_{\,1}=c_{\,1}^{\star\,\circ}}\,\bigg(\,k^{\star}_{\,1}\moins k_{\,1}^{\star\,\circ}\,\bigg)\,\bigg(\,c^{\star}_{\,1}\moins c_{\,1}^{\star\,\circ}\,\bigg)\plus  \\ \nonumber &
\mathcal{O}\,\Biggl(\,\mathrm{max}\bigg\{\bigg(\,k^{\star}_{\,1}\moins k_{\,1}^{\star\,\circ}\,\bigg)^3\,,\bigg(\,c^{\star}_{\,1}\moins c_{\,1}^{\star\,\circ}\,\bigg)^3\,\bigg\}\,\Biggr)\,. \nonumber
\end{align}

\end{answer}
\newpage
\begin{answer}
\section{Numerical model equations of the temperature and sensitivity coefficients.}
\label{sec:DF_pt2}
For the nonlinear case, the solution is calculated with the following expression:
\begin{align}
\label{eq:num_solu} 
u_{\,j}^{\,n+1} \egal \nu_{\,1}\cdot u_{\,j+1}^{\,n} \plus \nu_{\,2}\cdot u_{\,j-1}^{\,n} \plus \nu_{\,3}\cdot u_{\,j}^{\,n-1}\,, 
\end{align}
where
\begin{align*}
&\nu_{\,1} \egal \frac{\lambda_{\,1}}{\lambda_{\,0} \plus \lambda_{\,3}}\,,
&&\nu_{\,2} \egal \frac{\lambda_{\,2}}{\lambda_{\,0} \plus \lambda_{\,3}}\,,
&&&\nu_{\,3} \egal \frac{\lambda_{\,0} \moins \lambda_{\,3}}{\lambda_{\,0} \plus \lambda_{\,3}}\,,
\end{align*}
and
\begin{align*}
&\lambda_{\,0} \ \eqdef \ 1\,, 
&&\lambda_{\,3} \ \eqdef \ \frac{\Delta t^{\star}}{\Delta x^{\star\,2}}\,\frac{\mathrm{Fo}}{c^{\star}_{\,j}}\,\bigl(\,k^{\star}_{j\plus\frac{1}{2}}\plus k^{\star}_{j\moins\frac{1}{2}}\,\bigr)\,\\
&\lambda_{\,1} \ \eqdef \ \frac{2\,\Delta t^{\star}}{\Delta x^{\star\,2}}\,\frac{\mathrm{Fo}}{c^{\star}_{\,j}}\,k^{\star}_{j\plus\frac{1}{2}}\,, 
&& \lambda_{\,2} \ \eqdef \ \frac{2\,\Delta t^{\star}}{\Delta x^{\star\,2}}\,\frac{\mathrm{Fo}}{c^{\star}_{\,j}}\,k^{\star}_{j\moins\frac{1}{2}}\,. 
\end{align*}
The nonlinear coefficients are approximated by:
\begin{align}
k^{\star}_{j\,\pm\,\frac{1}{2}} \egal k^{\star}\,\biggl(\,\frac{x^{\star}_{\,j}\,+\,x^{\star}_{\,j\,\pm\,1}}{2}\,\biggr)\,.
\end{align}

One may obtain the numerical scheme for computing sensitivity coefficient $X_{\,k^{\star}_{\,1}}$  of the model output $u\,(\,x^{\star}\,,t^{\star}\,)$ with respect to parameter $k^{\star}_{\,1}$, by partially differentiating each term of Eq.~\eqref{eq:num_solu}:
\begin{align}
X_{\,k^{\star}_{\,1_{\,j}}}^{\,n+1} \egal & \nu_{\,1} \cdot X_{\,k^{\star}_{\,1_{\,j+1}}}^{\,n} \plus \nu_{\,2} \cdot X_{\,k^{\star}_{\,1_{\,j-1}}}^{\,n} \plus \nu_{\,3} \cdot X_{\,k^{\star}_{\,1_{\,j}}}^{\,n-1} \plus \\
&{} \pd{\nu_{\,1}}{k^{\star}_{\,1}} \cdot u_{\,j+1}^{\,n} \plus \pd{\nu_{\,2}}{k^{\star}_{\,1}} \cdot u_{\,j-1}^{\,n} \plus \pd{\nu_{\,3}}{k^{\star}_{\,1}} \cdot u_{\,j}^{\,n-1}\,. \nonumber
\end{align}
Similarly,the  expression for sensitivity coefficient $X_{\,c^{\star}_{\,1}}$ is as follows:
\begin{align}
X_{\,c^{\star}_{\,1_{\,j}}}^{\,n+1} \egal & \nu_{\,1} \cdot X_{\,c^{\star}_{\,1_{\,j+1}}}^{\,n} \plus \nu_{\,2} \cdot X_{\,c^{\star}_{\,1_{\,j-1}}}^{\,n} \plus \nu_{\,3} \cdot X_{\,c^{\star}_{\,1_{\,j}}}^{\,n-1} \plus \\
& \pd{\nu_{\,1}}{c^{\star}_{\,1}} \cdot u_{\,j+1}^{\,n} \plus \pd{\nu_{\,2}}{c^{\star}_{\,1}} \cdot u_{\,j-1}^{\,n} \plus \pd{\nu_{\,3}}{c^{\star}_{\,1}} \cdot u_{\,j}^{\,n-1}\,. \nonumber
\end{align}

Second order partial differentiation of Eq.~\eqref{eq:num_solu} with respect to a parameter $k^{\star}_{\,1}$ results as the numerical scheme for $X_{\,k^{\star}_{\,1}\,k^{\star}_{\,1}}$:
\begin{align}
X_{\,k^{\star}_{\,1}\,k^{\star}_{\,1_{\,j}}}^{\,n+1} \egal 
& \nu_{\,1} \cdot X_{\,k^{\star}_{\,1}\,k^{\star}_{\,1_{\,j+1}}}^{\,n} 
\plus \nu_{\,2} \cdot X_{\,k^{\star}_{\,1}\,k^{\star}_{\,1_{\,j-1}}}^{\,n} 
\plus \nu_{\,3} \cdot X_{\,k^{\star}_{\,1}\,k^{\star}_{\,1_{\,j}}}^{\,n-1} \plus \\
& 2 \cdot \pd{\nu_{\,1}}{k^{\star}_{\,1}} \cdot X_{\,k^{\star}_{\,1_{\,j+1}}}^{\,n} 
\plus 2 \cdot \pd{\nu_{\,2}}{k^{\star}_{\,1}} \cdot X_{\,k^{\star}_{\,1_{\,j-1}}}^{\,n} 
\plus 2 \cdot \pd{\nu_{\,3}}{k^{\star}_{\,1}} \cdot X_{\,k^{\star}_{\,1_{\,j}}}^{\,n-1} \plus \nonumber \\
& \frac{\partial^2{\nu_{\,1}}}{\partial{k^{\star\,2}_{\,1}}} \cdot u_{\,j+1}^{\,n} 
\plus \frac{\partial^2{\nu_{\,2}}}{\partial{k^{\star\,2}_{\,1}}} \cdot u_{\,j-1}^{\,n} 
\plus \frac{\partial^2{\nu_{\,3}}}{\partial{k^{\star\,2}_{\,1}}} \cdot u_{\,j}^{\,n-1}\,. \nonumber 
\end{align}
For the sensitivity coefficient $X_{\,c^{\star}_{\,1}\,c^{\star}_{\,1}}\,$, we have:
\begin{align}
X_{\,c^{\star}_{\,1}\,c^{\star}_{\,1_{\,j}}}^{\,n+1} \egal 
& \nu_{\,1} \cdot X_{\,c^{\star}_{\,1}\,c^{\star}_{\,1_{\,j+1}}}^{\,n} 
\plus \nu_{\,2} \cdot X_{\,c^{\star}_{\,1}\,c^{\star}_{\,1_{\,j-1}}}^{\,n} 
\plus \nu_{\,3} \cdot X_{\,c^{\star}_{\,1}\,c^{\star}_{\,1_{\,j}}}^{\,n-1} \plus \\
& 2 \cdot \pd{\nu_{\,1}}{c^{\star}_{\,1}} \cdot X_{\,c^{\star}_{\,1_{\,j+1}}}^{\,n} 
\plus 2 \cdot \pd{\nu_{\,2}}{c^{\star}_{\,1}} \cdot X_{\,c^{\star}_{\,1_{\,j-1}}}^{\,n} 
\plus 2 \cdot \pd{\nu_{\,3}}{c^{\star}_{\,1}} \cdot X_{\,c^{\star}_{\,1_{\,j}}}^{\,n-1} \plus \nonumber \\
& \frac{\partial^2{\nu_{\,1}}}{\partial{c^{\star\,2}_{\,1}}} \cdot u_{\,j+1}^{\,n} 
\plus \frac{\partial^2{\nu_{\,2}}}{\partial{c^{\star\,2}_{\,1}}} \cdot u_{\,j-1}^{\,n} 
\plus \frac{\partial^2{\nu_{\,3}}}{\partial{c^{\star\,2}_{\,1}}} \cdot u_{\,j}^{\,n-1}\,. \nonumber 
\end{align}
The coupled effects between parameters $k^{\star}_{\,1}$ and $c^{\star}_{\,1}$ is given through the numerical scheme for $X_{\,k^{\star}_{\,1}\,c^{\star}_{\,1}}$ and demonstrated below:
\begin{align}
X_{\,k^{\star}_{\,1}\,c^{\star}_{\,1_{\,j}}}^{\,n+1} \egal 
& \nu_{\,1} \cdot X_{\,k^{\star}_{\,1}\,c^{\star}_{\,1_{\,j+1}}}^{\,n} 
\plus \nu_{\,2} \cdot X_{\,k^{\star}_{\,1}\,c^{\star}_{\,1_{\,j-1}}}^{\,n} 
\plus \nu_{\,3} \cdot X_{\,k^{\star}_{\,1}\,c^{\star}_{\,1_{\,j}}}^{\,n-1} \plus \\
&  \pd{\nu_{\,1}}{c^{\star}_{\,1}} \cdot X_{\,k^{\star}_{\,1_{\,j+1}}}^{\,n} 
\plus \pd{\nu_{\,2}}{c^{\star}_{\,1}} \cdot X_{\,k^{\star}_{\,1_{\,j-1}}}^{\,n} 
\plus \pd{\nu_{\,3}}{k^{\star}_{\,1}} \cdot X_{\,k^{\star}_{\,1_{\,j}}}^{\,n-1}\,\plus \nonumber \\
& \pd{\nu_{\,1}}{k^{\star}_{\,1}} \cdot X_{\,c^{\star}_{\,1_{\,j+1}}}^{\,n} 
\plus \pd{\nu_{\,2}}{k^{\star}_{\,1}} \cdot X_{\,c^{\star}_{\,1_{\,j-1}}}^{\,n} 
\plus \pd{\nu_{\,3}}{k^{\star}_{\,1}} \cdot X_{\,c^{\star}_{\,1_{\,j}}}^{\,n-1}\,\plus \nonumber \\
& \frac{\partial^2{\nu_{\,1}}}{\partial{k^{\star}_{\,1}} \cdot \partial{c^{\star}_{\,1}}} \cdot u_{\,j+1}^{\,n} 
\plus \frac{\partial^2{\nu_{\,2}}}{\partial{k^{\star}_{\,1}} \cdot \partial{c^{\star}_{\,1}}} \cdot u_{\,j-1}^{\,n} 
\plus \frac{\partial^2{\nu_{\,3}}}{\partial{k^{\star}_{\,1}} \cdot \partial{c^{\star}_{\,1}}} \cdot u_{\,j}^{\,n-1}\,.  \nonumber
\end{align}
\end{answer}

\end{document}